\documentclass[twocolumn, twocolappendix, times]{aastex631}

\usepackage{amsmath, amssymb}
\usepackage{graphicx} % Required for inserting images
\usepackage{footnote}
\usepackage{url}
\usepackage{comment}
\usepackage{booktabs}
\usepackage{longtable}
\usepackage{enumitem}

\usepackage{soul} % To allow for the use of \hl.
\usepackage{xcolor}
\usepackage{diagbox}
\usepackage{footmisc}

\usepackage{hyperref}

\newcommand{\dd}{{\rm d}}

\newcommand{\tr}{\mathop{\mathrm{tr}}}
\newcommand{\mat}{\mathbf}
\newcommand{\boldvec}{\boldsymbol}

\newcommand{\half}{\tfrac12}
 
\newcommand{\e}{\mathrm{e}}

\newcommand{\rHU}{\mathrm{HU}}
\newcommand{\Myr}{\mathrm{Myr}}
\newcommand{\pc}{\mathrm{pc}}
\newcommand{\kpc}{\mathrm{kpc}}
\newcommand{\Msun}{M_{\odot}}
\newcommand{\km}{\mathrm{km}}
\newcommand{\s}{\mathrm{s}}

\newcommand{\krios}{{\sf KRIOS}}
\newcommand{\nbody}{{\tt NBODY6++GPU}}
\newcommand{\cmc}{{\tt CMC}}
\newcommand{\cosmic}{{\tt cosmic}}
\newcommand{\galpy}{{\tt galpy}}

\DeclareMathAlphabet\mathbfcal{OMS}{cmsy}{b}{n}

\defcitealias{tep2025}{Paper~I}
\defcitealias{fardal2015}{Fardal15}
\defcitealias{chen2025}{Chen25}
\defcitealias{binney2008}{B\&T}
\defcitealias{giesers2019}{G19}
\defcitealias{Farrah2023a}{F23}

\newcommand\soutpars[1]{\let\helpcmd\sout\parhelp#1\par\relax\relax}

\begin{document}

\title{Modeling Globular Cluster Stellar Streams with a Basis-Expansion $N$-body Code}

\correspondingauthor{Brian T.~Cook}
\email{btcook@unc.edu}

\author[0000-0003-0341-6928]{Brian T.~Cook}
\affiliation{\unc}

\author[0009-0002-8012-4048]{Kerwann~Tep}
\affiliation{\unc}
\affiliation{\strasbourg}

\author[0000-0003-4175-8881]{Carl L.~Rodriguez}
\affiliation{\unc}

\author{Leah English}
\affiliation{\iit}

\author[0000-0003-2539-8206]{Tjitske~Starkenburg}
\affiliation{\northwestern}
\affiliation{\ciera}
\affiliation{\skai}

\author[0000-0003-3939-3297]{Robyn~Sanderson}
\affiliation{\upenn}

\author[0000-0002-9660-9085]{Newlin~C.~Weatherford}
\affiliation{\carnegie}

\author[0000-0003-0256-5446]{Sarah~Pearson}
\affiliation{\affNBI}
\affiliation{\dtu}

\author[0000-0001-5214-8822]{Nondh~Panithanpaisal}
\affiliation{\carnegie}
\affiliation{\tapir}

\newcommand{\unc}{Department of Physics and Astronomy, University of North Carolina at Chapel Hill, 120 E. Cameron Ave, Chapel Hill, NC, 27514, USA}
\newcommand{\strasbourg}{Observatoire Astronomique de Strasbourg, UMR 7550, 11 rue de l'Université, Strasbourg 67000, France}
\newcommand{\upenn}{Department of Physics \& Astronomy, University of Pennsylvania, 209 S 33rd St, Philadelphia, PA 19104, USA}
\newcommand{\cca}{Center for Computational Astrophysics, Flatiron Institute, 162 5th Ave, New York, NY 10010, USA}
\newcommand{\carnegie}{Observatories of the Carnegie Institution for Science, 813 Santa Barbara St, Pasadena, CA 91101, USA}
\newcommand{\tapir}{TAPIR, Mailcode 350-17, California Institute of Technology, Pasadena, CA 91125, USA}
\newcommand{\affNBI}{DARK, Niels Bohr Institute, University of Copenhagen, Jagtvej 155A, 2200 Copenhagen, Denmark}
\newcommand{\dtu}{DTU Space, Technical University of Denmark, Elektrovej 327, DK- 2800 Kgs. Lyngby, Denmark}
\newcommand{\cmu}{McWilliams Center for Cosmology, Carnegie Mellon University, 5000 Forbes Ave, Pittsburgh, PA, 15213, USA}
\newcommand{\ciera}{Center for Interdisciplinary Exploration and Research in Astrophysics (CIERA), Northwestern University, 1800 Sherman Ave, Evanston IL 60201, USA}
\newcommand{\northwestern}{Department of Physics and Astronomy, Northwestern University, 2145 Sheridan Rd, Evanston IL 60208, USA}
\newcommand{\skai}{NSF-Simons AI Institute for the Sky (SkAI), 172 E. Chestnut St., Chicago, IL 60611, USA}
\newcommand{\iit}{Department of Physics, Illinois Institute of Technology, 3105 S Dearborn St, Chicago, IL, 60616, USA}



\begin{abstract}
Globular cluster stellar streams probe galaxy-formation processes and can potentially reveal the distribution of dark matter in galaxies. In many theoretical studies, streams are modeled with particle-spray or direct {$N$-body} codes. But particle-spray methods abstract away the internal dynamics of the progenitor by making strong assumptions about the escape physics, while direct {$N$-body} is prohibitively expensive for realistic ($N\!>\!10^5$) systems. In this paper, we present the stream-modeling capabilities of \krios, a new basis-expansion {$N$-body} code for collisional stellar dynamics, that bridges this runtime vs.~accuracy gap. We show that \krios\ reproduces \nbody\ cluster models, and their associated streams, more accurately than particle spray in a fraction of the \nbody\ wall-clock time. We then compare \krios\ to various particle-spray methods on 10 orbits similar to known Milky Way streams. The morphology and kinematics of these streams most disagree when the progenitor is tightly bound to the host, as these systems are often subject to stronger tidal forces. Finally, we discuss which elements of the progenitor physics are most important for modeling stellar streams and how these might be incorporated into particle-spray methods.
\end{abstract}


%% and \citet commands to identify citations. The citations are

\section{Introduction}
\label{sec:introduction}

There are lingering discrepancies between Local-Group-scale $\Lambda$CDM predictions and observations \citep{bullock2017} that near-field cosmologists are eager to rectify. The Milky Way (MW) and its population of globular clusters (GCs) could provide key insight into these tensions. GCs can either form {\it in situ} as a mode of star formation in galaxies, or can be accreted from other galaxies during mergers \citep{katz2013, belokurov2024}. Both formation channels can help constrain the MW's formation history \citep[e.g.,][]{katz2013, kruijssen2020a, malhan2022, chen2024}. Conversely, the host environment of the MW plays a key role in the dynamical evolution of its GCs. As the Galactic potential changes along a cluster's orbit, the loss of stars to the tidal field can significantly accelerate the destruction of Galactic GCs \citep[e.g.,][]{gnedin1997}. These lost stars go on to create debris such as stellar streams \citep[e.g.,][]{lyndenbell1995} or shells \citep[e.g.,][]{penarrubia2009}, which probe the Galactic potential. Accurate, long-term models of GC evolution that include interactions with their host galaxy \citep[e.g.,][]{riley2020} would thus improve both galaxy-formation \citep[e.g.,][]{cole2000, bullock2005} and GC studies.

%The {\it Gaia} data \citep{gaia2016} has dramatically expanded the scope of this research area \citep{deason2024, hunt2025, bonaca2025}. These faint systems are identifiable if the constituent positions or velocities are known with respect to the host reference frame \citep[e.g.,][]{aganze2025}. The energy and angular momentum component along the host's symmetry axis are integrals of the escapers' motion \citep{bonaca2021, nibauer2022} if the host potential is sufficiently static and axisymmetric \citep[e.g.,][]{brooks2024}. This establishes the importance of modeling tidal stripping correctly, as it affects the initial conditions of the stream stars' coordinates in the host's phase space.

\begin{figure*}
    \centering\includegraphics[width=\linewidth]{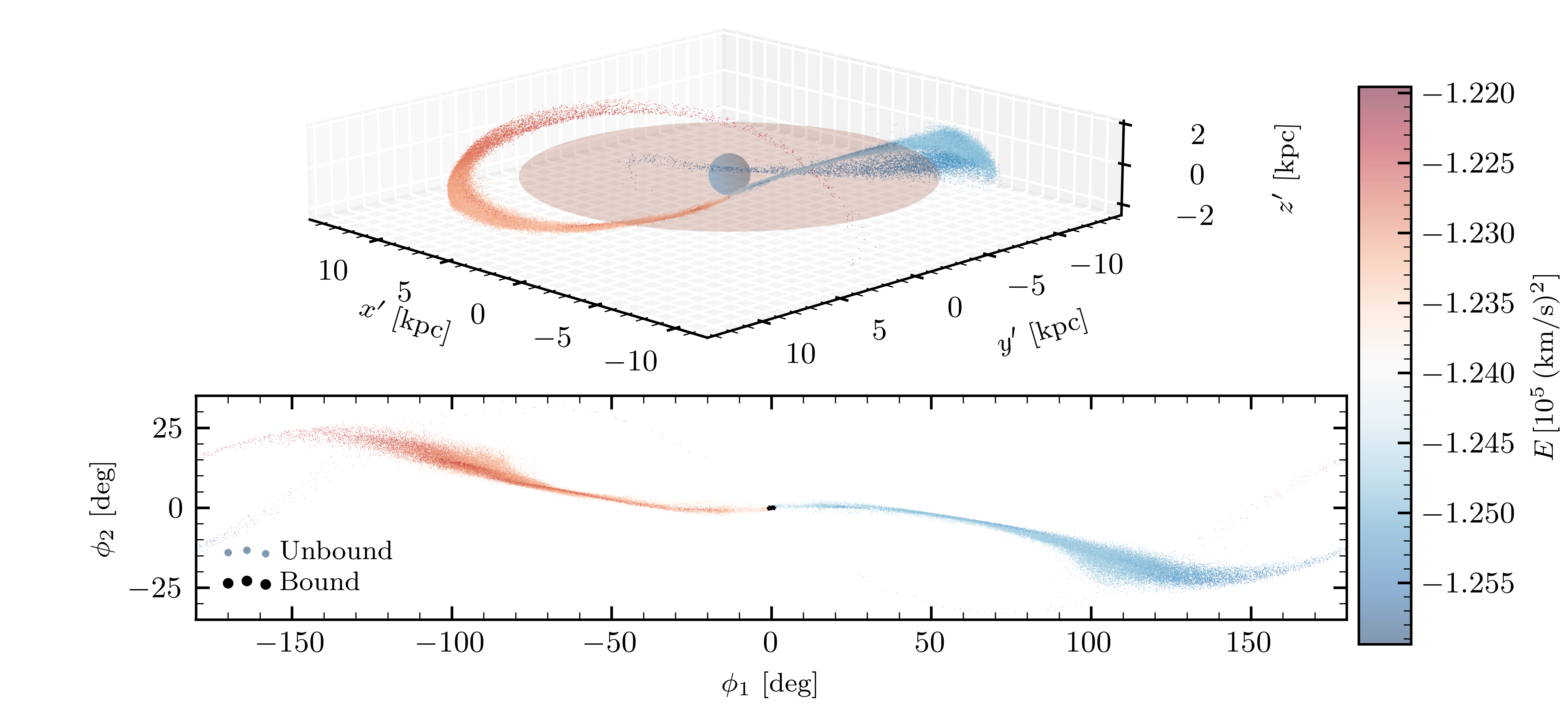}
    \caption{A model produced by the \krios\ hybrid {$N$-body} code (Orbit 3, Table~\ref{tab:orbit_info}), where we see both stream-like and shell-like features at ${t\!=\!5\,{\rm Gyr}}$. The unbound particles are color coded by their energy with respect to the host $E$, which is mostly conserved for escapers modulo small perturbations from the cluster potential. The bound particles (Section~\ref{subsec:non_rf_subsec}) are shown in black. {\it Top panel:} The stream shown in the MW's reference frame, with a bulge and disk added for illustrative purposes. {\it Bottom panel:} The stream shown in the great-circle reference frame for an observer stationed at the Galactic center. The angle $\phi_{1}$ subtends an element of the stream track and $\phi_{2}$ subtends the angle out of the progenitor's instantaneous orbital plane (see Figure~\ref{fig:reference_frames} for the axis definitions). The stream progenitor (i.e., the cluster) is at ${\phi_{1}\!=\!\phi_{2}\!=\!0^{\circ}}$. Both the leading (blue) and trailing (red) tails contain epicyclic density fluctuations \citep{kupper2010} and energy feathering \citep{amorisco2015}, both well-known features of streams in axisymmetric potentials \citep{bonaca2025}. The density fluctuations are most clear near the progenitor; see Figure~\ref{fig:krios_ps_comp_3} for a direct comparison.}
    \label{fig:example_krios_snapshot}
\end{figure*}

The dynamically cold streams from GCs, in particular, are highly sensitive to perturbations from substructure within the host galaxy \citep[see][for a review]{bonaca2025}, including both baryonic substructure \citep[e.g.,][]{pearson2017, amorisco2017} and dark matter subhalos \citep[e.g.,][]{yoon2011, carlberg2012, erkal2016}. The presence of dark subhalos, which are clumps of dark matter too small to form stars, in galaxies is a key prediction of $\Lambda$CDM and a potential clue to the nature of dark matter itself \citep[e.g][]{spergel2000,bullock2017}.  Several GC streams in the MW, including the GD-1 stream \citep{grillmair2006}, display features that may have been the result of an encounter with a subhalo \citep[e.g.,][]{bonaca2019}. The kinematic temperature of GC streams is also thought to depend on the subhalo mass-concentration relation \citep{penarrubia2006, bovy2017, carlberg2025}. However, substructure in GC streams can also be the result of the GC progenitor's properties and internal evolution \citep[e.g.,][]{weatherford2025}, interactions with baryonic substructure in the host galaxy, or larger perturbations in the host potential itself \citep[e.g.,][]{weerasooriya2025, guillaume2026}. Distinguishing which features of a stream are caused by its progenitor from those caused by the baryonic or DM substructure of the host galaxy is crucial to successfully probing dark matter physics with stellar streams.

The first approach for modeling dense GCs is often a direct $N$-body integration, where the forces between pairs of stars are directly summed every timestep. 
The gold standard for this is the \nbody\ code \citep{aarseth1999, wang2015}. A variation of the same code allows the GC to be coupled to an external tidal field as well \citep[{\tt nbody6tt},][]{renaud2015}, following both the progenitor and resultant stream self-consistently within a time-independent host potential. But the direct summation of forces required to solve $N$ coupled ordinary differential equations (ODEs) requires $\mathcal{O}(N^{2})$ operations, making this technique computationally expensive. Despite significant algorithmic enhancements (e.g.,~ regularization of close encounters and block timesteps) and hardware acceleration (e.g.,~GPUs), direct summation still has not been used to model a GC with a density and initial $N$ typical of Galactic GCs over a full Hubble time. Direct $N$-body methods are thus ill-suited to generating the large grids of model streams necessary to explore the dynamics within and external to GCs that affect stream properties relevant to probing Galactic substructure.
%\nw{[I rephrased this to emphasize the ``impracticality" and clarify the issue is not with modeling a GC with $N>10^5$ but for large-parameter space studies relevant to GCs. The original version is below.]}
%GC simulations with $N\!\gtrsim\!10^5$ remain computationally impractical with direct $N$-body.

%To sidestep this computational bottleneck, approximate techniques are often used when studying collisionless stellar streams.
%\nw{[Eliminated passive voice and shortened since collisionless is unnecessarily specific; approximate techniques dominate ALL stellar stream studies]}
To hasten computation, stream models rely on more approximate methods. Particle-spray codes \citep[e.g.,][]{kupper2012, gibbons2014, fardal2015, erkal2019, grondin2022, chen2025, palau2025} generate stream models far more rapidly than direct {$N$-body} by approximating the escape process, directly injecting stars near the Lagrange points of the GCs as they orbit their host. For example, \cite[][hereafter Fardal15]{fardal2015} and \cite[][hereafter Chen25]{chen2025} sample the rate and kinematic properties of escaping stars from distributions tuned to mimic $N$-body simulations of tidal disruption \citep{dehnen2000, stadel2001, dehnen2002}. While these methods excel in generating large ensembles of models, they assume that the details of the tidal-stripping process do not substantially affect stream properties such as its density and velocity dispersion profiles. Furthermore, most particle-spray studies in the literature have assumed a fixed progenitor \citep{grondin2022, kuzma2025} or no progenitor at all \citep[otherwise known as orbit fitting, e.g.,][]{koposov2010, malhan2019b}, both of which limit their ability to resolve realistic stream production over long timescales where the cluster has a time-dependent size and mass.

What is needed is an approach that can accurately model an evolving, collisional GC and its ejecta in a fraction of the direct $N$-body wall-clock time. In numerical studies of GCs, this is often done with Monte Carlo methods, where the collisional evolution of dense star clusters can be approximated by statistical techniques. Leading codes in this area include {\tt MOCCA} \citep{giersz2013} and \cmc\ \citep{rodriguez2022}, both of which use the \cite{henon1971} method to approximate two-body relaxation \citep{spitzer1987} via effective ``super-encounters'' with a nearby neighbor. The H\'enon method scales as $\mathcal{O}(N\log N)$, satisfying the need for star-by-star GC simulations with $\mathcal{O}(<\!N^{2})$ complexity that can capture the cluster's gravothermal evolution and conserve the integrals of motion. As an example, \cite{rodriguez2016} showed that \cmc\ produces models with ${N\!=\!10^{6}}$ that agree with \nbody\ on important structural and stellar population properties in about a day; by comparison, the same cluster model integrated with \nbody\ required approximately six months of wall-clock time \citep{wang2016}.

However, H\'{e}non-style Monte Carlo codes like \cmc\ assume that the cluster is spherically symmetric, with all the relevant dynamical processes occurring on a relaxation timescale \citep{rodriguez2022}. In reality, the tidal force is anisotropic in the cluster reference frame and can do work on the cluster on a dynamical timescale. The ${\psi_{\rm eff}(\mat{r})\!=\!E_{J}}$ surface through which a bound particle with Jacobi integral ${E_{J}\!=\!{v^{2}\over 2} + \psi_{\rm eff}}$ cannot pass \cite[i.e., the zero-velocity surface,][hereafter B\&T]{binney2008} is not spherically symmetric. Particles escape primarily in the vicinity of the $L_{1}$/$L_{2}$ Lagrange points, especially those whose Jacobi integral is only marginally above the escape threshold \citep[e.g.,][]{fukushige2000, weatherford2024}. Stars ejected at low energies can only escape through openings in the zero-velocity surface near the $L_{1}$/$L_{2}$ Lagrange points, even after the star has become energetically unbound from the system. As such, it typically requires many dynamical times for a star to actually exit the cluster \citep{fukushige2000,weatherford2024}. This description of stellar escape, which comes from the restricted three-body problem, breaks down when the cluster is subject to a varying tidal field (e.g., eccentric orbits, disk shocks). The zero-velocity surface and phase-space properties of $E_{J}\!>\!0$ cluster stars vary on the same timescale as the tidal field, which changes the distribution of escapers. To account for this, most Monte Carlo codes adopt prescriptions that approximate the tidal radius of the cluster \citep[e.g.,][]{giersz2008,Chatterjee2010} and introduce delays to the stripping of particles from the system \citep[e.g.,][]{giersz2013,sollima2014}. These prescriptions must be tuned to direct $N$-body simulations of clusters on specific orbits, and are not necessarily reliable for other orbits in a different, or time-varying, host potential. Furthermore, because escapers have to be removed from \cmc\ and its spherical collisional dynamics before evolving their orbits (collisionlessly) in a realistic asymmetric tidal field, these Monte Carlo methods require significant post-processing to model tidal debris in a Galactic context \citep[e.g.,][]{weatherford2024, weatherford2025, panithanpaisal2025}.

This combination of requirements has motivated the development of \krios, a new {$N$-body} code introduced by
\cite[][hereafter Paper I]{tep2025}.
Figure~\ref{fig:example_krios_snapshot} provides an illustrative \krios\ model of a GC stellar stream in a MW-like host potential. \krios\ integrates each particle in a self-consistent gravitational field \citep[SCF;][]{cluttonbrock1973, hernquist1992, zhao1996, lowing2011, vasiliev2015, fouvry2021} that adapts as the cluster evolves \citepalias[][]{tep2025}, while two-body relaxation is modeled using a 3D version of H\'enon's original effective encounters. The force calculations are $\mathcal{O}(N)$ and parallelizable, rendering complete forward models considerably faster than direct {$N$-body} methods. \krios\ replicates key dynamical properties of non-rotating clusters (such as core collapse in clusters with varying levels of initial velocity anisotropy), as well as rotating clusters (such as the collisional evolution of a cluster's rotation curve, and the emergence of the radial orbit instability in highly anisotropic clusters). See \citetalias[][]{tep2025} for details. Because \krios\ integrates the stellar orbits in a non-spherical cluster on a dynamical timescale, it is well suited to exploring the production of tidal debris from disrupting star clusters in a fraction of the direct $N$-body wall-clock time. 
%\nw{[Uniquely? Other codes are capable of doing the following, too, and even faster. You can rephrase to more precisely state what is unique about \krios.]}

In Section~\ref{sec:methods}, we introduce the scaffolding that helps \krios\ resolve mass loss due to external tidal fields. In Section~\ref{sec:results}, we show comparisons between the streams produced by \krios\ and \nbody\ to validate our method, as well as comparisons to various particle-spray methods, in order to explore the regions of Galactic orbital space where the reliability of the latter breaks down. The discussion (Section~\ref{sec:discussion}) highlights the benefits of using \krios\ design elements in modeling stellar streams, and describes key features relevant to future particle-spray studies

\section{Methods}
\label{sec:methods}

\krios\ decomposes the gravitational interactions in a star cluster into collisionless and collisional components. The collisionless gravitational potential is modeled by expanding the cluster's mass density with a set of basis functions:
\begin{subequations}
\begin{align}
\label{eq:total_potential} \psi_{\rm SCF}(\boldvec{r}) &= \sum_{n\ell m} a_{n \ell m} \, \psi^{(n\ell m)}(\boldvec{r}), \\
\label{eq:total_dens} \rho_{\rm SCF}(\boldvec{r}) &= \sum_{n\ell m} a_{n \ell m} \, \rho^{(n\ell m)}(\boldvec{r}).
\end{align}
\end{subequations}
The basis functions are separable in spherical coordinates:
\begin{subequations}
\begin{align}
\label{eq:psi_nlm} \psi^{(n\ell m)}(\boldvec{r}) &= U_{n}^{\ell}(r)\,Y_{\ell}^{m}(\theta, \phi), \\
\rho^{(n\ell m)}(\boldvec{r}) &= D_{n}^{\ell}(r)\,Y_{\ell}^{m}(\theta,\phi),
\end{align}
\end{subequations}

\noindent where ${n,\,\ell}$ index the radial basis functions ${\{U_{n}^{\ell}(r), D_{n}^{\ell}(r)\}}$ and ${\ell,\,m}$ index the spherical harmonics $Y_{\ell}^{m}(\theta,\phi)$. In the standard self-consistent field approach, $\psi^{(n\ell m)}$ and $\rho^{(n\ell m)}$ are bi-orthogonal sets where each mode constitutes an independent solution to Poisson's equation such that ${\nabla^{2}\psi^{(n\ell m)}\!=\!4\pi G \, \rho^{(n\ell m)}}$. The bi-orthogonality relation can be used to isolate each basis coefficient $a_{n\ell m}$:

\begin{subequations}
\begin{align}
\label{eq:a_nlm} a_{n \ell m} &= -\int \dd \boldvec{r} \,\rho(\boldvec{r}) \, \psi^{\star(n\ell m)}(\boldvec{r}),\\
&\simeq -\sum_k \,m_k \, \psi^{\star(n\ell m)}(\boldvec{r}_k),
\label{eqn:sumanlm}
\end{align}
\end{subequations}

\noindent where $\psi^{(n\ell m)\star}$ is the complex conjugate of $\psi^{(n\ell m)}$ and we have estimated the cluster's mass density from the particle data as ${\rho(\boldvec{r}) \!\simeq\! \sum_{k} m_{k} \, \delta^{3}(\boldvec{r}-\boldvec{r}_{k})}$.

The individual stars are then integrated forward in the global $\psi_{\rm SCF}(\boldvec{r})$ potential. This reduces the $N^2$ calculation of direct $N$-body to $N$ individual, embarrassingly parallel integrations. In general, the greatest computational bottleneck for the SCF method is the large number of modes that are required when the basis functions are not well fit to the cluster's mass density \citep{hernquist1992}. To address this, \krios\ uses a tunable set of basis functions from \cite{zhao1996}, where the zeroth-order mode of the potential has the following functional form:

\begin{equation}
U_{0}^{0}(r)\!\propto\!\frac{1}{\left(1+\left(\frac{r}{b}\right)^{1/\alpha}\right)^{\alpha}},
\label{eqn:u00}
\end{equation}

\noindent \citepalias[Appendix A of][]{tep2025}. \krios\ tunes the exponent $\alpha$ and scale length $b$ such that the true cluster potential is well approximated by the zeroth-order mode of the SCF even as the density profile becomes cuspy near core collapse \citepalias[Figure 9 of][]{tep2025}. This, combined with a filtering procedure that ignores modes where ${a_{n\ell m}\ll a_{000}}$, allows \krios\ to maintain an optimized description of the cluster potential over many relaxation times. See Section 2.2 of \citetalias{tep2025} for details.

Finally, \krios\ models the relaxation of star clusters with the \cite{henon1971} prescription, where the cumulative effect of many two-body encounters are modeled as a single effective ``super-encounter'' between a star and a suitably chosen neighbor. We approximate the local number density of particles, a key ingredient for computing the relaxation timescale \citep[Section~2.1.1 of][]{rodriguez2022}, using ${n_{\rm neigh}\!=\!30}$ neighbors. We then update each particle's 3D velocity based on the relevant two-body super-encounter at each system timestep. These scattering events diffuse energy from the cluster's core to its halo, and have been shown to reproduce the expected gravothermal evolution and core collapse of star clusters over many relaxation times; see \citetalias{tep2025}.  Note that we do not apply the energy conservation scheme from \cite{stodolkiewicz1982} in this study.  This scheme is often used to correct the energy drift observed in Monte Carlo codes \citep[e.g.,][]{rodriguez2022} that arises from sampling new orbital positions and velocities (post two-body relaxation) in the old cluster potential from the previous timestep.  In this study, however, the cluster potential is updated much more frequently (see Section \ref{subsec:non_rf_subsec}).  As a result, our models conserve energy in the host frame of the Galaxy $E$ ($\Delta E/E_0\lesssim 10^{-3}$) and the $z'$ component of the angular momentum $L_{z'}$ ($\Delta L_{z'}/L_{z',0}\lesssim 10^{-2}$) at acceptable levels over the 5~Gyr integrations presented here.

\subsection{\krios\ Reference Frames}
\label{subsec:reference_frames}

\begin{figure*}
    \centering
    \includegraphics[width=0.8\linewidth]{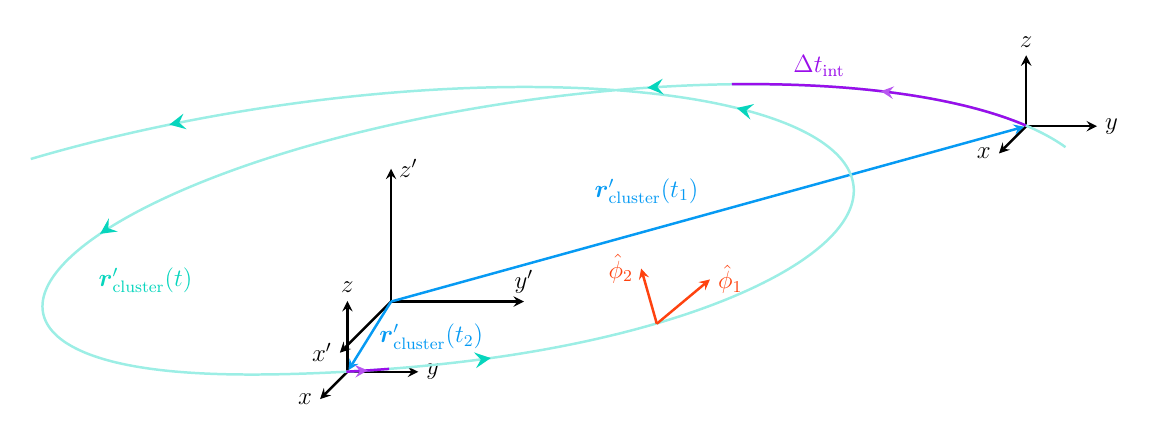}
    \caption{The time-dependent SCF reference frame (black, unprimed) with respect to the fixed host reference frame (black, primed). Integration along the orbit (green) is broken up into substeps (purple) that adapt to the host potential's tidal tensor (Equation~\ref{eq:tidal_tensor}). The unit vectors that define the great-circle reference frame (bottom panel of Figure~\ref{fig:example_krios_snapshot}, red axes) are ${\hat{\phi}_{1}'\!\parallel\!(\boldsymbol{\Omega}_{\rm cluster}'\times\boldvec{r}_{\rm cluster}')}$ and $\hat{\phi}_{2}'\!\parallel\!\boldvec{\Omega}_{\rm cluster}'$, where ${\boldvec{\Omega}_{\rm cluster}'\!\parallel\!\boldvec{r}_{\rm cluster}'\times\boldvec{v}_{\rm cluster}'}$.}
    \label{fig:reference_frames}
\end{figure*}

\krios\ uses accelerating, non-rotating reference frames for the SCF, which mitigates the need to include rotational pseudo-forces into the equations of motion explicitly \citep{renaud2011}. Our framework is illustrated in Figure~\ref{fig:reference_frames}. For a static host potential, we set the fixed inertial frame to coincide with the center of the bulge, disk, and dark matter halo components (the primed coordinates in Figure~\ref{fig:reference_frames}). 

\krios\ has separate reference frames for the host potential and the cluster. The position of the cluster frame (unprimed coordinates, Figure~\ref{fig:reference_frames}) is set to the location that maximizes the $a_{000}$ component of the basis expansion (using a similar procedure for finding the optimal $\alpha$ and $b$ in Equation~\ref{eqn:u00}). Because the SCF basis does not contain any information about the velocity of the system, the velocity for the reference frame is determined by taking a mass-weighted average of the particle velocities in the cluster's core, where ${r_{\rm core} \!=\! \sqrt{\sum_{i} \rho_{\rm SCF}^{2}(\boldvec{r}_{i}) r_{i}^{2} / \sum_{i}\rho_{\rm SCF}^{2}(\boldvec{r}_{i})}}$.\footnote{This is similar to the definition used in traditional $N$-body integrators \citepalias[][]{tep2025} where we substitute the computationally expensive calculation of the full particle density with the mass density of the SCF.} 

In order to return accurate solutions to each particle's equations of motion, the ODE solver must be able to determine each particle's location with respect to both the SCF and host at arbitrary times during each integration (purple segments in Figure~\ref{fig:reference_frames}). As such, each star must have access to the position of the cluster at arbitrary times along its orbit. To that end, we integrate a test particle---located at the cluster's center of mass\footnote{There is typically a small difference between the cluster's center of mass and the optimal location for the SCF; we assume that separation vector, ${\Delta \boldvec{r}'\!\equiv\boldvec{r}_{\rm COM}' - \boldvec{r}_{\rm SCF}'}$, to be constant during each integration.}---in the host potential, and collect position and velocity samples along that orbit.  These samples are then used to create a quintic Hermite spline \citep{finn2004, rehman2014}, to determine the position of the cluster at arbitrary times. Each star can then be integrated forward in the host reference frame individually, without the need to synchronize the center of the cluster after every individual particle integration step.  

%Because the orbit of an individual star can be significantly 
%To that end, \krios\ collects a set of phase-space samples from a test-particle integration in the host potential. Test-particle velocities and accelerations furnish quintic Hermite splines \citep{finn2004, rehman2014} for each of the primed coordinates in Figure~\ref{fig:reference_frames}. With these interpolators, the ODE solver can correctly iterate each particle forward in phase space.

\subsection{Cluster Timescales in an External Potential}
\label{subsec:non_rf_subsec}

For the collisional models in \citetalias{tep2025}, the \krios\ system timestep was chosen to be a fraction of the relaxation time of the cluster.  After the particles were integrated forward in the potential, we performed two-body relaxation and updated the SCF potential using the new positions of the particles, on the assumption that the potential of the cluster could only change significantly on a relaxation timescale.  However, for clusters in realistic tidal fields, the associated mass loss can occur faster than the typical relaxation timescale in the core.  Furthermore, the external potential of the host can do work on the cluster on a dynamical timescale, e.g.~compressive shocks from the Galactic disk \citep[Section 5.2.a of][]{spitzer1987} and during perigalacticon passages like $\boldvec{r}_{\rm cluster}'(t_{2})$ in Figure~\ref{fig:reference_frames}.  To resolve this, we introduce separate timesteps for performing two-body relaxation and updating the cluster potential:

\begin{itemize}
\item The \textbf{system timestep}, $\Delta t_{\rm sys}$, where we perform two-body relaxation and completely recompute the cluster SCF, as in Equation~(14) of \citetalias{tep2025}.
\item The \textbf{integration timestep}, $\Delta t_{\rm int}$, where we update the SCF potential on a timescale set by the changing tidal field experienced by the cluster.  
\end{itemize}

To calculate $\Delta t_{\rm int}$, we evaluate the tidal tensor ($\boldvec{T}$, the Hessian of the potential) at every point along the orbit of the cluster test particle described in Section~\ref{subsec:reference_frames}. Following \cite{grudic2020}, we define the integration timestep as 

\begin{equation}
 \Delta t_{\rm int} = \sqrt{\eta}~\left({1\over 6}||\boldsymbol{T}(\boldvec{r}'(t))||^{2}\right)^{-1/4}, 
 \label{eq:tidal_timestep}
\end{equation}

\noindent where $||\cdot||$ is the Frobenius norm of the tidal tensor, computed with respect to the primed coordinates,

\begin{equation}
T_{ij}(\boldvec{r}'(t))= \left(-\partial_{x_{i}'}\partial_{x_{j}'}\, \psi_{\rm host}\right)\Big|_{\boldvec{r}'=\boldvec{r}'(t)},
 \label{eq:tidal_tensor}
\end{equation}

\noindent and $\eta$ is a free parameter (${\eta\!=\!1/400}$ in this study, chosen from the values discussed in \cite{grudic2020} after validation tests). This allows \krios\ to resolve changes to the cluster mass and potential that occur faster than the two-body relaxation timescale.  \krios\ evaluates Equation~\eqref{eq:tidal_tensor} directly from the various host potential components; see Appendix~\ref{app:static_host_potentials_krios}.

\krios\ updates the SCF in two ways: ``complete" refreshes, where $\alpha$ and $b$ in Equation~\eqref{eqn:u00} are retuned and the SCF is reevaluated from scratch, and ``intermediate" refreshes, where only particles that have become unbound from (or been recaptured by) the cluster are subtracted from (or added to) $a_{n \ell m}$ via the sum in Equation~\eqref{eqn:sumanlm}.  In this study, \krios\ executes a complete SCF refresh every 5 integration timesteps. We use an energy-based analog to the Lagrange radii, i.e., $r_{x}$ is the smallest radius that encloses $x\%$ of the energetically-bound particles\footnote{We say the $i^{\rm th}$ particle is bound to the cluster if, via the standard $N$-body energy calculation in the cluster frame, \begin{equation*}
v_{i}^{2} - \sum_{j\neq i} \frac{G m_{j}}{\left|\boldvec{r}_{i}-\boldvec{r}_{j}\right| } < 0.
\end{equation*}As this calculation is only performed every integration timestep, its contribution to the runtime is marginal.}, to determine which particles are included in the Equation~\eqref{eqn:sumanlm} calculation. We only include particles where ${r \!\leq\! r_{99.9}}$ and refer to $r_{100}$, the smallest radius that encloses all energetically-bound particles, as the ``bound radius" $r_{b}$. %testing showed this to reduce noise in the SCF modes with ${\ell\!\neq0\!}$ from nearly unbound particles far from the SCF's center.  

Finally, the SCF is susceptible to finite-$N$ noise \citep{hernquist1992, vasiliev2015}; this can be reduced by sampling multiple positions along the particles' orbits during each integration.  In \citetalias{tep2025}, this was done with an equal number of samples per system timestep (based only on the relaxation timescale). As we are interested in changes that can occur on a dynamical timescale, here we collect orbital samples based on the cluster's instantaneous dynamical time, 
\begin{align}
N_{{\rm samples}} &= {\rm max}\left(1, \left\lceil n_{{\rm samples, \,} t_{\rm dyn}} \, \times{\Delta t_{\rm int} \over t_{\rm dyn}}\right\rceil\right),
\end{align}
where $n_{{\rm samples,} \, t_{\rm dyn}}$ is the desired number of samples per the instantaneous half-mass dynamical time \citepalias{binney2008}, i.e. 
\begin{align}
\left({t_{\rm dyn} \over 4.3451 \, {\rm Myr}}\right) = \left({M_{\rm cluster} \over  10^{5} \, M_{\odot}}\right)^{-1/2} \, \left({r_{\rm hm} \over  10 \, {\rm pc}}\right)^{3/2}.
\end{align}
In this study, we set ${n_{{\rm samples,} \, t_{\rm dyn}}\!=\!1}$.%, and calculate the meaning we collect at  permits approximating a smooth mass-density field with a mass-weighted sum of Dirac delta functions. %The basis coefficients can then be computed using all particles within the boundedness Lagrange radius $r_{x}$ chosen by the user:
%
%\begin{align}
%\label{eq:ap_update} a_{n \ell m} &\simeq -\hspace{-3mm}\sum_{i, \,r_{i} \leq r_{x}} \hspace{-1mm} {m_{i}\over N_{\rm samples}}\hspace{-2mm} \sum_{j=0}^{N_{\rm samples}-1} \hspace{-2mm} \psi^{(n\ell m)\star}\left(\boldvec{r}_{i}(t_{j})\right),
%\end{align}
%
%where ${N_{\rm samples}\!=\! \sum_{k=0}^{k_{\rm max}-1} N_{{\rm samples, k}}}$. 

\subsection{Sampling Orbits Consistent with MW GCs}
\label{subsec:sampling}

\begin{figure}
    \centering
    \includegraphics[width=0.95\linewidth]{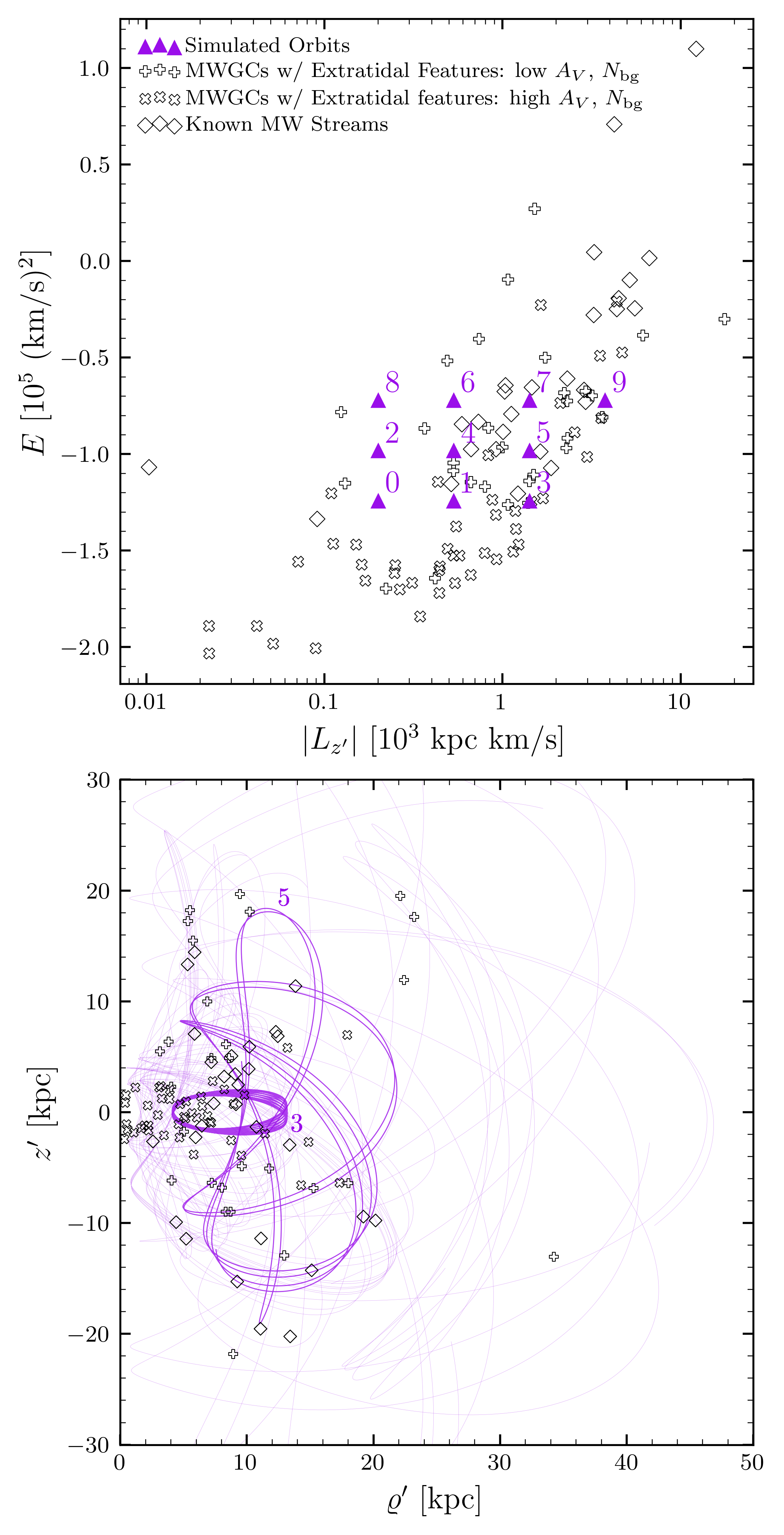}
    \caption{{\it Top panel}: The integrals of motion for the second set of Table~\ref{tab:orbit_info} orbits compared to the known population of MW GCs \citep[MWGCs;][]{harris2010, vasiliev2021, chen2025b} and stellar streams
    \citep[Table~1 of][]{malhan2022}. \cite{chen2025c} classifies MWGCs based on extinction $A_{V}$ and background density $N_{\rm bg}$. {\it Bottom panel}: The sampled orbits compared to the MWGCs and streams in the meridional plane, showing that our sample is biased toward orbits in the MW halo. Orbits 3 (closer to ${z'\!=\!0}$) and 5 are boldened for illustrative purposes, as they are the stream models shown explicitly in Sections~\ref{sec:results} and \ref{sec:discussion}.}
\label{fig:sampled_orbits}
\end{figure}

We must determine a set of orbit initial conditions consistent with the known population of MWGCs \citep[e.g.,][]{harris2010, piatti2020, vasiliev2021, chen2025b} and stellar streams \citep[e.g.,][]{malhan2022, mateu2023, bonaca2025} before highlighting the ways in which \krios\ models disagree with particle spray. The eccentricity and inclination are not necessarily conserved for orbits in non-Keplerian potentials, so we use conserved quantities that are readily accessible from observational data instead. Our set of orbits is shown in Figure~\ref{fig:sampled_orbits}. Recall that the energy $E$ and $z'$-component of the angular momentum $L_{z'}$ are integrals of the motion for a test particle in a static and axisymmetric host potential. The energy can be expressed in terms of the angular momentum and meridional plane coordinates \citepalias[Section 3.2.1 of][Figure~\ref{fig:sampled_orbits}'s bottom panel]{binney2008}:
\begin{subequations}
\begin{align}
\label{eq:energy_eff} E &= {1\over 2}(v_{\varrho'}^{2} + v_{z'}^{2}) + \psi_{\rm eff}(\varrho', z'), \\
\psi_{\rm eff}(\varrho', z') &\equiv \psi_{\rm host}(\varrho', z') + {L_{z'}^{2} \over 2\varrho'^{2}}.
\end{align}
\end{subequations}
Therefore, if we start with a grid of ${(|L_{z'}|, E)}$ values (Figure~\ref{fig:sampled_orbits}'s top panel), we can identify nearby MWGCs in that space if we use {\tt MilkyWayPotential2022} \citep{price-whelan2024} to estimate $\psi_{\rm host}$. We sample the initial Galactocentric distance of each orbit, $r_{\rm init}'$, based on the apogalacticon estimates---$\boldvec{r}_{\rm cluster}'(t_{1})$ in Figure~\ref{fig:reference_frames}---for the nearest MWGCs:
%($\boldvec{r}_{\rm cluster}'(t_{1})$, Figure~\ref{fig:reference_frames}) for the nearest MWGCs:
${r_{\rm init}'\!\sim\!\mathcal{N}(\mu_{r_{\rm apo}'}, \sigma_{r_{\rm apo}'}^{2})}$. This ensures that the cluster is not subject to substantive mass loss at the moment of initialization. The phase-space initial conditions, for a given set of ${(L_{z'},E,r_{\rm init}')}$ values, can then be sampled in the following way:
\begin{subequations}
\begin{align}
\alpha &\sim \mathcal{U}(-\pi/2, \pi/2), \hspace{3mm} \beta \sim \mathcal{U}(0,2\pi) \\
\varrho_{\rm init}' &= r_{\rm init}'  \cos\alpha, \\
\varphi_{\rm init}'&\sim\mathcal{U}(0,2\pi), \\
z_{\rm init}' &= r_{\rm init}'\sin\alpha, \\
v_{\varphi', {\rm init}} &= L_{z'} / \varrho_{\rm init}', \\
v_{\varrho', {\rm init}} &= v \cos\beta, \\
v_{z', {\rm init}} &= v\sin\beta,
\end{align}
\end{subequations}
where ${v \!\equiv\! \sqrt{2\left(E \!-\! \psi_{\rm eff}(\varrho_{\rm init}', z_{\rm init}')\right)}}$. We only create retrograde orbits \citep[i.e., ${L_{z'}\!>\!0}$;][]{bonaca2025} for simplification, which is permissible due to the axisymmetry of the host potential. An orbit with ${t_{\rm end} \!=\! 5 \, {\rm Gyr}}$ and timestep ${\Delta t \!=\! 0.1 \, {\rm Myr}}$ is integrated using {\tt gala} \citep{price-whelan2024} to confirm that ${1 \, {\rm kpc} \!\leq \!r_{\rm peri}'}$ and ${10 \, {\rm kpc} \!\leq\! r_{\rm apo}' \!\leq\! 60 \, {\rm kpc}}$; see the bottom panel of Figure~\ref{fig:sampled_orbits}. If there are no ${(\varrho',z')}$ pairs that satisfy ${E\!\geq\!\psi_{\rm eff}(\varrho_{\rm init}',z_{\rm init}')}$ and the ${(r_{\rm peri}', \,r_{\rm apo}')}$ constraints, then the ${(|L_{z'}|, E)}$ pair is discarded.

While our method for sampling orbits starts directly from the integrals of motion, it is still useful to describe orbits in terms of their circularity \citep{abadi2003} and eccentricity, which can be defined in the following way:
\begin{subequations}
\begin{align}
\label{eq:circularity} \varepsilon &\equiv {L_{z'} \over L_{z', {\rm circ}}(E)}, \\
e &\equiv {r_{\rm apo}' - r_{\rm peri}' \over r_{\rm apo}' + r_{\rm peri}'},
\end{align}
\end{subequations}
where ${L_{z',{\rm circ}} \!=\! \varrho'\sqrt{2(E \!-\! \psi_{\rm host}(\varrho',0))}}$ and $\varrho'$ is the radius at which a circular orbit in the disk has energy $E$. The inclination of the cluster's orbital plane with respect to the disk, ${\cos i\!=\!L_{z'}/|\boldvec{L}|}$, is not necessarily constant for a non-Keplerian potential. The properties of the orbits in Figure~\ref{fig:sampled_orbits} can be found in Table~\ref{tab:orbit_info}.

\begin{table*}
    \scriptsize
    \centering
    \caption{The 13 orbits considered in this study. The first set is used for \krios\ validation against \nbody\ in the {\tt MWPotential2014} potential (which was already present in the $N$-body codebase) while the second set are for \krios\ comparisons to particle spray in the more up-to-date {\tt MilkyWayPotential2022} potential. We report each orbit's integrals of motion, circularity $\varepsilon$, eccentricity $e$, mean/standard deviation of the inclination $i$, and for the {\tt MilkyWayPotential2022} orbits the times at which they reach perigalacticon and apogalacticon for the last time before the end of the simulation. We subtract off $\psi_{{\tt MW2014},\,\infty} \neq 0$ (Appendix~\ref{app:static_host_potentials_krios}) when reporting the energies of the {\tt MWPotential2014} orbits such that bound orbits have negative energies. See Table~\ref{tab:exact_phase_space} for the initial phase-space coordinates accurate to eight decimal places.}
    \begin{tabular}{ccccccccccc}
    \toprule
    Orbit ID & $r_{\rm peri}'$ [kpc] & $r_{\rm apo}'$ [kpc] & $t_{\rm last\,peri}$ [Gyr] & $t_{\rm last\,apo}$ [Gyr] & $\varepsilon$ & $e$ & $i$ [deg] & $L_{z'}$ [$10^{3}$ kpc km/s] & $E$ [$10^{5}$ (km/s)$^{2}$] \\
    \midrule
Circular 0 Test & 5.00 & 5.00 & -- & -- & 1.00 & 0.00 & $0.0\!\pm\!0.0$ & 1.13 & -1.30 \\
Circular 1 Test & 20.00 & 20.00 & -- & -- & 1.00 & 0.00 & $0.0\!\pm\!0.0$ & 3.95 & -0.72 \\
Eccentric Test & 20.00 & 5.00 & -- & -- & 0.20 & 0.60 & $79.27\!\pm\!0.68$ & 0.33 & -0.87 \\
\midrule
0 & 1.55 & 14.25 & 4.87 & 4.95 & 0.10 & 0.80 & $75.00\!\pm\!3.45$ & 0.20 & -1.24 \\
1 & 2.56 & 13.83 & 4.89 & 4.98 & 0.26 & 0.69 & $60.76\!\pm\!3.68$ & 0.53 & -1.24 \\
2 & 3.88 & 25.76 & 4.69 & 4.85 & 0.06 & 0.74 & $82.87\!\pm\!0.45$ & 0.20 & -0.98 \\
3 & 4.10 & 13.20 & 4.83 & 4.91 & 0.91 & 0.53 & $18.20\!\pm\!3.29$ & 1.41 & -1.24 \\
4 & 4.85 & 25.18 & 4.79 & 4.95 & 0.17 & 0.68 & $73.87\!\pm\!0.70$ & 0.53 & -0.98 \\
5 & 9.47 & 22.03 & 4.74 & 4.91 & 0.47 & 0.40 & $60.63\!\pm\!0.48$ & 1.41 & -0.98 \\
6 & 11.72 & 46.87 & 4.59 & 4.92 & 0.09 & 0.60 & $82.63\!\pm\!0.05$ & 0.53 & -0.72 \\
7 & 12.50 & 46.36 & 4.97 & 4.64 & 0.23 & 0.58 & $70.88\!\pm\!0.11$ & 1.41 & -0.72 \\
8 & 18.49 & 41.90 & 4.43 & 4.77 & 0.04 & 0.39 & $87.86\!\pm\!0.01$ & 0.20 & -0.72 \\
9 & 28.68 & 32.88 & 4.90 & 4.56 & 0.61 & 0.07 & $52.74\!\pm\!0.10$ & 3.76 & -0.72 \\
    \bottomrule
    \end{tabular}
    \label{tab:orbit_info}
\end{table*}
 
\section{Results}
\label{sec:results}
%\nw{[Topic / introductory sentence? This section starts rather abruptly, especially since the initial content is still about model setup as opposed to results.]}
%The $N$ equal-mass point particles of each progenitor cluster are drawn from a \cite{king1966} distribution with concentration parameter $W_{0}\!=\!5$ \citep{baumgardt2003, lamers2010}. The total cluster mass is ${M_{\rm cluster} \!=\! 10^{5} \, M_{\odot}}$ and the half-mass radius is ${r_{\rm hm}\!=\!10 \, {\rm pc}}$. We use initial conditions generated by the {\tt COSMIC} population synthesis code \citep{breivik2020} for each \krios\ and \nbody\ run. We perform ensembles of \krios\ and \nbody\ simulations, where each set of King model initial conditions were generated with a different random seed. 
We now run and compare ensembles of \krios, \nbody, and particle-spray models of stellar streams on several different orbits in our Galaxy. In each ensemble, the stream's progenitor cluster has an initial mass ${M_{\rm cluster} \!=\! 10^{5} \, M_{\odot}}$ and half-mass radius ${r_{\rm hm}\!=\!10 \, {\rm pc}}$. The progenitor cluster in each \krios\ and \nbody\ simulation is initialized with equal-mass point particles sampled---using the {\tt COSMIC} population synthesis code \citep{breivik2020}---from a \cite{king1966} distribution with concentration parameter $W_{0}\!=\!5$ \citep[e.g.,][]{baumgardt2003, lamers2010}. We generate several initial particle distributions for each cluster orbit and evolve each system forward for 5~Gyr. See Appendix~\ref{app:nbody6++} for the technical details of the \nbody\ runs, as well as a description of the necessary corrections to the source files to use the {\tt MWPotential2014} potential \citep{bovy2015}.  

%\sout{For validation (Section~\ref{subsec:validation_with_nbody}), ${N\!=\!5\times10^{4}}$ and ${\psi_{\rm host}\!=\!{\tt MWPotential2014}}$; for our comparisons with particle-spray models (Section~\ref{subsec:comparisons}), ${N\!=\!2\times10^{5}}$ and ${\psi_{\rm host}\!=\!{\tt MilkyWayPotential2022}}$. The \nbody\ validation runs took $(29,38,56)$ hours (see Appendix~\ref{app:nbody6++} for more details), compared to \krios's $(6.6,6.6,21.1)$ hours on 128 CPU cores. The ${N\!=\!2\times10^{5}}$ \krios}
For our comparisons between \nbody\, particle spray, and \krios\ (Section~\ref{subsec:comparisons}), we choose ${N\!=\!5\times10^{4}}$ and ${\psi_{\rm host}\!=\!{\tt MWPotential2014}}$, and compare 1 \krios\ against an ensemble of 10 \nbody\ runs and 10 particle-spray runs (to account for statistical fluctuations in our initial conditions and $N$-body integrations).  For our detailed comparison between \krios\ and particle spray, we choose and ${N\!=\!2\times10^{5}}$ and ${\psi_{\rm host}\!=\!{\tt MilkyWayPotential2022}}$, and compare 2 \krios\ runs to 2 particle-spray runs. Note that we use {\tt MWPotential2014} for our comparisons to \nbody\ (as it was already implemented in that codebase), while our comparisons to particle spray use the more up-to-date {\tt MilkyWayPotential2022}.  

Using 128 CPU cores, the \krios\ runs took $(6.6,6.6,21.1)$~hr versus $(29,38,56)$~hr for \nbody\ (see Appendix~\ref{app:nbody6++} for more details). The \krios\ runs with ${N\!=\!2\times10^{5}}$
took between 5.9 and 22.1~hr on 192 CPU cores, with a median wall-clock time of 11~hr. Including a host potential in \krios\ runs introduces some computational overhead \citepalias[Figure 14 of][]{tep2025}, but \krios\ is still a much faster alternative to direct {$N$-body} methods, and the speed advantage will only improve for larger $N$. Runtime optimizations in \krios\ are ongoing and we anticipate the code being made publicly available in 2027.

\subsection{\krios\ Validation against \nbody\ and Particle Spray}
\label{subsec:validation_with_nbody}

We consider three orbits in the {\tt MWPotential2014} potential for validation: two circular orbits in the Galactic disk with radii ${r_{0}'\!=\!5\,{\rm kpc}}$ and ${r_{0}'\!=\!20\,{\rm kpc}}$ and one eccentric orbit that is misaligned with the disk. The initial tidal radii\footnote{For generic orbits, the tidal radius can be approximated as $r_t=\left(GM_c/\lambda_{1,e}\right)^{1/3}$, where $\lambda_{1,e}= \lambda_1 - 0.5(\lambda_2+\lambda_3)$, $\lambda_1$ is the largest eigenvalue of the tidal tensor (Equation~\ref{eq:tidal_tensor}), and $0.5(\lambda_2+\lambda_3)$ approximates the centrifugal term.  See Appendix C of \cite{pfeffer2018}. \label{rt_footnote}} on these circular orbits are  ${r_{t} \!=\! 33.3\,{\rm pc}}$ and ${r_{t} \!=\! 101.9\,{\rm pc}}$, respectively. We refer to Table~\ref{tab:orbit_info} for the validation orbits' properties, Table~\ref{tab:exact_phase_space} for complete information on the phase-space initial conditions, and Appendix~\ref{app:static_host_potentials_krios} for more details on the host potential.

\begin{figure}
    \centering\includegraphics[width=\linewidth]{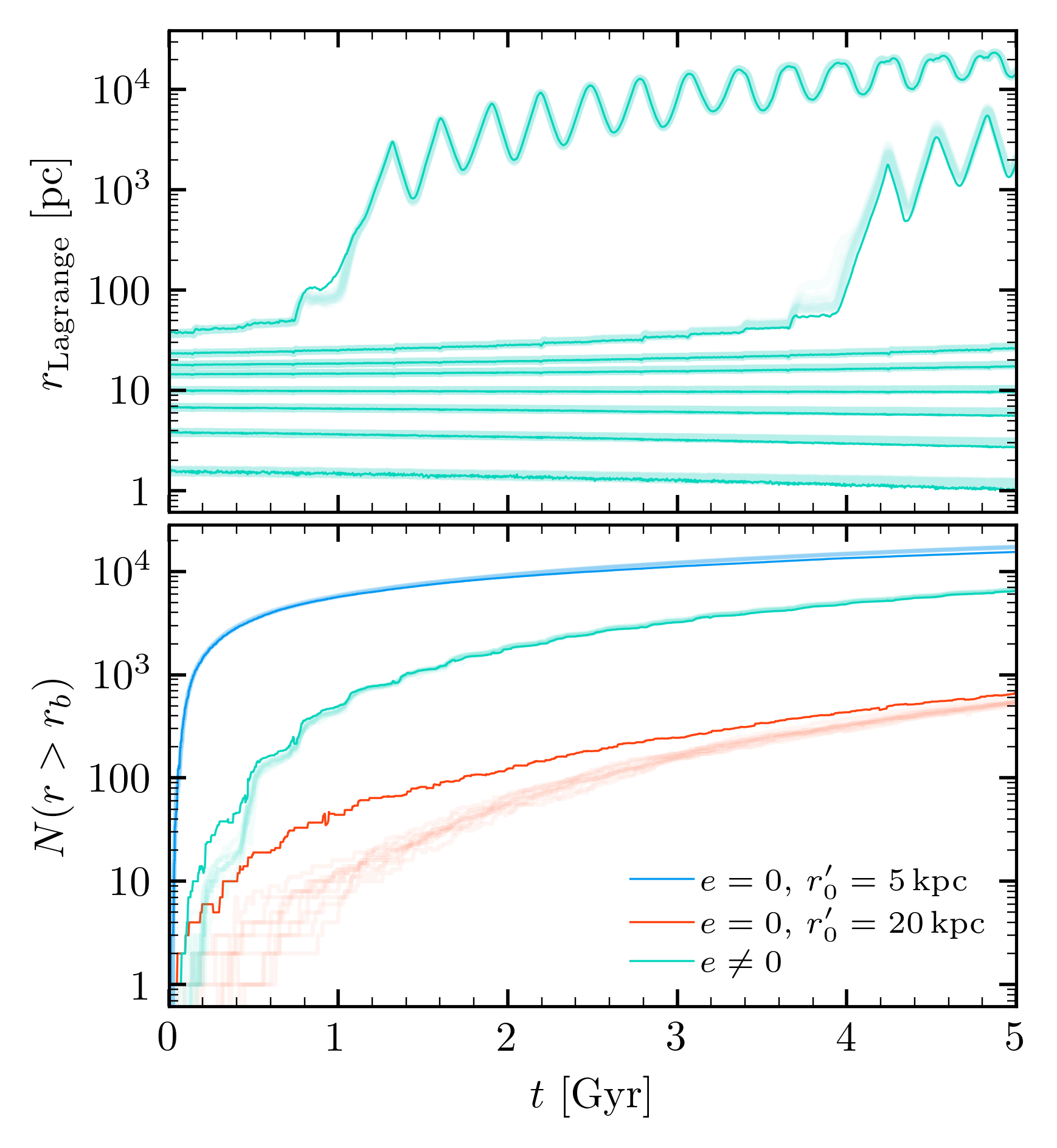}
    \caption{{\it Top panel}: The cluster's Lagrange radii (i.e., the smallest radii that enclose ${\{1\%, 10\%, 30\%, 50\%, 70\%, 80\%, 90\%, 99\%\}}$ of the cluster mass) as a function of time for the eccentric validation orbit. The solid line is the \krios\ result and the shaded region represents the ensemble of \nbody\ runs (10 for each orbit). {\it Bottom panel}: The number of particles outside the bound radius (defined in Section~\ref{subsec:non_rf_subsec}) as a function of time, color coded by orbit.}% The Coulomb logarithm parameter $\gamma\!=\!0.6$ is tuned such that \krios\ best agrees with \nbody\ on the eccentric orbit.} 
    \label{fig:cluster_validation}
\end{figure}

\begin{figure*}
\centering
\includegraphics[width=\linewidth]{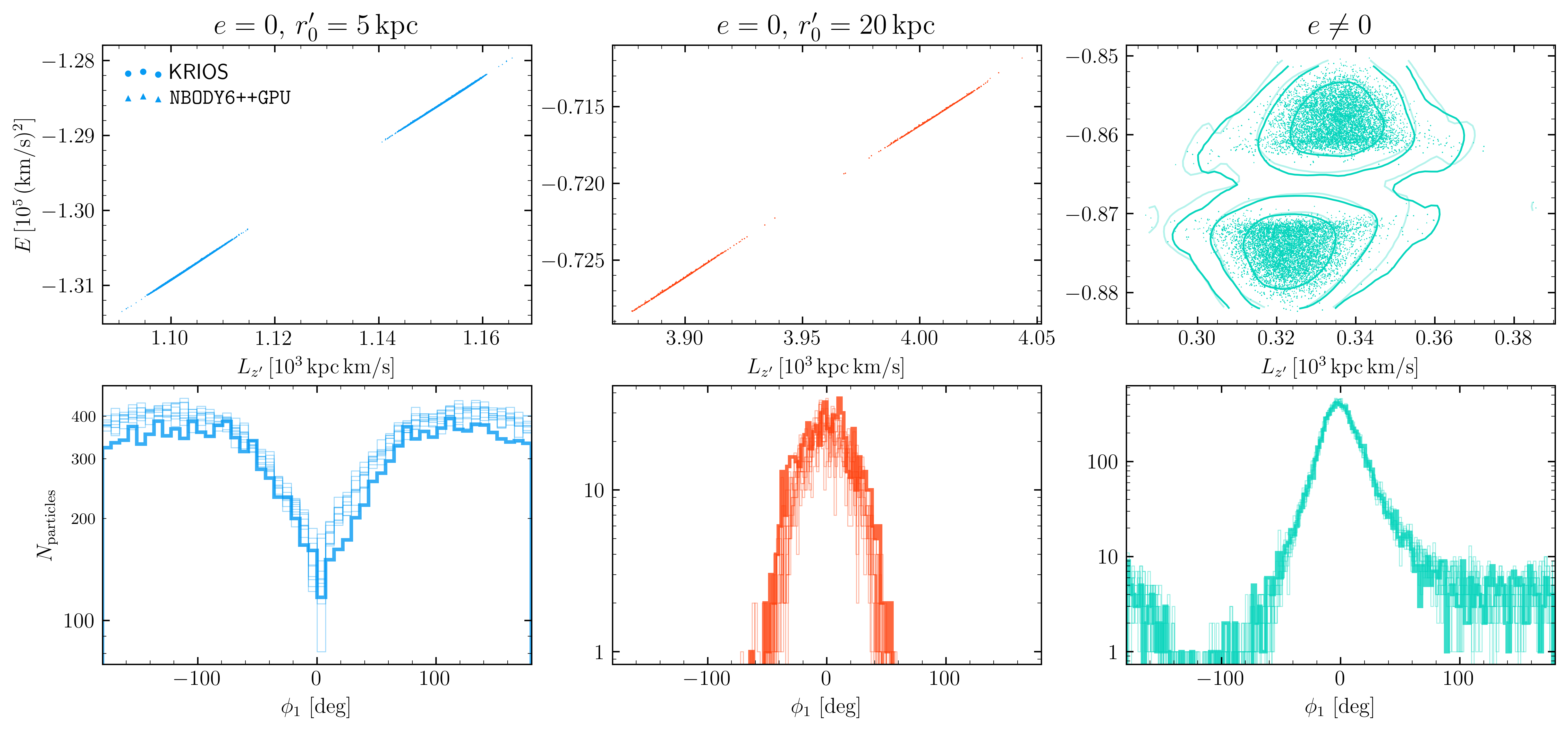}
    \caption{Integrals-of-motion and stream-track information from the last snapshot of the validation models. {\it First row}: a scatter plot of the stream particles' integrals of motion; contours encase $1\sigma$, $2\sigma$, and $3\sigma$ of the particles for the ${e\!\neq\!0}$ test. {\it Second row:} Particle counts along the stream track. The solid lines represent the \krios\ result in each panel, while in the top-right panel the translucent contours represent the entire \nbody\ ensemble. In the bottom row, each of the ensemble models are displayed separately. There are small discrepancies in the circular orbit tests consistent with the mass-loss rates shown in Figure~\ref{fig:cluster_validation}, whereas there is good agreement for the eccentric orbit validation test.}
\label{fig:stream_validation}
\end{figure*}

%\nw{[Bound radius $r_{b}$ hasn't been defined yet; I think the original definition was deleted. Since you do define boundedness, though, and only refer to the bound radius here and in Figure~5, you could just eliminate all references to $r_b$ (including in the vertical axis of Figure 5's bottom panel) and instead only mention the number of unbound bodies, as determined by your boundedness criterion.]} 

We start by examining the cluster's internal evolution: in Figure~\ref{fig:cluster_validation}, we show the cluster's Lagrange radii (top panel) and the number of unbound particles (a direct proxy for mass loss; bottom panel) for each of the validation orbits. There is good agreement in Lagrange radii between the \krios\ simulation and the \nbody\ ensemble for the eccentric orbit. The radii enclosing 90\% and 99\% of the cluster mass, for example, oscillate in accordance with the orbit. The connection between these panels is most clear at around ${t\!=\!1\,{\rm Gyr}}$, where $N(r\!>\!r_{b})\!\approx\!500$ and the 99\% Lagrange radius grows rapidly. The beginning of core collapse is evident, although this is a secondary effect for this cluster on this orbit. The timescale on which cluster orbits at the half-mass radius are modified by two-body relaxation is comparatively large for ${N\!=\!5\!\times\!10^{4}}$ and ${r_{\rm hm}\!=\!10\,{\rm pc}}$ \citep[e.g.,][]{spitzer1987}

\begin{equation}
T_{\rm rlx} = 0.138 \frac{N}{\ln\Lambda}\left(\frac{r_{\rm hm}^3}{GM}\right)^{1/2}~,
\label{eqn:rlx}
\end{equation}

\noindent where ${\ln \Lambda}$ is the well-known Coulomb Logarithm \citepalias[Equation 1.33b,][]{binney2008}. The argument of the logarithm can be written in terms of a free parameter $\gamma$ and the particle number $N$: ${\Lambda\!=\!\gamma N}$ \citep[Chapter 14,][]{heggie2003}, with order-of-magnitude calculations using the virial theorem suggesting $\gamma=0.4$ \citep[][]{spitzer1987}. In practice, many Monte Carlo \citep[e.g.,][]{rodriguez2022} and Fokker-Planck \citep[e.g.,][]{cohn1980, takahashi2012} codes treat $\gamma$ as a free parameter set by tuning to direct $N$-body integrations.  ${\gamma\!=\!0.11}$ has been shown to produce good agreement for isolated Plummer spheres without an external tidal field, while significantly smaller values produce good agreement when considering clusters with realistic stellar masses \citep[e.g., ${\gamma\!=\!0.01}$][]{Rodriguez2018}. For a King model of single-mass particles in a strong MW-like potential, we found \krios\ runs with ${\gamma\!=\!0.6}$ match the mass-loss rate of \nbody\ for the eccentric orbit, and produces good agreement for the two circular orbits as well. %\carl{it might be worth seeing if any of the EMOSAICS papers mention this; can discuss during meeting} \bc{In Pfeffer's paper they only mention the Coulomb logarithm in the context of estimating the cluster's dynamical friction timescale wrt the host, where they use ${\Lambda\!=\!1+M(r_{c})/m_{c}}$} %With these settings, \krios\ is well suited for modeling streams on orbits where the cluster is subject to strong tidal forces during perigalacticon and disk passages.

It is worth noting that these tests have a limited scope due to the approximation of $N$ equal-mass point particles in a static host potential. This setup is also ignorant to the beginning and end of a GC's life cycle, as the initial conditions represent a truncated isothermal sphere. GC dissolution is relevant in this area as several well-known MW streams from GC progenitors have no identifiable progenitor remaining \citep[e.g., GD-1,][]{grillmair2006}. \krios\ is reliant on a large number of GC stars via the mass-density term $\rho(\boldvec{r})$ in the integrand of Equation~\eqref{eq:a_nlm} and is, therefore, poorly suited for simulating the complete dissolution of GCs due to an external tidal field. In future work, we plan on constructing an interface such that data from a \krios\ snapshot can be ported into a \cite{barnes1986}-style tree code or REBOUND \citep{rein2012}.

Having demonstrated good agreement between \krios\ and \nbody\ for the internal evolution of the progenitor in the large-$N$ regime, we now compare the streams themselves. The first row of Figure~\ref{fig:stream_validation} shows the distribution of two of the integrals of motion\footnote{Strictly speaking, an isolated test particle in this host potential conserves energy with respect to the host $E$ and angular momentum $L_{z'}$, while escapers from the GC are subject to small perturbations from the cluster potential.} of the particles outside the bound radius of the cluster (defined in Section~\ref{subsec:non_rf_subsec}). The integrals of motion form a bimodal distribution, with the two peaks corresponding to the leading and trailing tails of the stream. The leading (trailing) tidal tail emanates from near the $L_{1}$ ($L_{2}$) Lagrange point \citep{fukushige2000}, placing the escapers at lower (higher) $L_{z'}$ and $E$. The structure in the integral-of-motion distribution persists even when the stream traces a complete orbit and the tails are harder to distinguish (e.g. second row, first and third columns, Figure~\ref{fig:stream_validation}).  The tidal tails from progenitors on circular orbits follow tracks with ${E\!\propto\!L_{z'}}$
(first and second panels, Figure~\ref{fig:stream_validation}), while the tidal tails from progenitors on eccentric and misaligned orbits present a more complicated picture (third panel, top row, Figure~\ref{fig:stream_validation}). This is attributable to variations in the escape process itself, as the zero-velocity surface and distribution of bound stars vary with the tidal field. The $1\sigma$, $2\sigma$, and $3\sigma$ contours of the \krios\ and \nbody\ distributions are in good agreement here, despite being more diffuse in integrals-of-motion space than their circular-orbit counterparts.

The second row of Figure \ref{fig:stream_validation} shows the distribution of particles along the stream track (i.e. in $\phi_1$) for \krios\ (darkest line) and each of the \nbody\ runs (lighter lines). There is a small but consistent offset in the distribution for the circular orbit, which is due to a difference in the mass-loss rate attributable to our ${\gamma\!=\!0.6}$ setting. The eccentric orbit, which best resembles realistic GC orbits, shows broad agreement at all $\phi_{1}$.

To more quantitatively compare stream models, we compare the probability distribution functions predicted by \krios\ and \nbody\ for ${(L_{z'}, E)}$ and ${(\phi_{1},\phi_{2})}$. The distribution functions are written as $\rho(\boldvec{x})$, where ${\boldvec{x}\!\in\!\boldvec{X}}$ is the space where the distribution function is defined (e.g., the sky plane in the great-circle reference frame) and ${\sum_{\boldvec{x}\in \boldvec{X}}\rho(\boldvec{x})\!=\!1}$. If these distributions were one-dimensional, we would be able to use more familiar statistics like the Kolmogorov-Smirnov test \citep{press2002} or variations in the mean and standard deviation. However, the spaces of interest are two-dimensional and therefore require more bespoke methods. We use two statistics, the Kullback-Leibler divergence \citep[KLD;][]{kullback1951, sanderson2015} and Earth Mover's distance \citep[EMD;][]{cohen1999}, to quantify the disagreement between a stream model (``A") and its \nbody\ counterpart (``NB"):
\begin{subequations}
\begin{align}
\label{eq:kld} {\rm KLD}_{A,{\rm NB }} &=  \sum_{\boldvec{x}\in\boldvec{X}} \rho_{\rm NB}(\boldvec{x}) \ln {\rho_{\rm NB}(\boldvec{x}) \over \rho_{A}(\boldvec{x})} \\
\label{eq:emd} {\rm EMD}_{A,{\rm NB}} &= \min_{f\in\mathcal{F}(\rho_{A},\rho_{\rm NB})} \, \sum_{ij} f_{ij} \, d(\boldvec{x}_{i}, \boldvec{x}_{j}),
\end{align}
\end{subequations}
where $f_{ij}$ is the ${(i,j)}$ element of the flow matrix $f$ that maps $\rho_{A}$ onto $\rho_{\rm NB}$ and $d(\boldvec{x}_{i},\boldvec{x}_{j})$ is the distance between those probability distribution elements. The KLD penalizes imprecision and inaccuracy separately\footnote{The functional form of the KLD between univariate Gaussians can be found in Proof P193 of \cite{soch2020} as an illustrative example.} when evaluating whether $\rho_{A}$ is a good approximation of the ground-truth distribution $\rho_{\rm NB}$. The EMD, on the other hand, measures the ``cost" of manipulating $\rho_{A}$ until it agrees with $\rho_{\rm NB}$. The key difference between these metrics is that the KLD is ignorant to the approximation's cost\footnote{This is especially true in cases where $\rho_{A}$ goes to zero and $\rho_{\rm NB}$ does not, as this causes undefined behavior in the KLD calculation.} in a way that is measured directly by the EMD.
When estimating $\rho$, $f$, and $d$, we take a random subsample from the more-populated model such that both distributions are estimated from the same number of particles. This allows us to use Monte Carlo approximations to the KLD and EMD in equations \ref{eq:kld} and \ref{eq:emd}. Each metric is normalized by the intrinsic scatter between different \nbody\ realizations, namely, ${\rm KLD_{{\rm NB},{\rm NB}}, \,{\rm EMD}_{{\rm NB}, {\rm NB}}}$.
%All \krios\ and particle-spray models presented here are in general agreement with the more expensive \nbody\ runs, with or without subsampling.  No need to say this here

\begin{figure*}
    \centering
    \includegraphics[width=\linewidth]{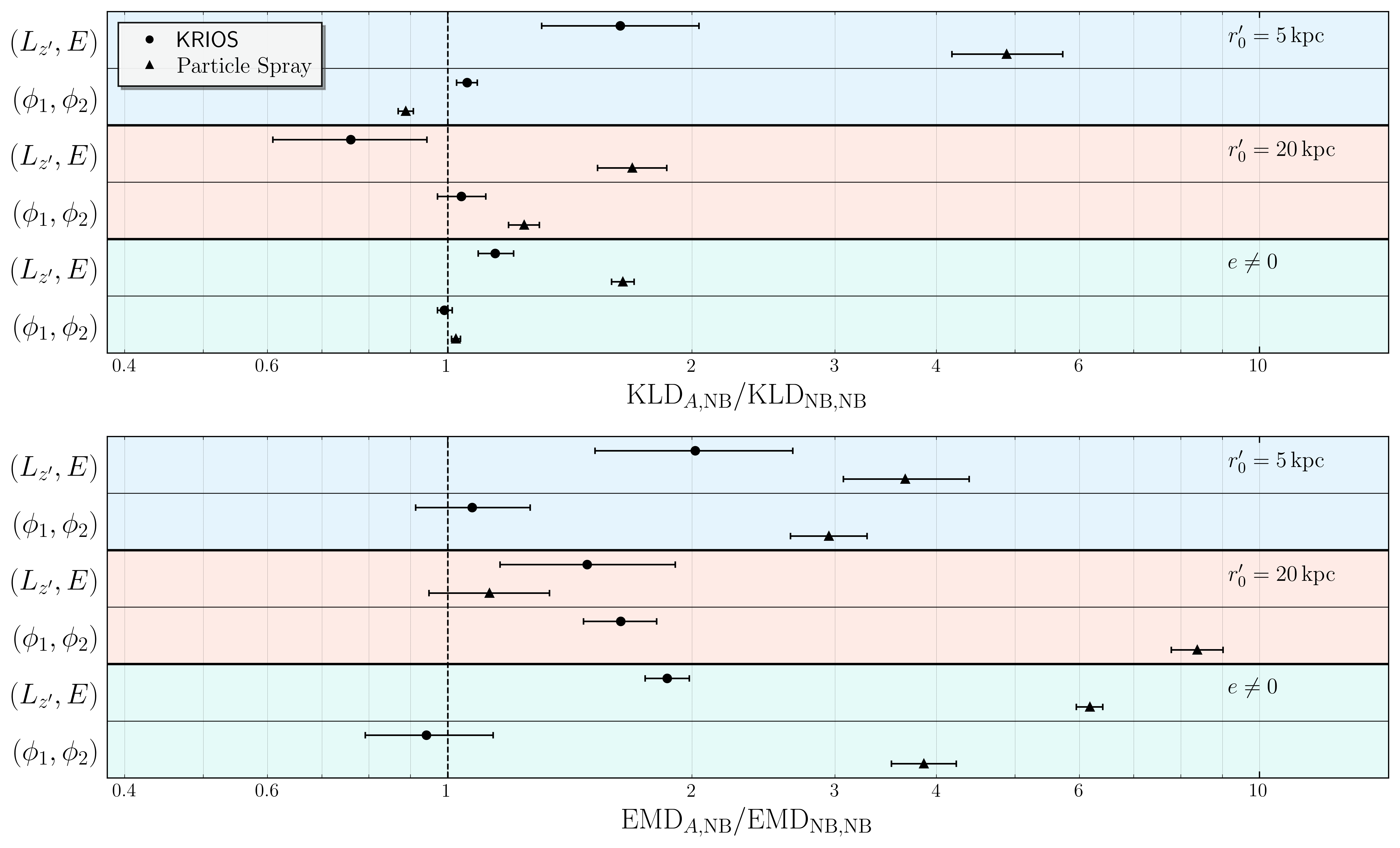}
    \caption{The
    Kullback-Leibler divergences and Earth Mover's distances measured in $(L_{z'},E)$ space and ${(\phi_{1},\phi_{2})}$ space for \krios\ and particle spray against \nbody. The reported values are normalized by the intrinsic scatter between independent \nbody\ realizations (dashed vertical line). 
    We show the 95\% confidence intervals for \krios\ against \nbody\ and particle spray (\citetalias{chen2025} with a Plummer progenitor) against \nbody. Each panel is color coded by orbit using the same scheme as Figures~\ref{fig:cluster_validation} and \ref{fig:stream_validation}: green for ${e\!\neq0}$, red for ${r_{0}\!=\!20\,{\rm kpc}}$, and blue for ${r_{0}\!=\!5\,{\rm kpc}}$. \krios\ and particle-spray scores that are close to one indicate the approximate model is accurate enough to be mistaken for an \nbody\ model in this context. \krios\ scores better than -- or is virtual indistinguishable from -- the ensemble of particle-spray models in all cases.}
    \label{fig:kld_emd_scores}
\end{figure*}

Figure~\ref{fig:kld_emd_scores} shows the 95\% confidence intervals for the KLD and EMD produced by comparing two approximate models against \nbody: we show (i) \krios\ in the first row, and (ii) the particle-spray results in the second row, for which we use a \cite{plummer1911} progenitor and the escaper distribution function from \cite{chen2025}. The comparisons are calculated for each of the three test orbits (shown in different colors), and for the predicted distributions in $(L_{z'},E)$ and ${(\phi_{1},\phi_{2})}$ (sets divided by solid black horizontal lines). The reported values are normalized by the intrinsic scatter between independent \nbody\ realizations.

In almost every case, \krios\ replicates the \nbody\ prediction for the $(L_{z'},E)$ distribution better than particle spray. This is more readily apparent in the EMD scores. In all but two cases---the KLD measured in ${(\phi_{1},\phi_{2})}$ space for $r_{0}'\!=\!5\,{\rm kpc}$ and the EMD measured in $(L_{z'},E)$ space for ${r_{0}'\!=\!20\,{\rm kpc}}$---\krios\ has a better mean score than the corresponding ensemble of particle-spray runs. The improvement is especially large for the predicted distribution of the integrals of motion in the eccentric orbit case, and for the distribution along the stream track in the case of circular orbits.  These results suggests that, overall, \krios\ can reproduce the relevant evolution and stream properties of full \nbody\ runs in a fraction of the time, with a significantly higher fidelity than equivalent particle-spray runs.
%The non-zero {$N$-body} scores represent the statistical scatter that is to be expected between distributions that are drawn from the same stochastic process. 
%More approximate methods (\krios, particle spray) produce acceptable results if their systematic variations with the {$N$-body} ensemble are consistent with this internal variation between {$N$-body} runs.

%More tailored pruning of the particle-spray models (\citetalias{chen2025} distribution function, \cite{plummer1911} progenitor) would require an {\it a priori} mass-loss function $\dot{M}(t)$ \citepalias{chen2025}. \nw{[The prior sentence seems unnecessary in the flow of this paragraph, and also out-of-place because the info about which particle spray models you're using should go earlier (or in the next subsection, but then you'd also have to avoid discussing the comparison to particle-spray codes until the next subsection, too. Otherwise, it would be a bit awkward throughout this subsection to keep referring to particle-spray runs, and showing results from them in Figure~7, when they haven't been described yet.]

\subsection{Statistical and Systematic Variations between \krios\ and Particle-Spray Models}
%\nw{[It's a bit clunky to talk about the particle spray stuff in Figure 7 during the subsection on the comparison to $N$-body rather than here. Changing the subsection titles or shifting things around would probably flow better. It seems like this subsection would be better framed around comparing the statistical variation.]}

\begin{figure}
    \centering
    \includegraphics[width=\linewidth]{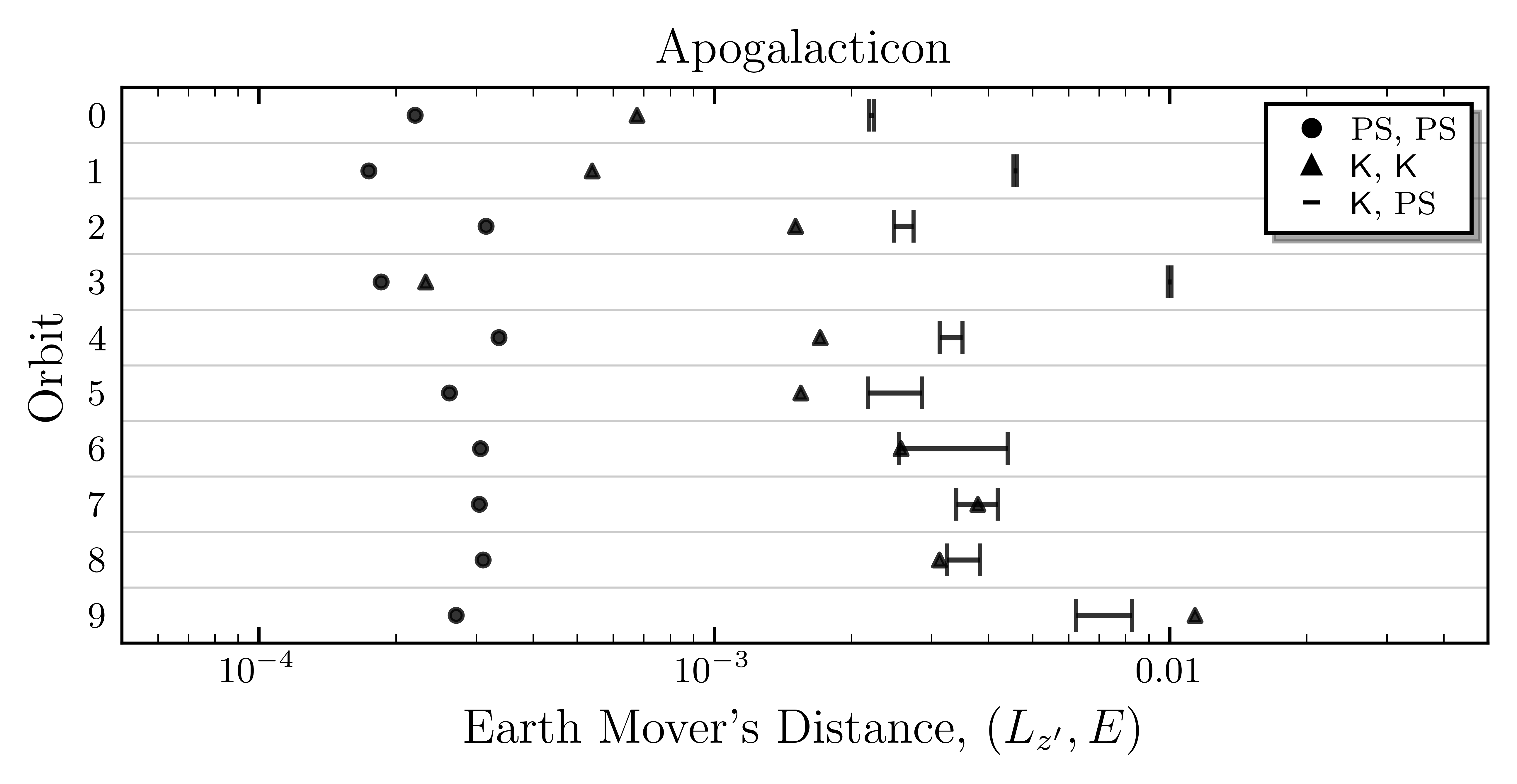}
    \caption{The integrals-of-motion EMD values for each orbit at the last apogalacticon passage. The circle markers represent particle spray's statistical variation (same distribution function, different seeds) and the triangle markers represent \krios's statistical variation (progenitor was initialized with a different random seed for each run). The errorbars show the minimum/maximum of the systematic variation between particle spray and \krios; there are ${2\!\times\!2\!=4}$ measurements.}
    \label{fig:results_emd_comp}
\end{figure}

%\nw{[and the Monte Carlo method?]} \bc{KRIOS isn't a Monte Carlo code it turns out because we do the superencounters in 3D space which is all known}) \nw{[Well, it's an orbit-following Monte Carlo code; there are two sources of ``Monte Carlo-ness'' and KRIOS only eliminates one of them (the orbit sampling part, as you point out). The other part of the Monte Carlo namesake comes from the random deflection angle drawn in each superencounter.]} \carl{Actually we eliminate that as well; since we have the full 3D positions and velocities, we just perform the encounter directly in the plane of the relative velocity.  There is actually no Monte Carlo sampling in the method at all.} 
% \sout{In \krios's case, further stochasticity derives from randomly sampling the deflection angle in each two-body superencounter (the ``Monte Carlo'' nature of an orbit-following Monte Carlo code like \krios).}

Having shown that \krios\ can produce streams with similar properties to those expected from \nbody\, we can now explore a wider range of orbits and compare the predictions from \krios\ to those from various particle-spray methods.  
While it is possible for sets of particle-spray models to match the properties of \nbody\ or \krios\ streams on average, it is not clear that they capture the statistical variation inherent in the $N$-body problem. All three stream-modeling methods exhibit inherent stochasticity due to the random sampling of the particles' initial positions and velocities. But in the case of particle-spray codes, the stochastic ``initial'' conditions are introduced at the time of escape, as random draws from a distribution function of escaper properties.  This ignores the well-known stochastic variation in the cluster evolution itself, which can produce changes in the cluster's mass and radii over time. 

To that end, we compare predictions for the integral of motion distribution made by \krios\ and particle-spray ensembles over the wider range of orbits drawn to span the space of known GC orbits (Section \ref{subsec:sampling}; Figure \ref{fig:sampled_orbits}).  Figure \ref{fig:results_emd_comp} shows the statistical variation between independent \krios\ (triangles) and \citetalias{chen2025} particle-spray (circles) runs, as well as the systematic differences when these models are compared to each other (dashes). The models are compared at apogalacticon and in integrals-of-motion space using the EMD, which we found to be the more sensitive of the two metrics used in Section \ref{subsec:comparisons}. Based on the results of those tests, we take the \krios\ prediction to be closer to the \nbody\ run. Large differences between particle-spray and \krios\ thus highlight which orbits require a more accurate treatment of the escape physics than is achievable with particle-spray methods. 

The EMD between two particle-spray models with different seeds (circle markers) is consistently lower than for \krios. This matches our expectation, as changing the random number generator's seed simply changes the starting point when drawing from a particle-spray model's constant distribution function, while in the \krios\ approach there is additional stochastic variation from the internal cluster dynamics. This suggests that ensembles of particle-spray models may underestimate stream-to-stream variability. On the other hand, an ensemble of \krios\ models is likely needed to confirm that an interesting feature of any particular stream is not an artifact of stochastic variation due to the cluster initial conditions or dynamical evolution. In future applications of {\sf KRIOS} where a stream model is consistent with a dark matter subhalo flyby, for example, such considerations are critical.

As in Figure \ref{fig:kld_emd_scores}, the EMD between \krios\ and particle-spray models (lines with errorbars) shown in Figure \ref{fig:results_emd_comp} should be compared against the statistical variations between \krios\ runs to gauge the difference between the predictions. If the ratio between the two is high, then particle-spray models systematically deviate from \krios\ in a way that calls into question whether particle spray is accurate for these orbits. 

\begin{figure*}
    \centering
    \includegraphics[width=\linewidth]{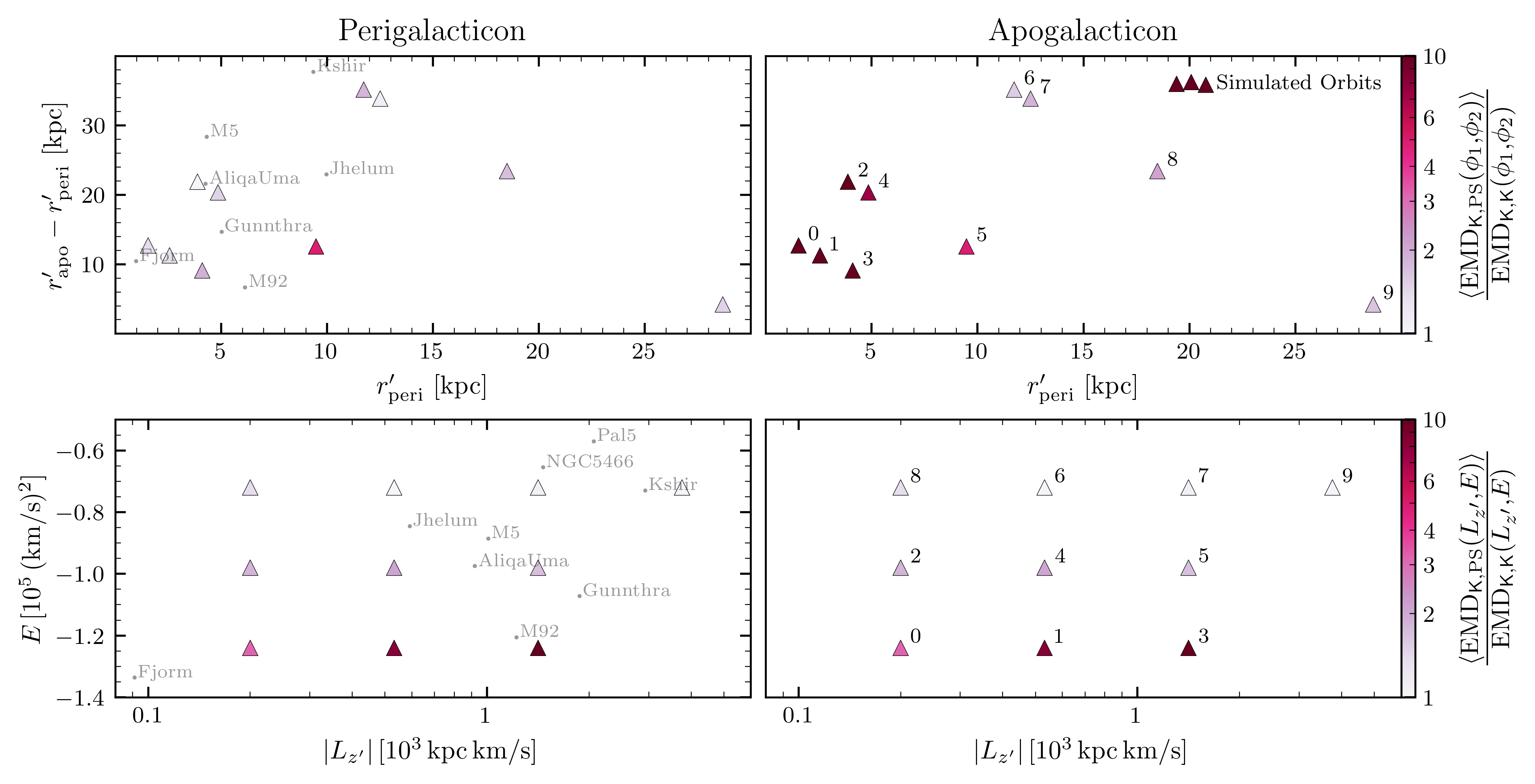}
    \caption{A comparison between the known MW stellar streams \citep[labeled in the left column,][]{malhan2022} and simulated orbits (labeled in the right column) in apogalacticon/perigalacticon space (top row) and integrals-of-motion space (bottom row). Each orbit is color coded by the ratio ${\langle{\rm EMD}_{{\sf K},{\rm PS}}\rangle/{\rm EMD}_{\sf{K},\sf{K}}}$, which controls for statistical variations expected from \krios\ runs with the same initial conditions. Particle spray models show poorer agreement to \krios\ when the progenitor is subject to strong tidal fields or is tightly bound to the host. }
    \label{fig:normalized_emd_comps}
\end{figure*}

\subsection{Analysis of Systematic Differences between \krios\ and Particle Spray for Realistic Orbits}
\label{subsec:comparisons}

We now compare the streams generated by \krios\ to those generated by particle spray across the 10 Figure~\ref{fig:sampled_orbits} orbits in the {\tt MilkyWayPotential2022} potential.  In Figure~\ref{fig:normalized_emd_comps}, we show the ratio of the EMD \krios\ streams vs.~particle-spray to the EMD between two different \krios\ runs in both ${(\phi_{1},\phi_{2})}$ (top row) and ${(L_{z'},E)}$ (bottom row) space.  This ratio allows us to focus on the systematic agreement between the streams by normalizing to the statistical variation expected between cluster models with different realizations of the same initial conditions.  We show these values calculated near their last perigalacticon passage (top row) and apogalacticon (bottom row).  Overall, these results suggest that the added computational expense of creating stream models with a full $N$-body calculation is necessary when the stream is subject to strong tidal forces (i.e., small perigalacticon) or when it is tightly bound to the host galaxy (low $E$), especially if it is near the effective-potential barrier (high $|L_{z'}|$). Any inferences made about the properties of mock stellar streams made using particle-spray models \citep[e.g.,][]{holmhansen2025} should be applied with caution to streams that have these properties. The stellar stream associated with the M92 globular cluster \citep{thomas2020}, for example, is susceptible both to modeling errors on the sky plane and in the integrals of motion. Streams like the Fj\"{o}rm \citep{ibata2019b} system are less susceptible to integrals-of-motion errors, but, at apogalacticon, sky-plane variations will be pronounced. The Gunnthr\'{a} stream \citep{ibata2021}, which may be debris from ${\omega\,\,{\rm Cen}}$ \citep{cloud2024}, appears to lie beyond the effective-potential barrier and is tightly bound to the Milky Way \citep[as is its sister Fimbulthul stream,][]{ibata2019}; disagreement in stream integrals-of-motion is high in this region. 

On the other hand, several scientifically interesting streams can be robustly modeled with particle-spray according to these metrics. The GD-1 stream \citep[Figure 1,][]{grillmair2006}, not shown in Figure \ref{fig:normalized_emd_comps}, is currently near perigalacticon \citep{koposov2010} and has one of the largest $|L_{z'}|$ values of all known stellar streams \citep{bonaca2021}. The Kshir stream, which has properties similar to GD-1 \citep{malhan2019a}, is shown in the bottom row of figures at high $|L_{z'}|$, similar to Orbit 9. Particle-spray models of streams on orbits of this kind will largely be consistent with \krios\ (and hence with \nbody). It is worth noting that the Kshir and GD-1 streams have completely dissolved progenitors. At the moment, neither \krios\ or particle spray natively handle the late stages of the GC's life cycle. In Section~\ref{subsec:validation_with_nbody}, we discuss a future interface between \krios\ and an integrator better suited for dissolving GCs.

\begin{figure*}[t]
    \centering
    \includegraphics[width=\linewidth]{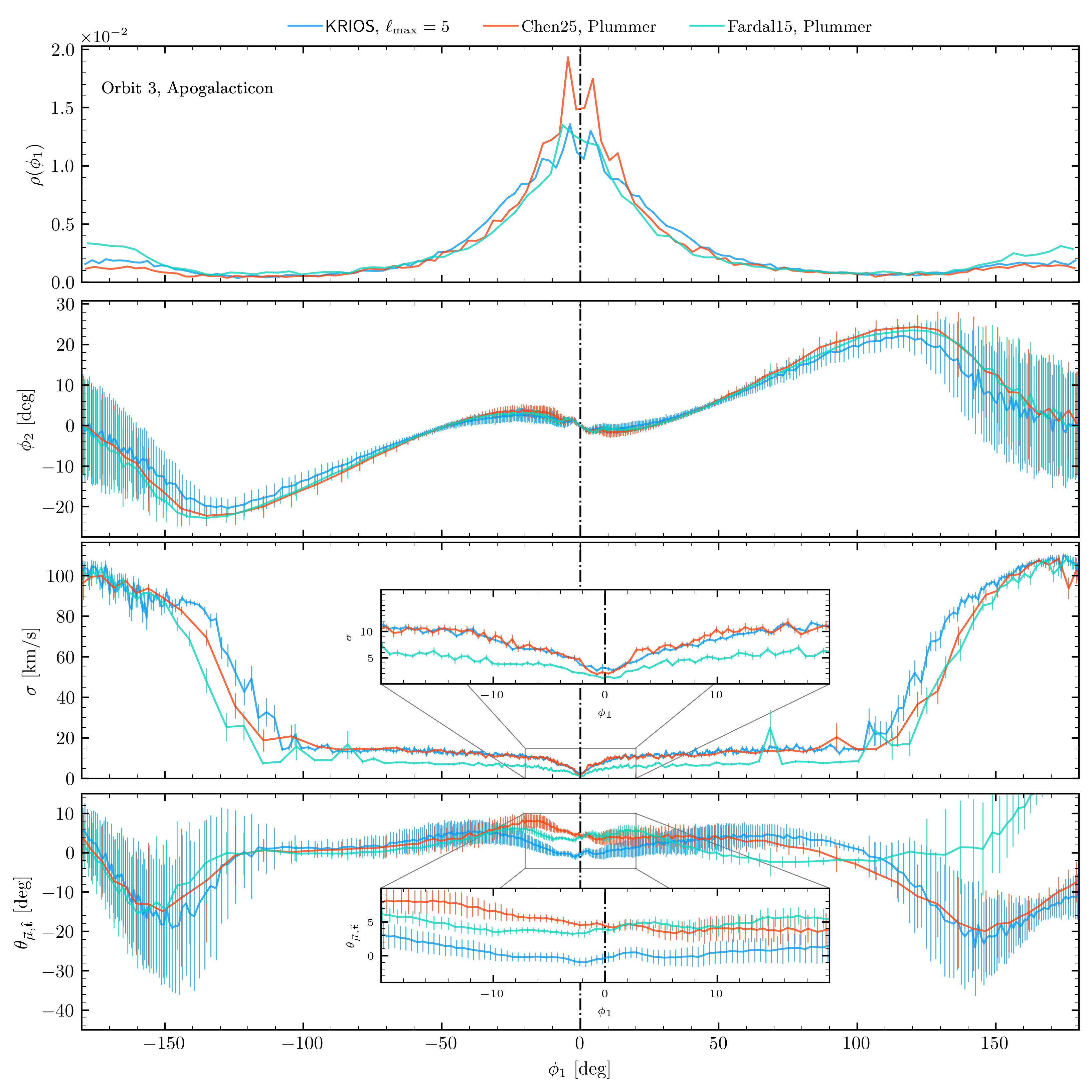}
\caption{The Orbit 3 stream morphology (normalized density along the stream track $\rho(\phi_{1})$, stream latitude $\phi_{2}$) and kinematic (velocity dispersion $\sigma$, misalignment angle $\theta_{\boldvec{\mu},\hat{\boldvec{t}}}$) information for each model near the last apogalacticon passage. The fiducial ($n_{\rm max}\!=\!10$, $\ell_{\rm max}\!=\!5$) \krios\ model is compared against particle-spray models with a Plummer progenitor and different escaper distribution functions \citepalias{fardal2015, chen2025}. The dashed line at ${\phi_{1}\!=\!0}$ indicates the progenitor location and inset panels zoom in to regions with notable differences between models. The largest disagreements are the density and misalignment angle near the progenitor, as well as the velocity dispersion and misalignment angle in the tails. This stream's velocity dispersion is systematically underestimated when using the \citetalias{fardal2015} distribution function and the \citetalias{chen2025} distribution function overestimates the central overdensity.}
    \label{fig:krios_ps_comp_3}
\end{figure*}

Interactions of streams with Galactic substructure, whether dark or baryonic, is thought to perturb the stream density $\rho(\phi_{1})$ and velocity dispersion $\sigma(\phi_{1})$ profiles along the stream track. It is critical that we understand which features of these profiles originate from the model assumptions, so that they are not mistaken for external perturbations. This is especially important given that several different escape prescriptions are commonly used in particle-spray models, yet there are very few tests of which prescriptions best reproduce $\rho(\phi_{1})$ and $\sigma(\phi_{1})$ for globular cluster streams.
%For more detailed analyses of the predicted streams (e.g., Figure~\ref{fig:krios_ps_comp_3}), 
To examine differences in the predicted profile, we select particles that are unbound from the cluster and bin them in $\phi_{1}$ from the inside out, starting at the progenitor, such that the number of particles in each bin $N_{\rm particles\,in\,bin}\!\geq\!100$ and the bin width ${\Delta\phi_{1,{\rm bin}}\!\geq\!0.5\,{\rm deg}}$. This binning strategy is chosen to accentuate differences in the $\rho(\phi_{1})$ and $\sigma(\phi_{1})$ profiles. Each distribution is computed from all particles in the stream (as opposed to Section~\ref{subsec:validation_with_nbody}, where the distributions were downsampled in order to calculate the KLD and EMD). We take many draws from each bin's particles, detrend their velocities by subtracting a linear term $\bar{v}(\phi_1)$ within the bin, and then estimate the local 1D velocity dispersion ${\sigma\!=\!\sqrt{\tr(\boldvec{\Sigma})/3}}$, where $\boldvec{\Sigma}$ is the covariance matrix of the detrended velocities. By the central-limit theorem, this nonparametric bootstrapping method \citep[Section~15.6.2 of ][]{press2002} reduces the uncertainty in $\sigma(\phi_{1})$ as we take more samples. %However, it assumes the particles in each bin are independent and identically distributed around the bin mean, which is a poor assumption for bins with large $\Delta \phi_{1,{\rm bin}}$ sizes unless the velocity gradient across the bin is removed by detrending before estimating the covariance matrix. We also could have increased the particle number $N$ for fixed cluster mass or used an ensemble of models to ensure sufficient sampling at each $\phi_{1}$ value as well. \carl{since we're not quantifying these issues there's no point in bringing them up}
We also calculate the average misalignment angle between the particles' proper motions and unit vector tangent to the stream track, i.e. ${\cos\theta_{\boldvec{\mu},\hat{\boldvec{t}}}(\phi_{1}) \!\equiv\!{1\over N}\sum_{i=1}^{N}(\hat{\boldsymbol{\mu}_{i}}\cdot\hat{\boldvec{t}}(\phi_{1,i})})$, which can also be used to probe the stream kinematics \citep[e.g.,][]{erkal2019}. We parameterize $\phi_{2}(\phi_{1})$ with a smoothed B-spline in order to calculate the tangent vector $\hat{\boldvec{t}}(\phi_{1})$ and hence the misalignment angle $\theta_{\boldvec{\mu},\hat{\boldvec{t}}}(\phi_{1})$.

%\nw{[Are you sure the deviations at the end of the stream aren't due to initial condition oddities? I would expect more high energy escapers at that time, which is what you see. You could check this by, e.g., cutting out the first 1~Gyr worth of escapers. This question also applies to Figure~14.]}

\begin{figure*}
    \centering
    \includegraphics[width=0.8\linewidth]{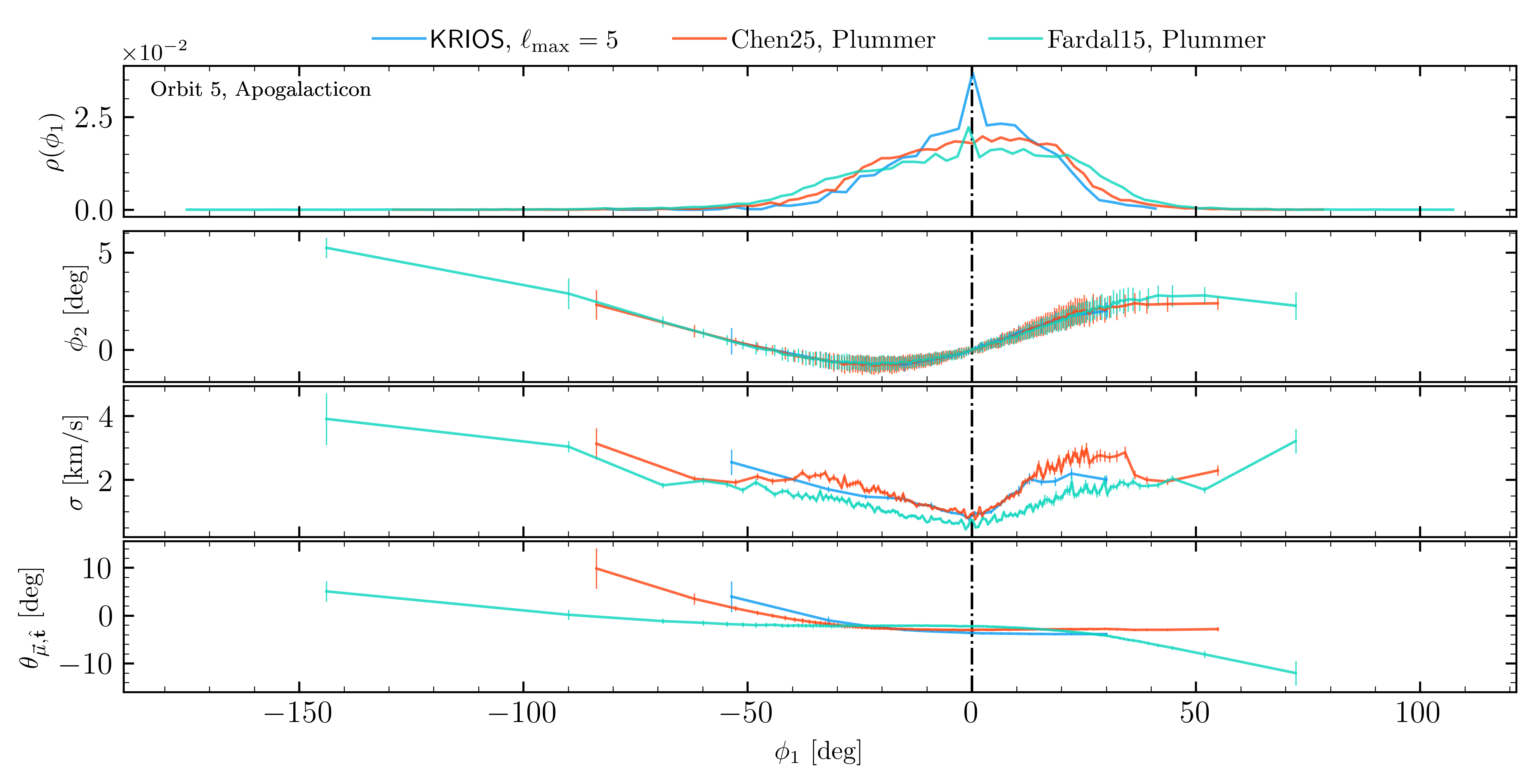}
    \caption{The same information as Figure~\ref{fig:krios_ps_comp_3}, but for the Orbit 5 stream at last apogalacticon. This is an orbit where particle spray and \krios\ are in moderate disagreement (see Figure~\ref{fig:normalized_emd_comps}). This stream is dynamically colder than the one displayed in Figure~\ref{fig:krios_ps_comp_3}; the velocity dispersion is ${\sigma\!\lesssim\!6\,{\rm km/s}}$ and the misalignment angle ${\theta_{\boldvec{\mu},\hat{\boldvec{t}}} \!\lesssim\! 10\,{\rm deg}}$ everywhere along the stream. The particle-spray are more diffuse longitudinally, leading to an underestimation of $\rho(\phi_{1})$ near the progenitor and variations in the $\theta(\phi_{1})$.}
    \label{fig:krios_ps_comp_5} 
\end{figure*}

As examples of stream structure that varies with the underlying modeling technique, we show the \krios\ and \citepalias{fardal2015, chen2025} particle-spray models for Orbit 3 (Figure~\ref{fig:krios_ps_comp_3}) and Orbit 5 (Figure~\ref{fig:krios_ps_comp_5}). \krios\ and particle spray disagree on these orbits, more so for Orbit 3 than Orbit 5 (see Figures~\ref{fig:results_emd_comp}, \ref{fig:normalized_emd_comps}). The latitude vs.~longitude trends agree, which suggests that \krios\ and particle spray are equally capable of integrating particles forward in the same host potential. Variations in $\rho(\phi_{1})$ are most notable near the progenitor, where different particle-spray distribution functions better replicate the density fluctuations predicted by \krios\ on different orbits. There is a notable demarcation between $\theta_{\boldsymbol{\mu}, \boldsymbol{t}}(\phi_{1})$ predictions near the progenitor for the Orbit 3 stream. The velocity dispersion profiles disagree as well, but this effect is seen most strongly in the tidal tails. The \citetalias{fardal2015} Orbit 3 stream, for example, systematically underestimates $\sigma(\phi_{1})$. Both particle-spray models for the Orbit 5 stream are more spread out in $\phi_{1}$ than the \krios\ model, which may be the source of the disagreement in $\sigma(\phi_{1})$ and $\theta_{\boldvec{\mu},\hat{\boldvec{t}}}(\phi_{1})$. These results illustrate that subtle differences between models are imprinted on the predicted kinematics of the stream.

\section{Discussion}
\label{sec:discussion}

%Numerical experiments are necessary to understand how the progenitor affects its stream. 
We showed in Section \ref{subsec:comparisons} that the predicted stream is sensitive to the implementation details of the underlying modeling technique. Particle-spray codes that sample escapers from some distribution function, for example, might ignore changes to the progenitor induced by the host potential or that mass loss is correlated with the strength of the tidal field. Additionally, stream-modeling techniques often assume the progenitor is spherically symmetric (e.g., \cmc's implementation of the H\'{e}non method), which substantially affect the mass-loss rate---and hence also the predicted stream. 
This is particularly important in light of several recent works that examine streams from GCs simulated with \cmc, ranging in complexity from GCs on circular orbits in a static and spherical MW \citep{weatherford2024, weatherford2025} to GCs on non-periodic orbits in an evolving and clumpy MW-like {\tt FIRE} cosmological simulation \citep{panithanpaisal2025}---see also \cite{grudic2023, rodriguez2023}. Here, we leverage \krios\ to explore potential discrepancies between comparable stream-modeling techniques more concretely.  We focus primarily on Orbit 3, as Figure \ref{fig:results_emd_comp} shows that this orbit shows the least stochastic scatter between similar \krios\ and particle-spray models, while showing significant disagreement between \krios\ and particle spray (as also seen in Figure \ref{fig:normalized_emd_comps}).

\subsection{A spherically symmetric progenitor is usually sufficient to model escaping stars}
\label{subsec:diff_progenitors}

Though \cmc\ simulations feature collisional dynamics and a particle-based GC potential, creation of model streams requires modeling the trajectories of escaping stars in post-processing since the full potential of the host galaxy is represented only by the tidal tensor. Bodies are removed from \cmc\ upon achieving an orbital energy or apocenter sufficient for escape (often in the GC's core), and then integrated in the mutual potential of the GC and its host galaxy. The GC potential is made analytic by fitting a Plummer sphere \citep{panithanpaisal2025}---or a three-component Plummer sphere to better reproduce the cores of evolved GCs  \citep{weatherford2024,weatherford2025}---to the raw \cmc\ potential and interpolating the fitting parameters coarsely in time (a few snapshots per half-mass relaxation time). The trajectories of stars forming a stream from a \cmc\ cluster model thus only experience a fairly rough approximation of the true potential at the boundary between the GC and host galaxy. By comparison, particle-spray models often assume a fixed Plummer progenitor, i.e. constant mass and half-mass radius. The self-consistent potential maintained by \krios, on the other hand (Equation~\ref{eq:total_potential}), is tuned to the GC's true configuration more accurately and at higher temporal resolution. Additionally, escapers are subject to the same SCF even after they leave the cluster. The model used in \krios\ is therefore more self-consistent and well-resolved on the relevant timescales within the cluster, and should be preferable to the \cmc\ approach. We can use the structure of the SCF to test the effect of the assumption of a spherically symmetric cluster on the accelerations felt by escaping stars directly, since setting $\ell_{\mathrm{max}}\!=\!0$ is equivalent to forcing the cluster potential to be spherically symmetric as it is in \cmc\ and many particle-spray codes.

%The mean-field approximation of the progenitor potential, whether the Plummer potential used in post-processing \cmc\ simulations or the SCF used in \krios\, seeks to replicate the potential of a self-gravitating system of $N$ Plummer-softened particles:
%
%\begin{align}
%\psi_{\rm cluster}(\boldvec{r}) \!=\! - G\sum_{j} {m_{j} \over \sqrt{|\boldvec{r}-%\boldvec{r}_{j}|^2+\epsilon^2}},
%\end{align}
%
%where the softening length $\epsilon$ smooths out discontinuities in the potential from close encounters \citep[${\epsilon\!\propto\!N^{-1/3}}$ is suitable for basis-expansion codes, see Section 6.1.2 of][]{dehnen2001}. The total acceleration of the $i$th particle, when coupled to an external tidal field, is then
%
%\begin{align}
%\label{eq:softened_acceleration} 
%\boldsymbol{a}_{i}\!=\!-G\sum_{j\neq i}{m_{j}\,\boldvec{r}_{ij}\over {(r_{ij}^{2}}+\epsilon^2 )^{3/2}} - \nabla_{\boldvec{r}_{i}'}\,\psi_{\rm host},
%\end{align}
%
%where ${\boldvec{r}_{ij}\!\equiv \!\boldvec{r}_{i}-\boldvec{r}_{j}}$. 

\begin{figure}
    \centering
    \includegraphics[width=\linewidth]{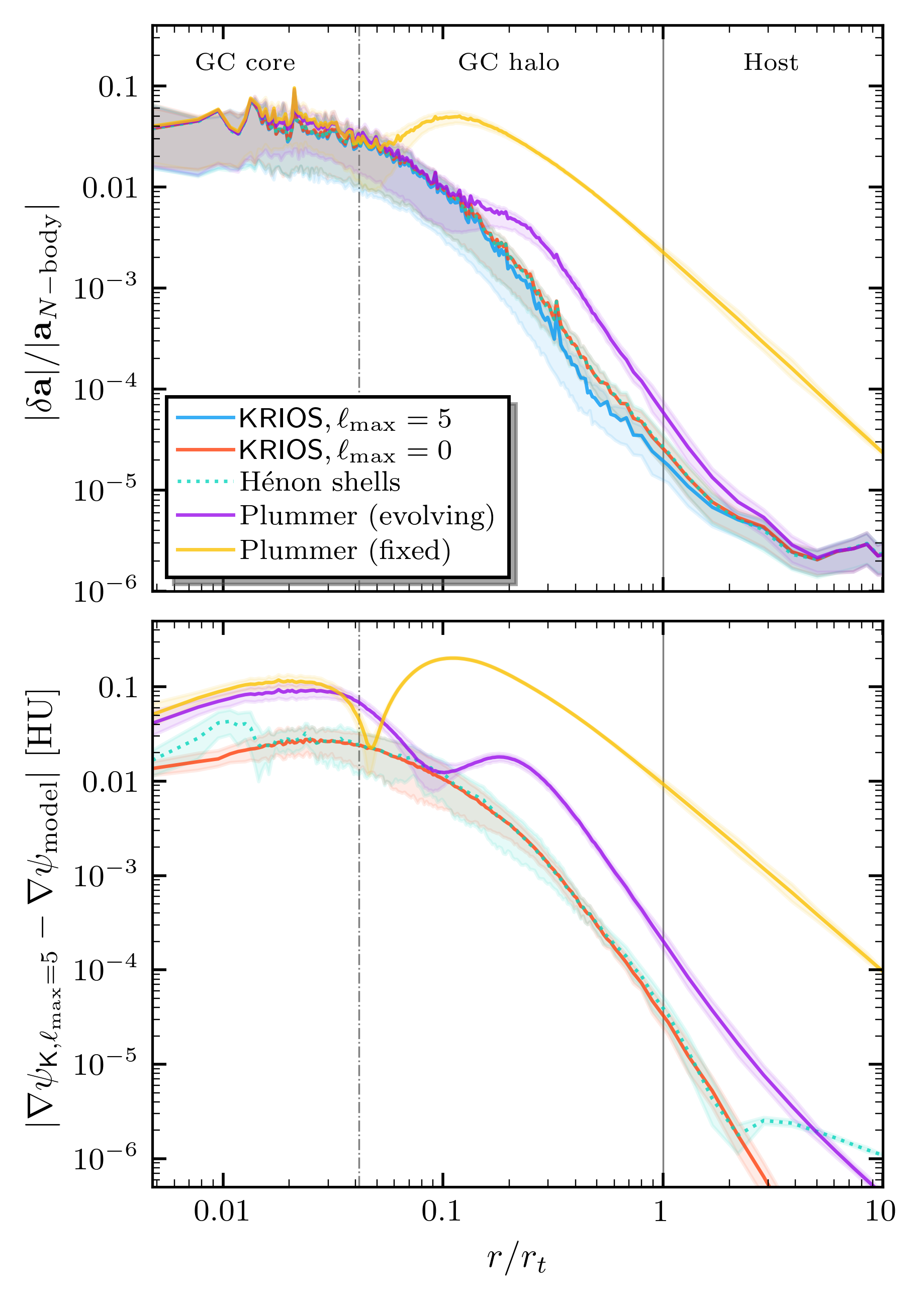}
    \caption{The residuals for the particle data from the last perigalacticon passage of the Orbit 3 system using different $\psi_{\rm cluster}$ models. The shaded regions show the $1\sigma$ spread for that bin. {\it Top panel:} The acceleration residuals when compared against direct {$N$-body} summation.%, i.e. Equation~\eqref{eq:softened_acceleration} and $\epsilon\!=\!0$. T
    The spherically asymmetric \krios\ potential is the only one capable of resolving tangential accelerations, and is generally the most successful at replicating the {$N$-body} acceleration profile outside of the core (black dashed line). This is especially important near the tidal boundary (black solid line\footref{rt_footnote}), where the effective-potential contributions from the host potential and centrifugal term are equally important. {\it Bottom panel:} The gradient of the fiducial cluster potential (\krios, ${\ell_{\rm max}\!=\!5}$) compared to other cluster potential models, where the variations between the various mean-field approximations are more clear. The fixed Plummer potential (yellow curve), which is often used in particle-spray models, is a comparatively inaccurate approximation.} 
    \label{fig:force_residuals}
\end{figure}

We can probe the effectiveness of each progenitor model by comparing the accelerations from each model to those calculated by direct $N$-body summation.  The top panel of Figure~\ref{fig:force_residuals} shows the acceleration residuals \citep[${\delta\boldvec{a}\!\equiv\!\boldvec{a}_{\rm model}-\boldvec{a}_{N-{\rm body}}}$,][]{mukherjee2021, arora2024} for different mean-field progenitor models, compared to the true values $N$-body accelerations.% calculatee---i.e., Equation~\eqref{eq:softened_acceleration} with ${\epsilon\!=\!0}$. 
The Plummer progenitor that does not take cluster evolution into account (yellow curve), which is often used in particle-spray methods, poorly captures the particle accelerations. We see significant improvement if we use the instantaneous cluster mass and scale length for the Plummer potential \citep[purple curve; e.g.,][]{panithanpaisal2025}. A multi-component Plummer potential fit would result in further improvement \citep{weatherford2024}, and using \cmc's exact spherical-shell potential ${\psi_{\rm cluster}\!=\!-{GM_{\rm enc}(<r)\over r}}$ (dotted green curve) is better still, agreeing well with the spherically symmetric \krios\ SCF (red curve). The fiducial \krios\ profile (${\ell_{\rm max}\!=\!5}$; blue curve) best reproduces the target accelerations. By construction, it is able to resolve tangential acceleration components in a way that spherically symmetric fields cannot. The bottom panel of Figure~\ref{fig:force_residuals} shows how the accelerations, expressed as the gradient of the potential in H\'{e}non units, compare to the fiducial model. Here, we see the Plummer-sphere approximations (both fixed and evolving) separate from the non-Plummer-sphere approximations in the GC's core, as well as how the zeroth-order mode of the SCF disagrees with the spherical-shell potential for ${r\!\lesssim\!0.5\,r_{\rm core}}$.

\begin{figure*}
    \centering
    \includegraphics[width=\linewidth]{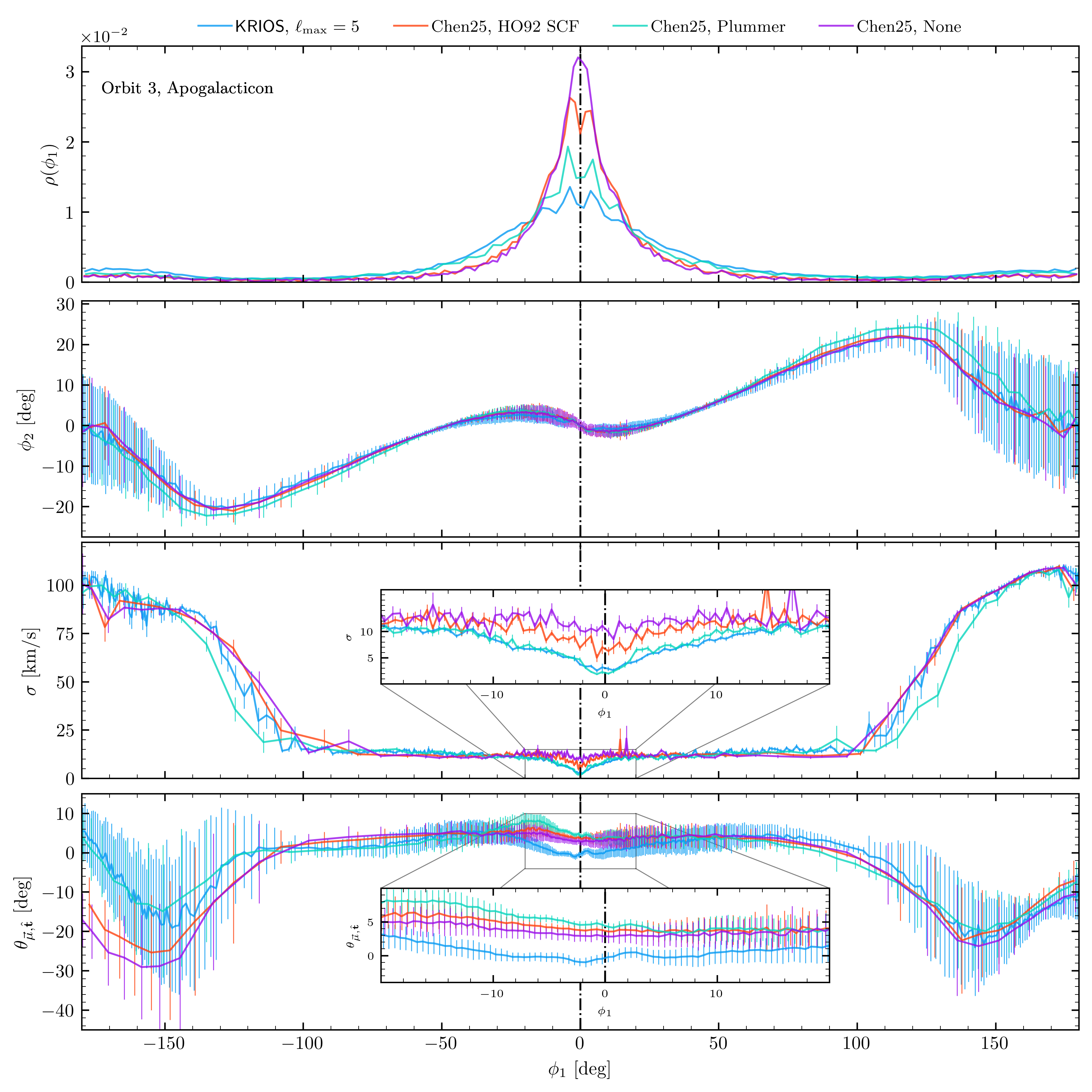}
    \caption{The fiducial \krios\ model for the Orbit 3 stream compared to three \citetalias{chen2025} models near the last apogalacticon passage: one with an SCF progenitor \citep[][${n_{\rm max}\!=\!\ell_{\rm max}\!=\!2}$]{hernquist1992}, one with a Plummer progenitor, and one with no progenitor. There does not appear to be immediate improvement in the modeling with improved treatment of the progenitor, assuming that the progenitor is fixed.}
    \label{fig:chen_comparisons}
\end{figure*}

Figure~\ref{fig:chen_comparisons} shows the impact of different progenitor prescriptions on the stream, which again are slight disagreements in $\rho(\phi_{1})$ and $\sigma(\phi_{1})$. There is no clear connection between improving the treatment of the progenitor potential---from no progenitor at all, to a Plummer model, to an SCF model---and the accuracy of stream density, morphology, or kinematics. Interestingly, it is the intermediate model (Plummer) that has the worst velocity dispersion agreement with \krios\ in the tails, but best replicates the density variations near the progenitor. The implication here is that there is an upper limit on the degree to which particle spray can be improved with more accurate initializations of a fixed progenitor. However, related improvements to modeling the progenitor's internal \textit{dynamics}, such as explicit treatment of strong encounters (in the process of being added to \krios), still can substantially affect stream properties \citep[e.g.,][]{gieles2021,weatherford2025}.

\subsection{Spherically (a)symmetric {\sf KRIOS} progenitors produce statistically similar streams}
\label{subsec:symmetry_breaking}

The restricted three-body problem's picture for stellar escape from GCs breaks down for eccentric orbits in a MW-like host potential (see Section~\ref{sec:introduction}). In particular, the zero-velocity surface is not necessarily spherical and escapers do not necessarily escape from near the L1/L2 Lagrange points. Figure 2 of \citetalias{chen2025}, which shows the normal distribution of escapers' spherical phase-space coordinates from {$N$-body} runs, provides further evidence of escape anisotropy.
%through \nw{Some people I've encountered had the impression that particles literally escape THROUGH L1/L2, whereas L1/L2 actually are foci that channel escapers through narrow bottlenecks in the effective potential. The highest density in the tail is actually on either side of each Lagrange point, not directly on it. Hopefully more precise wording will help stave off the misconception!}

\begin{figure}
    \centering
    \includegraphics[width=\linewidth]{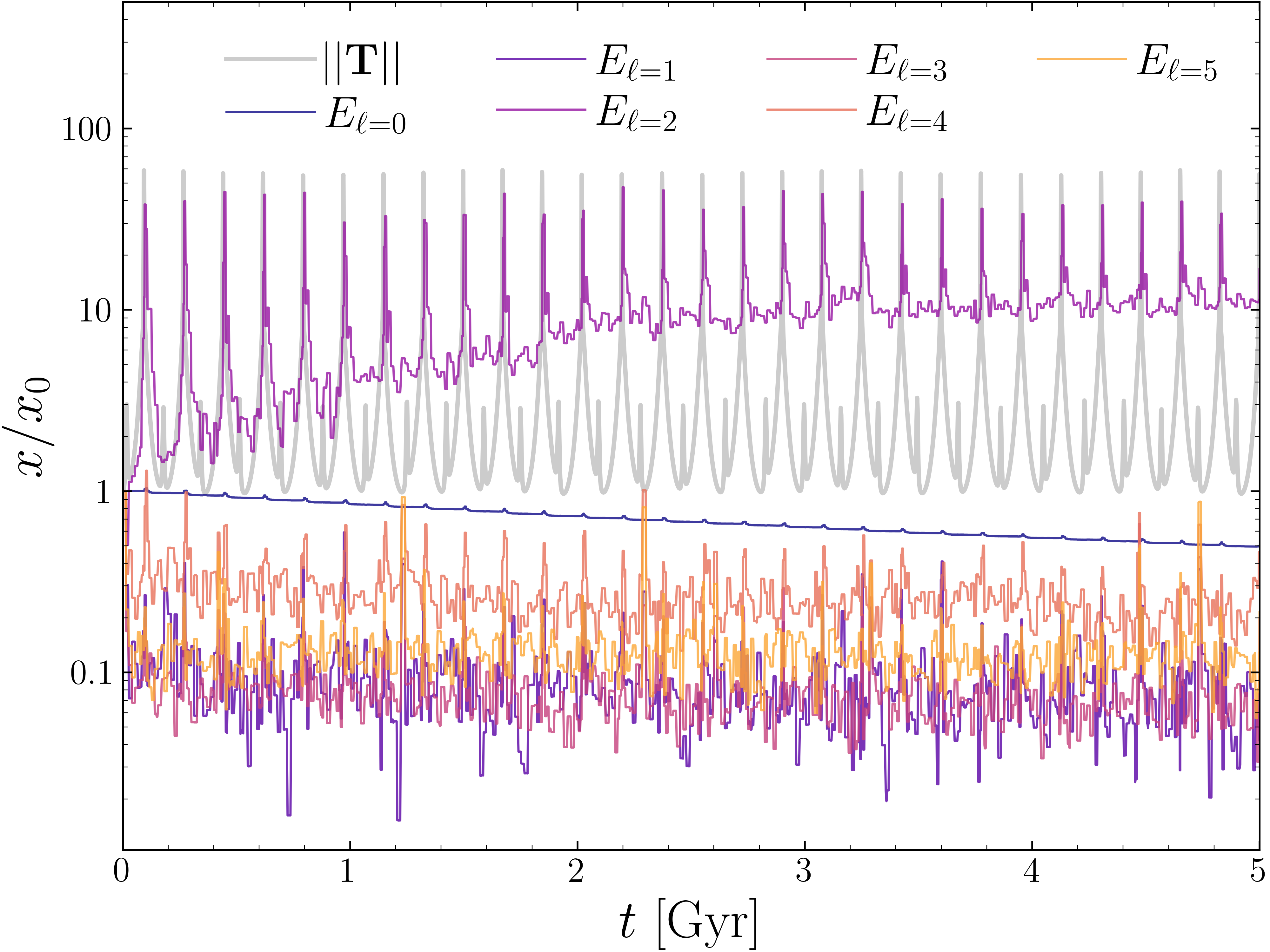}
    \caption{The power in each SCF $\ell$ mode (i.e., ${E_{\ell}\!=\!\sum_{nm}|a_{n\ell 
 m}|^{2}}$) compared to the tidal tensor's Frobenius norm for the Orbit 3 \krios\ run. The ${\ell\!=\!2}$ quadrupole term, which is related to the cluster's asphericity, is correlated with the strength of the tidal field. Discontinuities in the tidal field strength are attributable to perigalacticon and disk passages.}\label{fig:power_spectrum_time_series}
\end{figure}

The asphericity of the GC's mass distribution can be measured by the power in each of the $\ell$ modes, computed as ${E_{\ell} \!=\! \sum_{nm} |a_{n \ell m}|^{2}}$ \citepalias[Equation D36 of][]{tep2025}, where $\{a_{n\ell m}\}$ are the SCF's basis coefficients. This is especially important for MWGCs, many of which are slightly aspherical \citep{chen2010}. GCs dominated by tides are prolate spheroids, while rotation-dominated GCs are oblate \citep{bardeen1986, fitzpatrick2012}. The GC is spherically symmetric if ${E_{\ell}\!=\!0}$ for all ${\ell\!>\!0}$. Figure~\ref{fig:power_spectrum_time_series} compares the Orbit 3 model's SCF $\ell$-mode power spectrum and the tidal tensor's Frobenius norm $||\boldvec{T}||$, all normalized by their initial values, as a function of time. This time-series data demonstrates how \krios\ clusters evolve when coupled to external tidal fields. The ${\ell\!=\!0}$ monopole term loses power as the cluster loses mass. The ${\ell\!=\!1}$ dipole term measures how well the SCF is centered on the cluster\footnote{If the center of mass were used instead of the location that maximizes ${a_{n\ell m}\!=\!a_{000}}$, for example, there would be more power in the dipole term \citepalias[in contrast with what is stated in Section 2.4 of][]{binney2008}.}. The asphericity of the system is captured by the ${\ell\!\geq\!2}$ modes \citep{hernquist1992, weinberg1996, vasiliev2019b}. In particular, the ${\ell\!=\!2}$ quadrupole mode appears to be highly correlated with ${||\boldvec{T}||}$. Our models are initialized with no internal rotation, so in this context the SCF quadrupole appears to track the time-dependent prolateness of the progenitor.

Despite this measurable change in the cluster's asphericity, the integrals-of-motion and sky-plane EMD scores between the fiducial ${\ell_{\rm max}\!=\!5}$ models and the models where spherical symmetry is enforced (${\ell_{\rm max}=0}$) have the following ranges when evaluated at apogalacticon:
\begin{align*}
{\langle {\rm EMD}_{{\sf K}5, {\sf K}0} \rangle \over {\rm EMD}_{{\sf K}5, {\sf K}5}} \in  \begin{cases}
    (0.5, 1.4) &(L_{z'},E), \\
    (0.7, 1.9) &(\phi_{1},\phi_{2}). \\
\end{cases}
\end{align*}
This suggests that stream models produced by H\'{e}non-based {$N$-body} codes like \cmc\ will only suffer from mild disagreement due to the assumed spherical symmetry of the progenitor, as long as the progenitor is properly incorporated when calculating escaper dynamics (Section \ref{subsec:diff_progenitors}). We have not made direct comparisons between \cmc\ and \krios\ models, so the degree to which their predictions for stream substructure (e.g., gaps, spurs) agree warrants investigation in future studies.

\subsection{Mass-loss Rate Variability in Stream Modeling}

Our particle-spray models, to this point, have assumed a constant mass-loss rate (i.e., escapers drawn from the \citepalias[][]{fardal2015, chen2025} distribution functions at each timestep). Here, we consider the impact of a variable mass-loss rate. The third panel of Figure 2 in \cite{panithanpaisal2025} shows that the number of ejected particles peaks sharply during perigalacticon passages, which is corroborated by \krios\ (see the green curve in the second panel of Figure~\ref{fig:cluster_validation}). Variable mass-loss rates can be added to particle-spray models by passing an array of particle counts to be released during each timestep to the relevant {\tt gala} stream-generation method\footnote{\url{https://gala.adrian.pw/en/latest/dynamics/mockstreams.html}} or through post-processing of the stream data itself \citep[e.g.,][]{pearson2024}. Candidate variable mass-loss rates include ${\dot{M}(t)\!\propto\! M(t)^{a}\Omega(t)^{b}r_{\rm hm}(t)^{c}}$ \citep[fit to observational data in][]{chen2025b}, where $\Omega(t)$ increases during perigalacticon passages, and \citetalias{fardal2015}, which uses an analytic prescription for the mass-loss rate as a function of the orbit's radial phase to build its distribution function. The latter is more applicable to particle-spray methods that do not track the progenitor mass or half-mass radius. Neither prescription accounts for disk shocks that heat up the progenitor, and by default {\tt gala} assumes a constant mass-loss rate.

\begin{figure}
    \centering
    \includegraphics[width=\linewidth]{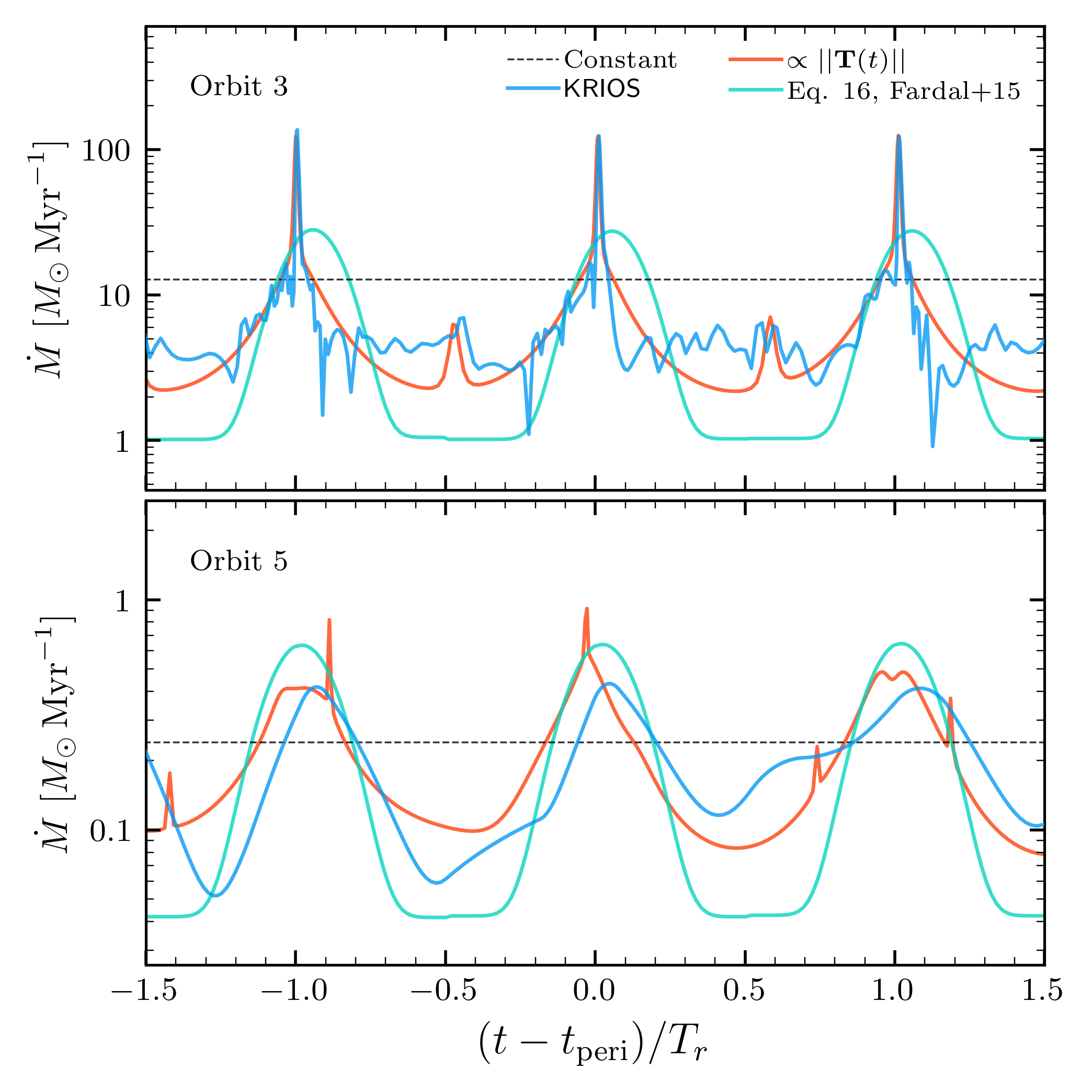}
    \caption{The mass-loss rate for three radial periods of the Orbit 3 and Orbit 5 \krios\ models (${t_{\rm peri}\sim2.5\,{\rm Gyr}}$). The {\sf KRIOS} mass-loss rate estimate is a spline fit to the ${\Delta M/ \Delta t}$ data from each integration timestep. The black dashed line represents the constant mass-loss rate typically used in particle-spray studies. The green curve shows Equation 16 of \citetalias{fardal2015}, where the peak mass-loss rate is slightly offset from perigalacticon. The red curve shows a hypothetical mass-loss rate proportional to the tidal tensor's Frobenius norm, which accommodates disk shocks.}
    \label{fig:mdot_diff_prescriptions}
\end{figure}

\begin{figure*}
    \centering
    \includegraphics[width=\linewidth]{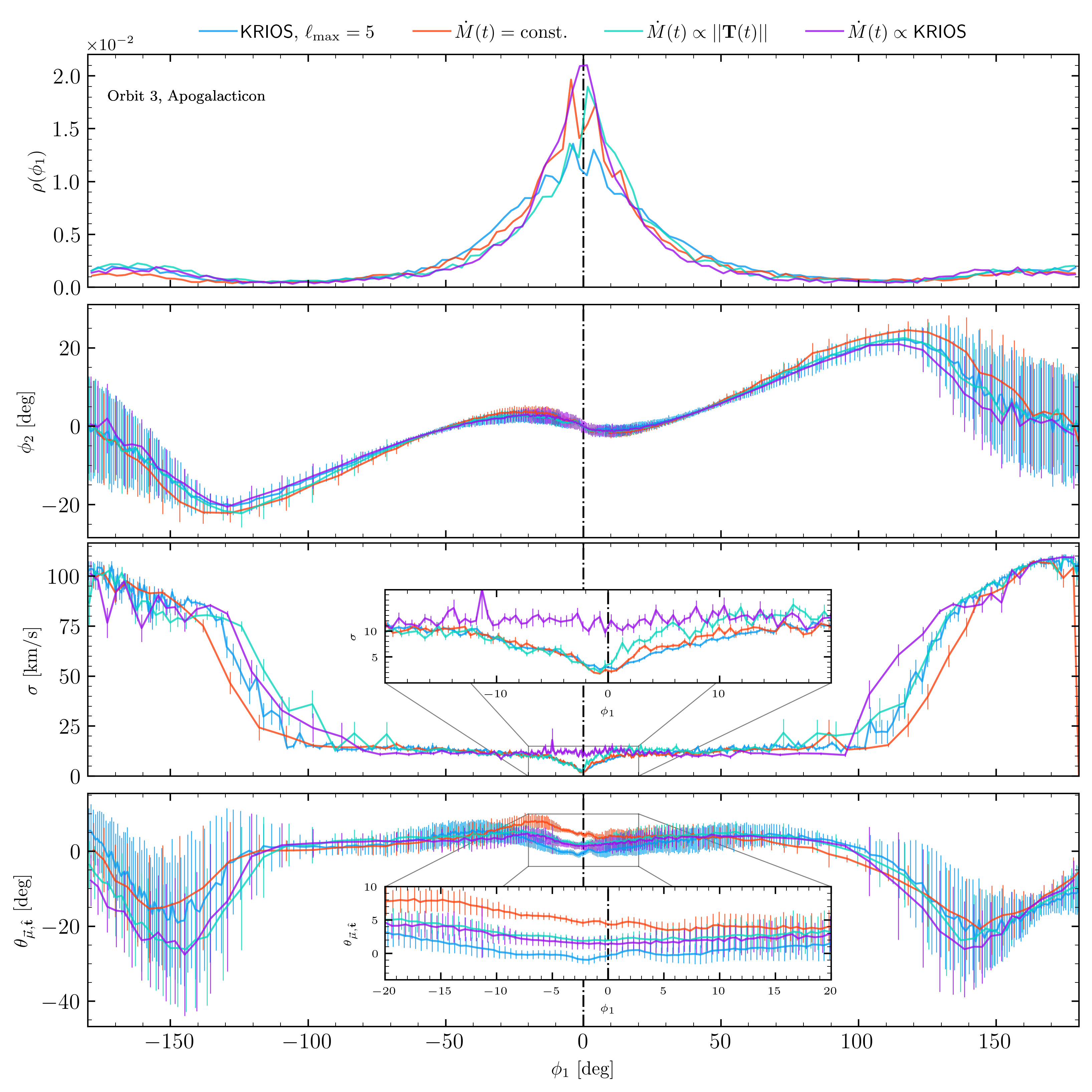}
    \caption{The fiducial \krios\ model for the Orbit 3 stream compared to three \citetalias{chen2025} models with a Plummer progenitor: one with a fixed mass-loss rate (red, see Figure~\ref{fig:krios_ps_comp_3}), one with a variable mass-loss rate proportional to the tidal tensor's Frobenius norm (green), and one with a variable mass-loss rate collected directly from the {\sf KRIOS} data (purple). There is a larger spread in the tail velocity dispersions than when the progenitor is modified (Figure~\ref{fig:chen_comparisons}) and the density fluctuations near the progenitor are not modeled correctly by particle spray even with more realistic mass-loss rates.}
    \label{fig:variable_mass_loss_streams}
\end{figure*}

Figure~\ref{fig:mdot_diff_prescriptions} shows the mass-loss rate in the Orbit 3 and Orbit 5 \krios\ simulations (blue) for three complete orbits and compares them to the following mass-loss prescriptions: constant (black dashed line), Equation 16 of \citetalias{fardal2015} (green), mass loss directly proportional to the tidal tensor's Frobenius norm (red), and mass loss proportional to the mass-loss history from the {\sf KRIOS} data (purple). A constant mass-loss rate is clearly a poor fit to the Orbit 3 data, as the peak mass-loss rate at perigalacticon (which coincides with a disk passage, see Figure~\ref{fig:sampled_orbits}) is ${\sim10\times}$ the average. Additionally, the mass-loss rate is suppressed after these peaks; this is likely attributable to the unavailability of easily stripped particles that are replenished through continued relaxation when the cluster is subject to comparatively weak tidal forces. For the Orbit 5 model, there are only $\lesssim\!5$ particles ejected per integration timestep, so a constant mass-loss rate heuristic is likely acceptable in this context. The tidal tensor captures disk shocks (see the peaks in the red curve), which might make it suitable as a generalized mass-loss rate prescription.

Figure~\ref{fig:variable_mass_loss_streams} shows how these mass-loss rates manifest in models of the Orbit 3 stream. The blue curve shows the fiducial \krios\ ${\ell_{\rm max}\!=\!5}$ model that is also shown in Figures \ref{fig:krios_ps_comp_3} and \ref{fig:chen_comparisons}. The green curve shows the particle-spray model with \citetalias{chen2025} distribution function and Plummer progenitor (see Figure~\ref{fig:chen_comparisons}) where the mass-loss rate is constant, and the red curve shows the same particle-spray model where the mass-loss rate is set by passing the $||\boldvec{T}(t)||$ time-series data to the stream-generation method. The purple curve's disagreement with the {\sf KRIOS} model is likely due to all of the other modeling differences (e.g., static Plummer sphere progenitor and escaper distribution function). Further investigation is needed into variable mass-loss rate prescriptions that better replicate the numerical experiments we have carried out with \krios\ in this study, e.g. ones that take the recent history of the tidal tensor along the orbit into account. Using escape and recapture events from \krios\ as training data for new particle-spray methods is left to future work as well.

\section{Conclusions}
\label{sec:conclusions}

In this paper, we expand the scope of the \krios\ hybrid {$N$-body} code that was recently introduced in \citetalias{tep2025} by modeling tidal debris from globular clusters orbiting a host galaxy. Realistic MW potential models are used to validate \krios\ against known, successful stream-modeling techniques. There is good agreement between \krios\ and \nbody\ when the respective codes model the internal evolution of GCs in an external tidal field, as well as their resultant debris. It is found that, on an orbit consistent with realistic MWGC orbits, \krios\ replicates the sky-plane and integrals-of-motions distributions from comparable \nbody\ runs.

%\subsection{Conclusions for Cluster Modelers}
%\label{subsec:gc_modeler_conclusions}

%One might think of \krios's design as a modification of \cmc, where the spherical symmetry and virial equilibrium assumptions are relaxed. This is executed with a self-consistent field that adapts to the instantaneous, and potentially aspherical, configuration of the cluster as it gets tidally distorted. The effective scattering events that replicate the effects of many weak two-body encounters are carried out, not by considering modifications to the energies and angular momenta, but rather through direct evaluations of the velocity perturbations in three-dimensional space. This combination renders non-orbit-averaged cluster models where escape events are easily accommodated and the potential adapts to the cluster on dynamical timescales. Reference frames are constructed in such a way that forces in the fixed inertial frame can be evaluated directly by an ODE solver, where the calculations are parallelizable and scale linearly with $N$.

\subsection{Conclusions for Stream Modelers}

\begin{comment}

my summary of takeaways for stream modelers

\begin{itemize}
    \item The ability of particle-spray models to accurately reproduce results from N-body simulations varies significantly with stream orbit (Section \ref{subsec:comparisons}; Figure \ref{fig:normalized_emd_comps}). In particular, streams with small pericenter distance and/or large $L_z$ are not well modeled by particle-spray and should instead be modeled with \krios\ or an N-body code. This region of orbit space includes many scientifically important known streams, with a few notable exceptions.
    \item Even when particle-spray does a decent job at reproducing the stream (as measured statistically by the KLD or EMD), its internal structure is affected by the choice of mass-loss model (section 3.2; figs 9 and 10). If particle-spray is used to model a stream, several different mass-loss prescriptions should be considered in order to eliminate this effect as an explanation for substructure. With \krios\ the mass-loss rate is computed self-consistently so no prescription is needed.
    \item Assuming a spherical model of the progenitor cluster, as done in CMC, is sufficient to model the escape of stars from the cluster (Section \ref{subsec:diff_progenitors}) and calculate the mass-loss rate (Section \ref{subsec:symmetry_breaking}) when predicting the density, velocity dispersion, and misaligment along the stream. This may not be the case for all cluster properties though---see section above on ``conclusions for GC modelers.''
\end{itemize}

\end{comment}

We use a collection of {\tt MilkyWayPotential2022} sample orbits, spanning the region of orbit space occupied by known MW streams, to test how \krios\ and particle-spray models might disagree in scenarios similar to those found in the observational data. Particle-spray results deviate from \krios\ in the following scenarios:

\begin{itemize}
    \item The GC is subject to strong tidal forces during its orbit, characterized by small perigalacticon and apogalacticon distances. The Fj\"{o}rm and M92 streams are examples of such systems. Differences in the stream's distribution on the sky plane are most apparent when they are near apogalacticon.
    \item The GC is tightly bound to the host galaxy. This applies to circular (M92, Gunnthr\'{a}) and radial (Fj\"{o}rm) orbits alike. Systems of this kind lead to disagreements in how well the integrals of motion of the stream stars are conserved.
\end{itemize}

The most common form of disagreement between \krios\ and particle-spray stream models are in the resolution of density fluctuations near the progenitor and the velocity dispersion of the tails far from the progenitor. Inferences made about the morphology of the GC's extratidal features or the kinematics of sparsely populated regions in the GC streams should consider whether the progenitor physics are appropriately taken into account. Even when particle spray reproduces the stream (as measured statistically by the KLD or EMD), its internal structure depends on the particle-spray implementation (Section~\ref{subsec:comparisons}; Figures~\ref{fig:krios_ps_comp_3} and \ref{fig:krios_ps_comp_5}). Several different escaper distribution functions and mass-loss prescriptions should be considered in order to eliminate these as explanations for substructure. The mass-loss rate is computed self-consistently with \krios, so no prescription is needed. We find that there are limitations to more sophisticated models of a fixed progenitor in particle-spray methods; rather, it is likely more important to incorporate time-dependent progenitors that accommodate variable mass-loss rates, and impacts of a realistic stellar mass function and ejection via strong encounters \citep[see, e.g.,][]{gieles2021,grondin2024,weatherford2025}. 

Assuming a spherical model of the progenitor cluster, as is done in \cmc, is sufficient, in most cases, to model the escape of stars from the cluster (Section \ref{subsec:diff_progenitors}) and resolve the internal cluster dynamics (Section \ref{subsec:symmetry_breaking}) when predicting the density, velocity dispersion, and misalignment along the stream. This may not be the case for all clusters, e.g.~if there is internal rotation \citep{chen2010, white2025}. In the future, we plan to implement time-dependent host potentials from cosmological simulations, as there are known defects with stream models in a smooth \citep[e.g.,][]{bonaca2014} or static \citep[e.g.,][]{arora2022} host potential. This will be the foundation for future cosmological stellar stream catalogs \citep[e.g.,][]{panithanpaisal2025, holmhansen2025}, where the progenitor's {$N$-body} code is designed specifically for this use case. 

\krios\ runs are terminated once there are only 100 particles in the core (i.e., core collapse, see Section~\ref{sec:methods}). ${N\!\leq\! 4}$ microphysics \citep{heggie2003} are not yet accounted for, e.g. the production of binary star systems \citep[e.g.,][]{atallah2024}, which are critical to modeling the post-collapse evolution of globular clusters. Figure 2 of \cite{weatherford2023} demonstrates that these phenomena have a tangible impact on stream production as well, which necessitates their inclusion in future implementations of \krios. In future work, we plan to include both three-body binary formation and strong encounters between single and binary stars.  We will also model stellar/binary evolution with a realistic initial mass function \citep{kroupa2001, baumgardt2023} using the {\tt COSMIC} population-synthesis code \citep{breivik2020}. Before \krios\ is made public, we will add non-static host potentials from cosmological simulations \citep[e.g.,][]{arora2022, panithanpaisal2025} and the option to model dark matter subhalo fly-bys.

\section*{Acknowledgments}

We thank Dany Atallah for helpful discussions and the anonymous referee for their insightful comments, both of which improved the quality of the manuscript.  This work was supported by the National Science Foundation under Grants Numbers AST-2510181 to the University of North Carolina, AST-2510183 to Northwestern University, and by NASA ATP Grant 80NSSC24K0687. CR also acknowledges support from an Alfred P.~Sloan Research Fellowship and a David and Lucile Packard Foundation Fellowship.  BTC was partially funded by the North Carolina Space Grant's Graduate Research Fellowship. TS gratefully acknowledge the support of the NSF-Simons AI-Institute for the Sky (SkAI) via grants NSF AST-2421845 and Simons Foundation MPS-AI-00010513. TS was also supported by NASA through grant 22-ROMAN22-0013. This work was supported by a research grant (VIL53081) from VILLUM FONDEN. CR, KT, and RS also thank the organizers of the workshop ``Interconnections between the Physics of Plasmas and Self-gravitating Systems'', supported by NSF Grant PHY-2309135 to the Kavli Institute for Theoretical Physics (KITP).  This work was also co-funded by the European Union (ERC, BeyondSTREAMS, 101115754) grant. Views and opinions expressed are however those of the author(s) only and do not necessarily reflect those of the European Union or the European Research Council. Neither the European Union nor the granting authority can be held responsible for them.

We would like to thank the University of North Carolina at Chapel Hill and the Research Computing group for providing computational resources and support that have contributed to these research results. We thank St\'ephane Rouberol for the smooth running of the Infinity cluster of the Institute of Astrophysics of Paris, where the \nbody\ runs were performed. 

\krios\ simulations, and their subsequent analyses, depend on the following software packages: GNU Compiler Collection\footnote{\url{https://gcc.gnu.org/}}, GNU Scientific Library \citep[GSL,][]{gough2009}, OpenMP \citep{dagum1998}, numba \citep{lam2015}, NumPy \citep{harris2020}, SciPy \citep{virtanen2020}, pandas \citep{mckinney2010}, Matplotlib \citep{hunter2007}, Astropy \citep{astropy2018}, gala \citep{price-whelan2024}, galpy \citep{bovy2015}, Python Optimal Transport \citep[POT,][]{flamary2021}.

\section*{Data Distribution}

There is a public GitHub repository (\url{https://github.com/BrianTCook/krios_ii_paper_supplement}) containing a subset of the data presented in this paper, as well as a Jupyter notebook that produces relevant observables from that data. Additional data (e.g., {\tt .csv} files containing particle data from snapshots or time-series data about the tidal tensor) will be made available upon request to the authors.

To facilitate the reproducibility of our \nbody\ results and allow its use of the
 {\tt MWPotential2014} external field, we created a fork of the original \nbody\ repository containing these corrections (\href{https://github.com/KerwannTEP/Nbody6ppGPU}{https://github.com/KerwannTEP/Nbody6ppGPU}, see commit \textsf{5d506fb}). Additionally, we provide pre- and post-processing scripts at \href{https://github.com/KerwannTEP/NB6-MW2014-Tools}{https://github.com/KerwannTEP/NB6-MW2014-Tools}.

\bibliography{krios_streams}

@ARTICLE{rein2012,
       author = {{Rein}, H. and {Liu}, S.-F.},
        title = "{REBOUND: an open-source multi-purpose N-body code for collisional dynamics}",
      journal = {\aap},
         year = 2012,
        month = jan,
       volume = {537},
          eid = {A128},
        pages = {A128},
          doi = {10.1051/0004-6361/201118085},
archivePrefix = {arXiv},
       eprint = {1110.4876},
 primaryClass = {astro-ph.EP},
       adsurl = {https://ui.adsabs.harvard.edu/abs/2012A&A...537A.128R}
}

@article{Chatterjee2010,
	title = {Monte {Carlo} {Simulations} of {Globular} {Cluster} {Evolution}. {V}. {Binary} {Stellar} {Evolution}},
	volume = {719},
	url = {http://adsabs.harvard.edu/cgi-bin/nph-data_query?bibcode=2010ApJ...719..915C&link_type=EJOURNAL},
	number = {1},
	journal = {ApJ},
	author = {Chatterjee, Sourav and Fregeau, John M and Umbreit, Stefan and Rasio, Frederic A},
	year = {2010},
	note = {Publisher: Cornell University Library
Place: Department of Physics and Astronomy, Northwestern University, Evanston, IL 60208, USA},
	pages = {915--930},
}

@article{Wang2016,
	title = {The dragon simulations: globular cluster evolution with a million stars},
	volume = {458},
	issn = {0035-8711},
	url = {https://ui.adsabs.harvard.edu/#abs/2016MNRAS.458.1450W/abstract},
	doi = {10.1093/mnras/stw274},
	number = {2},
	urldate = {2016-03-20},
	journal = {MNRAS},
	author = {Wang, Long and Spurzem, Rainer and Aarseth, Sverre and Giersz, Mirek and Askar, Abbas and Berczik, Peter and Naab, Thorsten and Schadow, Riko and Kouwenhoven, M. B. N.},
	month = may,
	year = {2016},
	pages = {1450--1465},
}

@BOOK{heggie2003,
       author = {{Heggie}, Douglas and {Hut}, Piet},
        title = "{The Gravitational Million-Body Problem: A Multidisciplinary Approach to Star Cluster Dynamics}",
         year = 2003,
       adsurl = {https://ui.adsabs.harvard.edu/abs/2003gmbp.book.....H}
}

@ARTICLE{cohn1980,
       author = {{Cohn}, H.},
        title = "{Late core collapse in star clusters and the gravothermal instability}",
      journal = {\apj},
         year = 1980,
        month = dec,
       volume = {242},
        pages = {765-771},
          doi = {10.1086/158511},
       adsurl = {https://ui.adsabs.harvard.edu/abs/1980ApJ...242..765C}
}

@book{bovy2026,
  author    = {Jo Bovy},
  title     = {Dynamics and Astrophysics of Galaxies},
  publisher = {Princeton University Press},
  year      = {2026},
  note      = {In press. Available at \url{https://galaxiesbook.org}},
}

@ARTICLE{smith2015,
       author = {{Smith}, R. and {Flynn}, C. and {Candlish}, G.~N. and {Fellhauer}, M. and {Gibson}, B.~K.},
        title = "{Simple and accurate modelling of the gravitational potential produced by thick and thin exponential discs}",
      journal = {\mnras},
         year = 2015,
        month = apr,
       volume = {448},
       number = {3},
        pages = {2934-2940},
          doi = {10.1093/mnras/stv228},
archivePrefix = {arXiv},
       eprint = {1502.00627},
 primaryClass = {astro-ph.GA},
       adsurl = {https://ui.adsabs.harvard.edu/abs/2015MNRAS.448.2934S}
}

@software{price-whelan2024,
       author = {{Price-Whelan}, Adrian and {Souchereau}, Harrison and {Wagg}, Tom and {Sip{\H{o}}cz}, Brigitta and {Starkman}, Nathaniel and {Chen}, Yingtian and {Lilleengen}, Sophia and {NGC} and {Lenz}, Daniel and {Greco}, Johnny and {Hart}, Akeem and {AlexKurek} and {Robert}, Cl{\'e}ment and {Foreman-Mackey}, Dan and {HNLala} and {Lim}, P.~L. and {Oh}, Semyeong and {Koposov}, Sergey and {Li}, Zhaozhou},
        title = "{adrn/gala: v1.9.1}",
         year = 2024,
        month = aug,
          eid = {10.5281/zenodo.13377376},
          doi = {10.5281/zenodo.13377376},
      version = {v1.9.1},
    publisher = {Zenodo},
       adsurl = {https://ui.adsabs.harvard.edu/abs/2024zndo..13377376P}
}

@ARTICLE{abadi2003,
       author = {{Abadi}, Mario G. and {Navarro}, Julio F. and {Steinmetz}, Matthias and {Eke}, Vincent R.},
        title = "{Simulations of Galaxy Formation in a {\ensuremath{\Lambda}} Cold Dark Matter Universe. II. The Fine Structure of Simulated Galactic Disks}",
      journal = {\apj},
         year = 2003,
        month = nov,
       volume = {597},
       number = {1},
        pages = {21-34},
          doi = {10.1086/378316},
archivePrefix = {arXiv},
       eprint = {astro-ph/0212282},
 primaryClass = {astro-ph},
       adsurl = {https://ui.adsabs.harvard.edu/abs/2003ApJ...597...21A}
}

@ARTICLE{hernquist1990,
       author = {{Hernquist}, Lars},
        title = "{An Analytical Model for Spherical Galaxies and Bulges}",
      journal = {\apj},
         year = 1990,
        month = jun,
       volume = {356},
        pages = {359},
          doi = {10.1086/168845},
       adsurl = {https://ui.adsabs.harvard.edu/abs/1990ApJ...356..359H}
}

@ARTICLE{gieles2021,
       author = {{Gieles}, Mark and {Erkal}, Denis and {Antonini}, Fabio and {Balbinot}, Eduardo and {Pe{\~n}arrubia}, Jorge},
        title = "{A supra-massive population of stellar-mass black holes in the globular cluster Palomar 5}",
      journal = {Nature Astronomy},
         year = 2021,
        month = jul,
       volume = {5},
        pages = {957-966},
          doi = {10.1038/s41550-021-01392-2},
archivePrefix = {arXiv},
       eprint = {2102.11348},
 primaryClass = {astro-ph.GA},
       adsurl = {https://ui.adsabs.harvard.edu/abs/2021NatAs...5..957G}
}

@ARTICLE{mateu2023,
       author = {{Mateu}, Cecilia},
        title = "{galstreams: A library of Milky Way stellar stream footprints and tracks}",
      journal = {\mnras},
         year = 2023,
        month = apr,
       volume = {520},
       number = {4},
        pages = {5225-5258},
          doi = {10.1093/mnras/stad321},
archivePrefix = {arXiv},
       eprint = {2204.10326},
 primaryClass = {astro-ph.GA},
       adsurl = {https://ui.adsabs.harvard.edu/abs/2023MNRAS.520.5225M}
}

@ARTICLE{bonaca2025,
       author = {{Bonaca}, Ana and {Price-Whelan}, Adrian M.},
        title = "{Stellar streams in the Gaia era}",
      journal = {\nar},
         year = 2025,
        month = jun,
       volume = {100},
          eid = {101713},
        pages = {101713},
          doi = {10.1016/j.newar.2024.101713},
archivePrefix = {arXiv},
       eprint = {2405.19410},
 primaryClass = {astro-ph.GA},
       adsurl = {https://ui.adsabs.harvard.edu/abs/2025NewAR.10001713B}
}

@ARTICLE{gnedin1997,
       author = {{Gnedin}, Oleg Y. and {Ostriker}, Jeremiah P.},
        title = "{Destruction of the Galactic Globular Cluster System}",
      journal = {\apj},
         year = 1997,
        month = jan,
       volume = {474},
       number = {1},
        pages = {223-255},
          doi = {10.1086/303441},
archivePrefix = {arXiv},
       eprint = {astro-ph/9603042},
 primaryClass = {astro-ph},
       adsurl = {https://ui.adsabs.harvard.edu/abs/1997ApJ...474..223G}
}

@ARTICLE{pearson2017,
       author = {{Pearson}, Sarah and {Price-Whelan}, Adrian M. and {Johnston}, Kathryn V.},
        title = "{Gaps and length asymmetry in the stellar stream Palomar 5 as effects of Galactic bar rotation}",
      journal = {Nature Astronomy},
         year = 2017,
        month = aug,
       volume = {1},
        pages = {633-639},
          doi = {10.1038/s41550-017-0220-3},
archivePrefix = {arXiv},
       eprint = {1703.04627},
 primaryClass = {astro-ph.GA},
       adsurl = {https://ui.adsabs.harvard.edu/abs/2017NatAs...1..633P}
}

@Article{         harris2020,
 title         = {Array programming with {NumPy}},
 author        = {Charles R. Harris and K. Jarrod Millman and St{\'{e}}fan J.
                 van der Walt and Ralf Gommers and Pauli Virtanen and David
                 Cournapeau and Eric Wieser and Julian Taylor and Sebastian
                 Berg and Nathaniel J. Smith and Robert Kern and Matti Picus
                 and Stephan Hoyer and Marten H. van Kerkwijk and Matthew
                 Brett and Allan Haldane and Jaime Fern{\'{a}}ndez del
                 R{\'{i}}o and Mark Wiebe and Pearu Peterson and Pierre
                 G{\'{e}}rard-Marchant and Kevin Sheppard and Tyler Reddy and
                 Warren Weckesser and Hameer Abbasi and Christoph Gohlke and
                 Travis E. Oliphant},
 year          = {2020},
 month         = sep,
 journal       = {Nature},
 volume        = {585},
 number        = {7825},
 pages         = {357--362},
 doi           = {10.1038/s41586-020-2649-2},
 publisher     = {Springer Science and Business Media {LLC}},
 url           = {https://doi.org/10.1038/s41586-020-2649-2}
}

@ARTICLE{kupper2010,
       author = {{K{\"u}pper}, Andreas H.~W. and {Kroupa}, Pavel and {Baumgardt}, Holger and {Heggie}, Douglas C.},
        title = "{Tidal tails of star clusters}",
      journal = {\mnras},
         year = 2010,
        month = jan,
       volume = {401},
       number = {1},
        pages = {105-120},
          doi = {10.1111/j.1365-2966.2009.15690.x},
archivePrefix = {arXiv},
       eprint = {0909.2619},
 primaryClass = {astro-ph.SR},
       adsurl = {https://ui.adsabs.harvard.edu/abs/2010MNRAS.401..105K}
}

@ARTICLE{riley2020,
       author = {{Riley}, Alexander H. and {Strigari}, Louis E.},
        title = "{The Milky Way's stellar streams and globular clusters do not align in a Vast Polar Structure}",
      journal = {\mnras},
         year = 2020,
        month = may,
       volume = {494},
       number = {1},
        pages = {983-1001},
          doi = {10.1093/mnras/staa710},
archivePrefix = {arXiv},
       eprint = {2001.11564},
 primaryClass = {astro-ph.GA},
       adsurl = {https://ui.adsabs.harvard.edu/abs/2020MNRAS.494..983R}
}

@ARTICLE{vasiliev2021,
       author = {{Vasiliev}, Eugene and {Baumgardt}, Holger},
        title = "{Gaia EDR3 view on galactic globular clusters}",
      journal = {\mnras},
         year = 2021,
        month = aug,
       volume = {505},
       number = {4},
        pages = {5978-6002},
          doi = {10.1093/mnras/stab1475},
archivePrefix = {arXiv},
       eprint = {2102.09568},
 primaryClass = {astro-ph.GA},
       adsurl = {https://ui.adsabs.harvard.edu/abs/2021MNRAS.505.5978V}
}

@ARTICLE{king1966,
       author = {{King}, Ivan R.},
        title = "{The structure of star clusters. III. Some simple dynamical models}",
      journal = {\aj},
         year = 1966,
        month = feb,
       volume = {71},
        pages = {64},
          doi = {10.1086/109857},
       adsurl = {https://ui.adsabs.harvard.edu/abs/1966AJ.....71...64K}
}

@ARTICLE{sanderson2015,
       author = {{Sanderson}, Robyn E. and {Helmi}, Amina and {Hogg}, David W.},
        title = "{Action-space Clustering of Tidal Streams to Infer the Galactic Potential}",
      journal = {\apj},
         year = 2015,
        month = mar,
       volume = {801},
       number = {2},
          eid = {98},
        pages = {98},
          doi = {10.1088/0004-637X/801/2/98},
archivePrefix = {arXiv},
       eprint = {1404.6534},
 primaryClass = {astro-ph.GA},
       adsurl = {https://ui.adsabs.harvard.edu/abs/2015ApJ...801...98S}
}

@INPROCEEDINGS{cohen1999,
  author={Cohen, S. and Guibasm, L.},
  booktitle={Proceedings of the Seventh IEEE International Conference on Computer Vision}, 
  title={The Earth Mover's Distance under transformation sets}, 
  year={1999},
  volume={2},
  number={},
  pages={1076-1083 vol.2},
  doi={10.1109/ICCV.1999.790393}}

@misc{finn2004,
  author       = {David L. Finn},
  title        = {MA 323 Geometric Modelling: Course Notes, Day 09 --- Quintic Hermite Interpolation},
  year         = {2004},
  note         = {December 13},
  url = {https://www.rose-hulman.edu/~finn/CCLI/Notes/day09.pdf}
}

@ARTICLE{facey2026,
       author = {{Facey}, Alexander E.~V. and {Petersen}, Michael S.},
        title = "{Fast, self-consistent stellar stream simulations with basis-function expansions in EXP}",
      journal = {\mnras},
         year = 2026,
        month = nov,
       volume = {552},
       number = {4},
          eid = {stag1818},
        pages = {stag1818},
          doi = {10.1093/mnras/stag1818},
       adsurl = {https://ui.adsabs.harvard.edu/abs/2026MNRAS.552g1818F}
}

@article{rehman2014,
  title = {Accuracy and Computational Cost of  Interpolation Schemes While Performing &amp;lt;i&amp;gt;N&amp;lt;/i&amp;gt;-Body Simulations},
  volume = {04},
  ISSN = {2161-1211},
  url = {http://dx.doi.org/10.4236/ajcm.2014.45037},
  DOI = {10.4236/ajcm.2014.45037},
  number = {05},
  journal = {American Journal of Computational Mathematics},
  publisher = {Scientific Research Publishing,  Inc.},
  author = {Rehman,  Shafiq Ur},
  year = {2014},
  pages = {446–454}
}

@ARTICLE{baumgardt2003,
       author = {{Baumgardt}, Holger and {Makino}, Junichiro},
        title = "{Dynamical evolution of star clusters in tidal fields}",
      journal = {\mnras},
         year = 2003,
        month = mar,
       volume = {340},
       number = {1},
        pages = {227-246},
          doi = {10.1046/j.1365-8711.2003.06286.x},
archivePrefix = {arXiv},
       eprint = {astro-ph/0211471},
 primaryClass = {astro-ph},
       adsurl = {https://ui.adsabs.harvard.edu/abs/2003MNRAS.340..227B}
}

@article{lamers2010,
    author = {Lamers, Henny J. G. L. M. and Baumgardt, Holger and Gieles, Mark},
    title = {Mass-loss rates and the mass evolution of star clusters},
    journal = {Monthly Notices of the Royal Astronomical Society},
    volume = {409},
    number = {1},
    pages = {305-328},
    year = {2010},
    month = {11},
    issn = {0035-8711},
    doi = {10.1111/j.1365-2966.2010.17309.x},
    url = {https://doi.org/10.1111/j.1365-2966.2010.17309.x},
    eprint = {https://academic.oup.com/mnras/article-pdf/409/1/305/18485868/mnras0409-0305.pdf},
}

@ARTICLE{baumgardt2023,
       author = {{Baumgardt}, H. and {H{\'e}nault-Brunet}, V. and {Dickson}, N. and {Sollima}, A.},
        title = "{Evidence for a bottom-light initial mass function in massive star clusters}",
      journal = {\mnras},
         year = 2023,
        month = may,
       volume = {521},
       number = {3},
        pages = {3991-4008},
          doi = {10.1093/mnras/stad631},
archivePrefix = {arXiv},
       eprint = {2303.01636},
 primaryClass = {astro-ph.GA},
       adsurl = {https://ui.adsabs.harvard.edu/abs/2023MNRAS.521.3991B}
}

@ARTICLE{belokurov2024,
       author = {{Belokurov}, Vasily and {Kravtsov}, Andrey},
        title = "{In-situ versus accreted Milky Way globular clusters: a new classification method and implications for cluster formation}",
      journal = {\mnras},
         year = 2024,
        month = feb,
       volume = {528},
       number = {2},
        pages = {3198-3216},
          doi = {10.1093/mnras/stad3920},
archivePrefix = {arXiv},
       eprint = {2309.15902},
 primaryClass = {astro-ph.GA},
       adsurl = {https://ui.adsabs.harvard.edu/abs/2024MNRAS.528.3198B}
}

@Article{hunter2007,
  Author    = {Hunter, J. D.},
  Title     = {Matplotlib: A 2D graphics environment},
  Journal   = {Computing in Science \& Engineering},
  Volume    = {9},
  Number    = {3},
  Pages     = {90--95},
  publisher = {IEEE COMPUTER SOC},
  doi       = {10.1109/MCSE.2007.55},
  year      = 2007
}

@InProceedings{mckinney2010,
  author    = { Wes McKinney },
  title     = { Data Structures for Statistical Computing in Python },
  booktitle = { Proceedings of the 9th Python in Science Conference },
  pages     = { 51 - 56 },
  year      = { 2010 },
  editor    = { St\'efan van der Walt and Jarrod Millman }
}

@INPROCEEDINGS{cloud2024,
       author = {{Cloud}, Aiden and {Carr}, Christopher and {Tavangar}, Kiyan and {Johnston}, Kathryn},
        title = "{Connecting the stellar stream Gunnthr{\'a} to the globular cluster {\ensuremath{\omega}}Cen}",
    booktitle = {American Astronomical Society Meeting Abstracts \#243},
         year = 2024,
       series = {American Astronomical Society Meeting Abstracts},
       volume = {243},
        month = feb,
          eid = {458.24},
        pages = {458.24},
       adsurl = {https://ui.adsabs.harvard.edu/abs/2024AAS...24345824C}
}

@ARTICLE{ibata2021,
       author = {{Ibata}, Rodrigo and {Malhan}, Khyati and {Martin}, Nicolas and {Aubert}, Dominique and {Famaey}, Benoit and {Bianchini}, Paolo and {Monari}, Giacomo and {Siebert}, Arnaud and {Thomas}, Guillaume F. and {Bellazzini}, Michele and {Bonifacio}, Piercarlo and {Caffau}, Elisabetta and {Renaud}, Florent},
        title = "{Charting the Galactic Acceleration Field. I. A Search for Stellar Streams with Gaia DR2 and EDR3 with Follow-up from ESPaDOnS and UVES}",
      journal = {\apj},
         year = 2021,
        month = jun,
       volume = {914},
       number = {2},
          eid = {123},
        pages = {123},
          doi = {10.3847/1538-4357/abfcc2},
archivePrefix = {arXiv},
       eprint = {2012.05245},
 primaryClass = {astro-ph.GA},
       adsurl = {https://ui.adsabs.harvard.edu/abs/2021ApJ...914..123I}
}

@ARTICLE{thomas2020,
       author = {{Thomas}, Guillaume F. and {Jensen}, Jaclyn and {McConnachie}, Alan and {C{\^o}t{\'e}}, Patrick and {Venn}, Kim and {Longeard}, Nicolas and {Carlberg}, Raymond and {Chapman}, Scott and {Cuillandre}, Jean-Charles and {Famaey}, Benoit and {Ferrarese}, Laura and {Gwyn}, Stephen and {Hammer}, Fran{\c{c}}ois and {Ibata}, Rodrigo A. and {Malhan}, Khyati and {Martin}, Nicolas F. and {Mei}, Simona and {Navarro}, Julio F. and {Reyl{\'e}}, C{\'e}line and {Starkenburg}, Else},
        title = "{The Hidden Past of M92: Detection and Characterization of a Newly Formed 17{\textdegree} Long Stellar Stream Using the Canada-France Imaging Survey}",
      journal = {\apj},
         year = 2020,
        month = oct,
       volume = {902},
       number = {2},
          eid = {89},
        pages = {89},
          doi = {10.3847/1538-4357/abb6f7},
archivePrefix = {arXiv},
       eprint = {2009.04487},
 primaryClass = {astro-ph.GA},
       adsurl = {https://ui.adsabs.harvard.edu/abs/2020ApJ...902...89T}
}

@article{tep2025,
doi = {10.3847/1538-4357/ae0478},
url = {https://doi.org/10.3847/1538-4357/ae0478},
year = {2025},
month = {oct},
publisher = {The American Astronomical Society},
volume = {993},
number = {2},
pages = {180},
author = {Tep, Kerwann and Cook, Brian T. and Rodriguez, Carl L. and Jolly, Jiya and Sawin, Eddie and Petersen, Michael S. and Gaffud, Christoph},
title = {KRIOS: A New Basis-expansion N-body Code for Collisional Stellar Dynamics},
journal = {The Astrophysical Journal}
}

@ARTICLE{grudic2020,
       author = {{Grudi{\'c}}, Michael Y. and {Hopkins}, Philip F.},
        title = "{A general-purpose time-step criterion for simulations with gravity}",
      journal = {\mnras},
         year = 2020,
        month = jul,
       volume = {495},
       number = {4},
        pages = {4306-4313},
          doi = {10.1093/mnras/staa1453},
archivePrefix = {arXiv},
       eprint = {1910.06349},
 primaryClass = {astro-ph.IM},
       adsurl = {https://ui.adsabs.harvard.edu/abs/2020MNRAS.495.4306G}
}

@ARTICLE{fardal2015,
       author = {{Fardal}, Mark A. and {Huang}, Shuiyao and {Weinberg}, Martin D.},
        title = "{Generation of mock tidal streams}",
      journal = {\mnras},
         year = 2015,
        month = sep,
       volume = {452},
       number = {1},
        pages = {301-319},
          doi = {10.1093/mnras/stv1198},
archivePrefix = {arXiv},
       eprint = {1410.1861},
 primaryClass = {astro-ph.GA},
       adsurl = {https://ui.adsabs.harvard.edu/abs/2015MNRAS.452..301F}
}

@ARTICLE{Rodriguez2018,
       author = {{Rodriguez}, Carl L. and {Pattabiraman}, Bharath and {Chatterjee}, Sourav and {Choudhary}, Alok and {Liao}, Wei-keng and {Morscher}, Meagan and {Rasio}, Frederic A.},
        title = "{A new hybrid technique for modeling dense star clusters}",
      journal = {Computational Astrophysics and Cosmology},
         year = 2018,
        month = nov,
       volume = {5},
       number = {1},
          eid = {5},
        pages = {5},
          doi = {10.1186/s40668-018-0027-3},
archivePrefix = {arXiv},
       eprint = {1511.00695},
 primaryClass = {astro-ph.IM},
       adsurl = {https://ui.adsabs.harvard.edu/abs/2018ComAC...5....5R}
}

@ARTICLE{pfeffer2018,
       author = {{Pfeffer}, Joel and {Kruijssen}, J.~M. Diederik and {Crain}, Robert A. and {Bastian}, Nate},
        title = "{The E-MOSAICS project: simulating the formation and co-evolution of galaxies and their star cluster populations}",
      journal = {\mnras},
         year = 2018,
        month = apr,
       volume = {475},
       number = {4},
        pages = {4309-4346},
          doi = {10.1093/mnras/stx3124},
archivePrefix = {arXiv},
       eprint = {1712.00019},
 primaryClass = {astro-ph.GA},
       adsurl = {https://ui.adsabs.harvard.edu/abs/2018MNRAS.475.4309P}
}

@ARTICLE{malhan2022,
       author = {{Malhan}, Khyati and {Ibata}, Rodrigo A. and {Sharma}, Sanjib and {Famaey}, Benoit and {Bellazzini}, Michele and {Carlberg}, Raymond G. and {D'Souza}, Richard and {Yuan}, Zhen and {Martin}, Nicolas F. and {Thomas}, Guillaume F.},
        title = "{The Global Dynamical Atlas of the Milky Way Mergers: Constraints from Gaia EDR3-based Orbits of Globular Clusters, Stellar Streams, and Satellite Galaxies}",
      journal = {\apj},
         year = 2022,
        month = feb,
       volume = {926},
       number = {2},
          eid = {107},
        pages = {107},
          doi = {10.3847/1538-4357/ac4d2a},
archivePrefix = {arXiv},
       eprint = {2202.07660},
 primaryClass = {astro-ph.GA},
       adsurl = {https://ui.adsabs.harvard.edu/abs/2022ApJ...926..107M}
}

@ARTICLE{weatherford2025,
       author = {{Weatherford}, Newlin C. and {Bonaca}, Ana},
        title = "{Kinematics of Stellar Streams from Globular Clusters Depend on Black Hole Retention and Star Mass: A Selection Effect for Dark Matter Inference}",
      journal = {arXiv e-prints},
         year = 2025,
        month = sep,
          eid = {arXiv:2509.15307},
        pages = {arXiv:2509.15307},
archivePrefix = {arXiv},
       eprint = {2509.15307},
 primaryClass = {astro-ph.GA},
       adsurl = {https://ui.adsabs.harvard.edu/abs/2025arXiv250915307W}
}

@ARTICLE{atallah2024,
       author = {{Atallah}, Dany and {Weatherford}, Newlin C. and {Trani}, Alessandro A. and {Rasio}, Frederic A.},
        title = "{On Binary Formation from Three Initially Unbound Bodies}",
      journal = {\apj},
         year = 2024,
        month = aug,
       volume = {970},
       number = {2},
          eid = {112},
        pages = {112},
          doi = {10.3847/1538-4357/ad5185},
archivePrefix = {arXiv},
       eprint = {2402.12429},
 primaryClass = {astro-ph.SR},
       adsurl = {https://ui.adsabs.harvard.edu/abs/2024ApJ...970..112A}
}

@ARTICLE{bonaca2019,
       author = {{Bonaca}, Ana and {Hogg}, David W. and {Price-Whelan}, Adrian M. and {Conroy}, Charlie},
        title = "{The Spur and the Gap in GD-1: Dynamical Evidence for a Dark Substructure in the Milky Way Halo}",
      journal = {\apj},
         year = 2019,
        month = jul,
       volume = {880},
       number = {1},
          eid = {38},
        pages = {38},
          doi = {10.3847/1538-4357/ab2873},
archivePrefix = {arXiv},
       eprint = {1811.03631},
 primaryClass = {astro-ph.GA},
       adsurl = {https://ui.adsabs.harvard.edu/abs/2019ApJ...880...38B}
}

@ARTICLE{guillaume2026,
       author = {{Guillaume}, Claire and {Renaud}, Florent and {Martin}, Nicolas F. and {Famaey}, Benoit and {Di Matteo}, Paola and {Thomas}, Guillaume F. and {Ferrone}, Salvatore and {Ibata}, Rodrigo and {Pagnini}, Giulia},
        title = "{Asymmetries in stellar streams induced by a galactic merger}",
      journal = {\aap},
         year = 2026,
        month = jan,
       volume = {705},
          eid = {A6},
        pages = {A6},
          doi = {10.1051/0004-6361/202557552},
archivePrefix = {arXiv},
       eprint = {2510.06329},
 primaryClass = {astro-ph.GA},
       adsurl = {https://ui.adsabs.harvard.edu/abs/2026A&A...705A...6G}
}

@ARTICLE{weerasooriya2025,
       author = {{Weerasooriya}, Sachi and {Starkenburg}, Tjitske and {Cunningham}, Emily C. and {Johnston}, Kathryn V},
        title = "{Dancing Streams In Merging Halos: Stellar Streams in a MW--LMC-like merger}",
      journal = {arXiv e-prints},
         year = 2025,
        month = may,
          eid = {arXiv:2505.14792},
        pages = {arXiv:2505.14792},
          doi = {10.48550/arXiv.2505.14792},
archivePrefix = {arXiv},
       eprint = {2505.14792},
 primaryClass = {astro-ph.GA},
       adsurl = {https://ui.adsabs.harvard.edu/abs/2025arXiv250514792W}
}

@ARTICLE{penarrubia2009,
       author = {{Pe{\~n}arrubia}, Jorge and {Walker}, Matthew G. and {Gilmore}, Gerard},
        title = "{Tidal disruption of globular clusters in dwarf galaxies with triaxial dark matter haloes}",
      journal = {\mnras},
         year = 2009,
        month = nov,
       volume = {399},
       number = {3},
        pages = {1275-1292},
          doi = {10.1111/j.1365-2966.2009.15027.x},
archivePrefix = {arXiv},
       eprint = {0905.0924},
 primaryClass = {astro-ph.GA},
       adsurl = {https://ui.adsabs.harvard.edu/abs/2009MNRAS.399.1275P}
}

@ARTICLE{stodolkiewicz1982,
       author = {{Stodolkiewicz}, J.~S.},
        title = "{Dynamical evolution of globular clusters. I}",
      journal = {\actaa},
         year = 1982,
        month = jan,
       volume = {32},
       number = {1-2},
        pages = {63-91},
       adsurl = {https://ui.adsabs.harvard.edu/abs/1982AcA....32...63S}
}

@ARTICLE{vasiliev2015,
       author = {{Vasiliev}, Eugene},
        title = "{A new Monte Carlo method for dynamical evolution of non-spherical stellar systems}",
      journal = {\mnras},
         year = 2015,
        month = jan,
       volume = {446},
       number = {3},
        pages = {3150-3161},
          doi = {10.1093/mnras/stu2360},
archivePrefix = {arXiv},
       eprint = {1411.1757},
 primaryClass = {astro-ph.GA},
       adsurl = {https://ui.adsabs.harvard.edu/abs/2015MNRAS.446.3150V}
}

@ARTICLE{bonaca2021,
       author = {{Bonaca}, Ana and {Naidu}, Rohan P. and {Conroy}, Charlie and {Caldwell}, Nelson and {Cargile}, Phillip A. and {Han}, Jiwon Jesse and {Johnson}, Benjamin D. and {Kruijssen}, J.~M. Diederik and {Myeong}, G.~C. and {Speagle}, Joshua S. and {Ting}, Yuan-Sen and {Zaritsky}, Dennis},
        title = "{Orbital Clustering Identifies the Origins of Galactic Stellar Streams}",
      journal = {\apjl},
         year = 2021,
        month = mar,
       volume = {909},
       number = {2},
          eid = {L26},
        pages = {L26},
          doi = {10.3847/2041-8213/abeaa9},
archivePrefix = {arXiv},
       eprint = {2012.09171},
 primaryClass = {astro-ph.GA},
       adsurl = {https://ui.adsabs.harvard.edu/abs/2021ApJ...909L..26B}
}

@ARTICLE{bovy2015,
       author = {{Bovy}, Jo},
        title = "{galpy: A python Library for Galactic Dynamics}",
      journal = {\apjs},
         year = 2015,
        month = feb,
       volume = {216},
       number = {2},
          eid = {29},
        pages = {29},
          doi = {10.1088/0067-0049/216/2/29},
archivePrefix = {arXiv},
       eprint = {1412.3451},
 primaryClass = {astro-ph.GA},
       adsurl = {https://ui.adsabs.harvard.edu/abs/2015ApJS..216...29B}
}

@ARTICLE{aarseth1999,
       author = {{Aarseth}, Sverre J.},
        title = "{From NBODY1 to NBODY6: The Growth of an Industry}",
      journal = {\pasp},
         year = 1999,
        month = nov,
       volume = {111},
       number = {765},
        pages = {1333-1346},
          doi = {10.1086/316455},
       adsurl = {https://ui.adsabs.harvard.edu/abs/1999PASP..111.1333A}
}

@BOOK{spitzer1987,
       author = {{Spitzer}, Lyman},
        title = "{Dynamical evolution of globular clusters}",
         year = 1987,
       adsurl = {https://ui.adsabs.harvard.edu/abs/1987degc.book.....S}
}

@BOOK{binney2008,
       author = {{Binney}, James and {Tremaine}, Scott},
        title = "{Galactic Dynamics: Second Edition}",
         year = 2008,
       adsurl = {https://ui.adsabs.harvard.edu/abs/2008gady.book.....B}
}

@ARTICLE{bullock2005,
       author = {{Bullock}, James S. and {Johnston}, Kathryn V.},
        title = "{Tracing Galaxy Formation with Stellar Halos. I. Methods}",
      journal = {\apj},
         year = 2005,
        month = dec,
       volume = {635},
       number = {2},
        pages = {931-949},
          doi = {10.1086/497422},
archivePrefix = {arXiv},
       eprint = {astro-ph/0506467},
 primaryClass = {astro-ph},
       adsurl = {https://ui.adsabs.harvard.edu/abs/2005ApJ...635..931B}
}

@misc{soch2020,
  title        = {Proof: Kullback-Leibler divergence for the normal distribution},
  author       = {Soch, Joram},
  howpublished = {The Book of Statistical Proofs, 2025 release},
  publisher    = {Zenodo},
  year         = {2025},
  doi          = {10.5281/zenodo.4305949},
  url          = {https://statproofbook.github.io/P/norm-kl.html},
  note         = {Proof ID~P193, shortcut \texttt{norm-kl};
                  original date 2020-05-05},
}

@article{spergel2000,
  title = {Observational Evidence for Self-Interacting Cold Dark Matter},
  author = {Spergel, David N. and Steinhardt, Paul J.},
  journal = {Phys. Rev. Lett.},
  volume = {84},
  issue = {17},
  pages = {3760--3763},
  numpages = {0},
  year = {2000},
  month = {Apr},
  publisher = {American Physical Society},
  doi = {10.1103/PhysRevLett.84.3760},
  url = {https://link.aps.org/doi/10.1103/PhysRevLett.84.3760}
}

@ARTICLE{bullock2017,
       author = {{Bullock}, James S. and {Boylan-Kolchin}, Michael},
        title = "{Small-Scale Challenges to the {\ensuremath{\Lambda}}CDM Paradigm}",
      journal = {\araa},
         year = 2017,
        month = aug,
       volume = {55},
       number = {1},
        pages = {343-387},
          doi = {10.1146/annurev-astro-091916-055313},
archivePrefix = {arXiv},
       eprint = {1707.04256},
 primaryClass = {astro-ph.CO},
       adsurl = {https://ui.adsabs.harvard.edu/abs/2017ARA&A..55..343B}
}

@ARTICLE{wang2015,
       author = {{Wang}, Long and {Spurzem}, Rainer and {Aarseth}, Sverre and {Nitadori}, Keigo and {Berczik}, Peter and {Kouwenhoven}, M.~B.~N. and {Naab}, Thorsten},
        title = "{NBODY6++GPU: ready for the gravitational million-body problem}",
      journal = {\mnras},
         year = 2015,
        month = jul,
       volume = {450},
       number = {4},
        pages = {4070-4080},
          doi = {10.1093/mnras/stv817},
archivePrefix = {arXiv},
       eprint = {1504.03687},
 primaryClass = {astro-ph.IM},
       adsurl = {https://ui.adsabs.harvard.edu/abs/2015MNRAS.450.4070W}
}

@ARTICLE{zhao1996,
       author = {{Zhao}, Hongsheng},
        title = "{Analytical models for galactic nuclei}",
      journal = {\mnras},
         year = 1996,
        month = jan,
       volume = {278},
       number = {2},
        pages = {488-496},
          doi = {10.1093/mnras/278.2.488},
archivePrefix = {arXiv},
       eprint = {astro-ph/9509122},
 primaryClass = {astro-ph},
       adsurl = {https://ui.adsabs.harvard.edu/abs/1996MNRAS.278..488Z}
}

@ARTICLE{ibata2019b,
       author = {{Ibata}, Rodrigo A. and {Malhan}, Khyati and {Martin}, Nicolas F.},
        title = "{The Streams of the Gaping Abyss: A Population of Entangled Stellar Streams Surrounding the Inner Galaxy}",
      journal = {\apj},
         year = 2019,
        month = feb,
       volume = {872},
       number = {2},
          eid = {152},
        pages = {152},
          doi = {10.3847/1538-4357/ab0080},
archivePrefix = {arXiv},
       eprint = {1901.07566},
 primaryClass = {astro-ph.GA},
       adsurl = {https://ui.adsabs.harvard.edu/abs/2019ApJ...872..152I}
}

@ARTICLE{ibata2019,
       author = {{Ibata}, Rodrigo A. and {Bellazzini}, Michele and {Malhan}, Khyati and
         {Martin}, Nicolas and {Bianchini}, Paolo},
        title = "{Identification of the long stellar stream of the prototypical massive globular cluster {\ensuremath{\omega}} Centauri}",
      journal = {Nature Astronomy},
         year = "2019",
        month = "Apr",
       volume = {3},
        pages = {667-672},
          doi = {10.1038/s41550-019-0751-x},
archivePrefix = {arXiv},
       eprint = {1902.09544},
 primaryClass = {astro-ph.GA},
       adsurl = {https://ui.adsabs.harvard.edu/abs/2019NatAs...3..667I}
}

@ARTICLE{virtanen2020,
  author  = {Virtanen, Pauli and Gommers, Ralf and Oliphant, Travis E. and
            Haberland, Matt and Reddy, Tyler and Cournapeau, David and
            Burovski, Evgeni and Peterson, Pearu and Weckesser, Warren and
            Bright, Jonathan and {van der Walt}, St{\'e}fan J. and
            Brett, Matthew and Wilson, Joshua and Millman, K. Jarrod and
            Mayorov, Nikolay and Nelson, Andrew R. J. and Jones, Eric and
            Kern, Robert and Larson, Eric and Carey, C J and
            Polat, {\.I}lhan and Feng, Yu and Moore, Eric W. and
            {VanderPlas}, Jake and Laxalde, Denis and Perktold, Josef and
            Cimrman, Robert and Henriksen, Ian and Quintero, E. A. and
            Harris, Charles R. and Archibald, Anne M. and
            Ribeiro, Ant{\^o}nio H. and Pedregosa, Fabian and
            {van Mulbregt}, Paul and {SciPy 1.0 Contributors}},
  title   = {{{SciPy} 1.0: Fundamental Algorithms for Scientific
            Computing in Python}},
  journal = {Nature Methods},
  year    = {2020},
  volume  = {17},
  pages   = {261--272},
  adsurl  = {https://rdcu.be/b08Wh},
  doi     = {10.1038/s41592-019-0686-2},
}

@ARTICLE{dagum1998,
  author={Dagum, L. and Menon, R.},
  journal={IEEE Computational Science and Engineering}, 
  title={OpenMP: an industry standard API for shared-memory programming}, 
  year={1998},
  volume={5},
  number={1},
  pages={46-55},
  doi={10.1109/99.660313}}

@ARTICLE{dehnen2000,
       author = {{Dehnen}, Walter},
        title = "{A Very Fast and Momentum-conserving Tree Code}",
      journal = {\apjl},
         year = 2000,
        month = jun,
       volume = {536},
       number = {1},
        pages = {L39-L42},
          doi = {10.1086/312724},
archivePrefix = {arXiv},
       eprint = {astro-ph/0003209},
 primaryClass = {astro-ph},
       adsurl = {https://ui.adsabs.harvard.edu/abs/2000ApJ...536L..39D}
}

@ARTICLE{sollima2014,
       author = {{Sollima}, A. and {Mastrobuono Battisti}, A.},
        title = "{Treatment of realistic tidal field in Monte Carlo simulations of star clusters}",
      journal = {\mnras},
         year = 2014,
        month = oct,
       volume = {443},
       number = {4},
        pages = {3513-3527},
          doi = {10.1093/mnras/stu1426},
archivePrefix = {arXiv},
       eprint = {1407.4169},
 primaryClass = {astro-ph.SR},
       adsurl = {https://ui.adsabs.harvard.edu/abs/2014MNRAS.443.3513S}
}

@ARTICLE{dehnen2002,
       author = {{Dehnen}, Walter},
        title = "{A Hierarchical <E10>O</E10>(N) Force Calculation Algorithm}",
      journal = {Journal of Computational Physics},
         year = 2002,
        month = jun,
       volume = {179},
       number = {1},
        pages = {27-42},
          doi = {10.1006/jcph.2002.7026},
archivePrefix = {arXiv},
       eprint = {astro-ph/0202512},
 primaryClass = {astro-ph},
       adsurl = {https://ui.adsabs.harvard.edu/abs/2002JCoPh.179...27D}
}

@ARTICLE{holmhansen2025,
       author = {{Holm-Hansen}, Colin and {Chen}, Yingtian and {Gnedin}, Oleg Y.},
        title = "{Catalog of Mock Stellar Streams in Milky Way-Like Galaxies}",
      journal = {arXiv e-prints},
         year = 2025,
        month = oct,
          eid = {arXiv:2510.09604},
        pages = {arXiv:2510.09604},
          doi = {10.48550/arXiv.2510.09604},
archivePrefix = {arXiv},
       eprint = {2510.09604},
 primaryClass = {astro-ph.GA},
       adsurl = {https://ui.adsabs.harvard.edu/abs/2025arXiv251009604H}
}

@ARTICLE{koposov2010,
       author = {{Koposov}, Sergey E. and {Rix}, Hans-Walter and {Hogg}, David W.},
        title = "{Constraining the Milky Way Potential with a Six-Dimensional Phase-Space Map of the GD-1 Stellar Stream}",
      journal = {\apj},
         year = 2010,
        month = mar,
       volume = {712},
       number = {1},
        pages = {260-273},
          doi = {10.1088/0004-637X/712/1/260},
archivePrefix = {arXiv},
       eprint = {0907.1085},
 primaryClass = {astro-ph.GA},
       adsurl = {https://ui.adsabs.harvard.edu/abs/2010ApJ...712..260K}
}

@ARTICLE{rodriguez2016,
       author = {{Rodriguez}, Carl L. and {Morscher}, Meagan and {Wang}, Long and {Chatterjee}, Sourav and {Rasio}, Frederic A. and {Spurzem}, Rainer},
        title = "{Million-body star cluster simulations: comparisons between Monte Carlo and direct N-body}",
      journal = {\mnras},
         year = 2016,
        month = dec,
       volume = {463},
       number = {2},
        pages = {2109-2118},
          doi = {10.1093/mnras/stw2121},
archivePrefix = {arXiv},
       eprint = {1601.04227},
 primaryClass = {astro-ph.IM},
       adsurl = {https://ui.adsabs.harvard.edu/abs/2016MNRAS.463.2109R}
}

@ARTICLE{giersz2013,
       author = {{Giersz}, Mirek and {Heggie}, Douglas C. and {Hurley}, Jarrod R. and {Hypki}, Arkadiusz},
        title = "{MOCCA code for star cluster simulations - II. Comparison with N-body simulations}",
      journal = {\mnras},
         year = 2013,
        month = may,
       volume = {431},
       number = {3},
        pages = {2184-2199},
          doi = {10.1093/mnras/stt307},
archivePrefix = {arXiv},
       eprint = {1112.6246},
 primaryClass = {astro-ph.GA},
       adsurl = {https://ui.adsabs.harvard.edu/abs/2013MNRAS.431.2184G}
}

@ARTICLE{lyndenbell1995,
       author = {{Lynden-Bell}, D. and {Lynden-Bell}, R.~M.},
        title = "{Ghostly streams from the formation of the Galaxy's halo}",
      journal = {\mnras},
         year = 1995,
        month = jul,
       volume = {275},
       number = {2},
        pages = {429-442},
          doi = {10.1093/mnras/275.2.429},
       adsurl = {https://ui.adsabs.harvard.edu/abs/1995MNRAS.275..429L}
}

@ARTICLE{palau2025,
       author = {{Palau}, Carles G. and {Wang}, Wenting and {Han}, Jiaxin},
        title = "{On the internal structure of stellar streams}",
      journal = {arXiv e-prints},
         year = 2025,
        month = aug,
          eid = {arXiv:2508.21408},
        pages = {arXiv:2508.21408},
          doi = {10.48550/arXiv.2508.21408},
archivePrefix = {arXiv},
       eprint = {2508.21408},
 primaryClass = {astro-ph.GA},
       adsurl = {https://ui.adsabs.harvard.edu/abs/2025arXiv250821408P}
}

@ARTICLE{kupper2012,
       author = {{K{\"u}pper}, Andreas H.~W. and {Lane}, Richard R. and {Heggie}, Douglas C.},
        title = "{More on the structure of tidal tails}",
      journal = {\mnras},
         year = 2012,
        month = mar,
       volume = {420},
       number = {3},
        pages = {2700-2714},
          doi = {10.1111/j.1365-2966.2011.20242.x},
archivePrefix = {arXiv},
       eprint = {1111.5013},
 primaryClass = {astro-ph.GA},
       adsurl = {https://ui.adsabs.harvard.edu/abs/2012MNRAS.420.2700K}
}

@ARTICLE{carlberg2025,
       author = {{Carlberg}, Raymond G.},
        title = "{GD-1 and the Milky Way Starless Dark Matter Subhalos}",
      journal = {\apj},
         year = 2025,
        month = aug,
       volume = {989},
       number = {1},
          eid = {38},
        pages = {38},
          doi = {10.3847/1538-4357/adec91},
archivePrefix = {arXiv},
       eprint = {2503.13290},
 primaryClass = {astro-ph.GA},
       adsurl = {https://ui.adsabs.harvard.edu/abs/2025ApJ...989...38C}
}

@ARTICLE{bonaca2014,
       author = {{Bonaca}, Ana and {Geha}, Marla and {K{\"u}pper}, Andreas H.~W. and {Diemand}, J{\"u}rg and {Johnston}, Kathryn V. and {Hogg}, David W.},
        title = "{Milky Way Mass and Potential Recovery Using Tidal Streams in a Realistic Halo}",
      journal = {\apj},
         year = 2014,
        month = nov,
       volume = {795},
       number = {1},
          eid = {94},
        pages = {94},
          doi = {10.1088/0004-637X/795/1/94},
archivePrefix = {arXiv},
       eprint = {1406.6063},
 primaryClass = {astro-ph.GA},
       adsurl = {https://ui.adsabs.harvard.edu/abs/2014ApJ...795...94B}
}

@ARTICLE{cole2000,
       author = {{Cole}, Shaun and {Lacey}, Cedric G. and {Baugh}, Carlton M. and {Frenk}, Carlos S.},
        title = "{Hierarchical galaxy formation}",
      journal = {\mnras},
         year = 2000,
        month = nov,
       volume = {319},
       number = {1},
        pages = {168-204},
          doi = {10.1046/j.1365-8711.2000.03879.x},
archivePrefix = {arXiv},
       eprint = {astro-ph/0007281},
 primaryClass = {astro-ph},
       adsurl = {https://ui.adsabs.harvard.edu/abs/2000MNRAS.319..168C}
}

@ARTICLE{erkal2019,
       author = {{Erkal}, D. and {Belokurov}, V. and {Laporte}, C.~F.~P. and {Koposov}, S.~E. and {Li}, T.~S. and {Grillmair}, C.~J. and {Kallivayalil}, N. and {Price-Whelan}, A.~M. and {Evans}, N.~W. and {Hawkins}, K. and {Hendel}, D. and {Mateu}, C. and {Navarro}, J.~F. and {del Pino}, A. and {Slater}, C.~T. and {Sohn}, S.~T. and {Orphan Aspen Treasury Collaboration}},
        title = "{The total mass of the Large Magellanic Cloud from its perturbation on the Orphan stream}",
      journal = {\mnras},
         year = 2019,
        month = aug,
       volume = {487},
       number = {2},
        pages = {2685-2700},
          doi = {10.1093/mnras/stz1371},
archivePrefix = {arXiv},
       eprint = {1812.08192},
 primaryClass = {astro-ph.GA},
       adsurl = {https://ui.adsabs.harvard.edu/abs/2019MNRAS.487.2685E}
}

@ARTICLE{gibbons2014,
       author = {{Gibbons}, S.~L.~J. and {Belokurov}, V. and {Evans}, N.~W.},
        title = "{`Skinny Milky Way please', says Sagittarius}",
      journal = {\mnras},
         year = 2014,
        month = dec,
       volume = {445},
       number = {4},
        pages = {3788-3802},
          doi = {10.1093/mnras/stu1986},
archivePrefix = {arXiv},
       eprint = {1406.2243},
 primaryClass = {astro-ph.GA},
       adsurl = {https://ui.adsabs.harvard.edu/abs/2014MNRAS.445.3788G}
}

@ARTICLE{pearson2024,
       author = {{Pearson}, Sarah and {Bonaca}, Ana and {Chen}, Yingtian and {Gnedin}, Oleg Y.},
        title = "{Forecasting the Population of Globular Cluster Streams in Milky Way{\textendash}type Galaxies}",
      journal = {\apj},
         year = 2024,
        month = nov,
       volume = {976},
       number = {1},
          eid = {54},
        pages = {54},
          doi = {10.3847/1538-4357/ad8348},
archivePrefix = {arXiv},
       eprint = {2405.15851},
 primaryClass = {astro-ph.GA},
       adsurl = {https://ui.adsabs.harvard.edu/abs/2024ApJ...976...54P}
}

@ARTICLE{chen2025b,
       author = {{Chen}, Yingtian and {Li}, Hui and {Gnedin}, Oleg Y.},
        title = "{Stellar Streams Reveal the Mass Loss of Globular Clusters}",
      journal = {\apjl},
         year = 2025,
        month = feb,
       volume = {980},
       number = {2},
          eid = {L18},
        pages = {L18},
          doi = {10.3847/2041-8213/adaf93},
archivePrefix = {arXiv},
       eprint = {2411.19899},
 primaryClass = {astro-ph.GA},
       adsurl = {https://ui.adsabs.harvard.edu/abs/2025ApJ...980L..18C}
}

@ARTICLE{astropy2018,
   author = {{Astropy Collaboration} and {Price-Whelan}, A.~M. and {Sip{\H o}cz}, B.~M. and 
	{G{\"u}nther}, H.~M. and {Lim}, P.~L. and {Crawford}, S.~M. and 
	{Conseil}, S. and {Shupe}, D.~L. and {Craig}, M.~W. and {Dencheva}, N. and 
	{Ginsburg}, A. and {VanderPlas}, J.~T. and {Bradley}, L.~D. and 
	{P{\'e}rez-Su{\'a}rez}, D. and {de Val-Borro}, M. and {Aldcroft}, T.~L. and 
	{Cruz}, K.~L. and {Robitaille}, T.~P. and {Tollerud}, E.~J. and 
	{Ardelean}, C. and {Babej}, T. and {Bach}, Y.~P. and {Bachetti}, M. and 
	{Bakanov}, A.~V. and {Bamford}, S.~P. and {Barentsen}, G. and 
	{Barmby}, P. and {Baumbach}, A. and {Berry}, K.~L. and {Biscani}, F. and 
	{Boquien}, M. and {Bostroem}, K.~A. and {Bouma}, L.~G. and {Brammer}, G.~B. and 
	{Bray}, E.~M. and {Breytenbach}, H. and {Buddelmeijer}, H. and 
	{Burke}, D.~J. and {Calderone}, G. and {Cano Rodr{\'{\i}}guez}, J.~L. and 
	{Cara}, M. and {Cardoso}, J.~V.~M. and {Cheedella}, S. and {Copin}, Y. and 
	{Corrales}, L. and {Crichton}, D. and {D'Avella}, D. and {Deil}, C. and 
	{Depagne}, {\'E}. and {Dietrich}, J.~P. and {Donath}, A. and 
	{Droettboom}, M. and {Earl}, N. and {Erben}, T. and {Fabbro}, S. and 
	{Ferreira}, L.~A. and {Finethy}, T. and {Fox}, R.~T. and {Garrison}, L.~H. and 
	{Gibbons}, S.~L.~J. and {Goldstein}, D.~A. and {Gommers}, R. and 
	{Greco}, J.~P. and {Greenfield}, P. and {Groener}, A.~M. and 
	{Grollier}, F. and {Hagen}, A. and {Hirst}, P. and {Homeier}, D. and 
	{Horton}, A.~J. and {Hosseinzadeh}, G. and {Hu}, L. and {Hunkeler}, J.~S. and 
	{Ivezi{\'c}}, {\v Z}. and {Jain}, A. and {Jenness}, T. and {Kanarek}, G. and 
	{Kendrew}, S. and {Kern}, N.~S. and {Kerzendorf}, W.~E. and 
	{Khvalko}, A. and {King}, J. and {Kirkby}, D. and {Kulkarni}, A.~M. and 
	{Kumar}, A. and {Lee}, A. and {Lenz}, D. and {Littlefair}, S.~P. and 
	{Ma}, Z. and {Macleod}, D.~M. and {Mastropietro}, M. and {McCully}, C. and 
	{Montagnac}, S. and {Morris}, B.~M. and {Mueller}, M. and {Mumford}, S.~J. and 
	{Muna}, D. and {Murphy}, N.~A. and {Nelson}, S. and {Nguyen}, G.~H. and 
	{Ninan}, J.~P. and {N{\"o}the}, M. and {Ogaz}, S. and {Oh}, S. and 
	{Parejko}, J.~K. and {Parley}, N. and {Pascual}, S. and {Patil}, R. and 
	{Patil}, A.~A. and {Plunkett}, A.~L. and {Prochaska}, J.~X. and 
	{Rastogi}, T. and {Reddy Janga}, V. and {Sabater}, J. and {Sakurikar}, P. and 
	{Seifert}, M. and {Sherbert}, L.~E. and {Sherwood-Taylor}, H. and 
	{Shih}, A.~Y. and {Sick}, J. and {Silbiger}, M.~T. and {Singanamalla}, S. and 
	{Singer}, L.~P. and {Sladen}, P.~H. and {Sooley}, K.~A. and 
	{Sornarajah}, S. and {Streicher}, O. and {Teuben}, P. and {Thomas}, S.~W. and 
	{Tremblay}, G.~R. and {Turner}, J.~E.~H. and {Terr{\'o}n}, V. and 
	{van Kerkwijk}, M.~H. and {de la Vega}, A. and {Watkins}, L.~L. and 
	{Weaver}, B.~A. and {Whitmore}, J.~B. and {Woillez}, J. and 
	{Zabalza}, V. and {Astropy Contributors}},
    title = "{The Astropy Project: Building an Open-science Project and Status of the v2.0 Core Package}",
  journal = {\aj},
archivePrefix = "arXiv",
   eprint = {1801.02634},
 primaryClass = "astro-ph.IM",
     year = 2018,
    month = sep,
   volume = 156,
      eid = {123},
    pages = {123},
      doi = {10.3847/1538-3881/aabc4f},
   adsurl = {https://ui.adsabs.harvard.edu/abs/2018AJ....156..123A}
}

@ARTICLE{kroupa2001,
       author = {{Kroupa}, Pavel},
        title = "{On the variation of the initial mass function}",
      journal = {\mnras},
         year = 2001,
        month = apr,
       volume = {322},
       number = {2},
        pages = {231-246},
          doi = {10.1046/j.1365-8711.2001.04022.x},
archivePrefix = {arXiv},
       eprint = {astro-ph/0009005},
 primaryClass = {astro-ph},
       adsurl = {https://ui.adsabs.harvard.edu/abs/2001MNRAS.322..231K}
}

@ARTICLE{chen2025,
       author = {{Chen}, Yingtian and {Valluri}, Monica and {Gnedin}, Oleg Y. and {Ash}, Neil},
        title = "{Improved Particle Spray Algorithm for Modeling Globular Cluster Streams}",
      journal = {\apjs},
         year = 2025,
        month = feb,
       volume = {276},
       number = {2},
          eid = {32},
        pages = {32},
          doi = {10.3847/1538-4365/ad9904},
archivePrefix = {arXiv},
       eprint = {2408.01496},
 primaryClass = {astro-ph.GA},
       adsurl = {https://ui.adsabs.harvard.edu/abs/2025ApJS..276...32C}
}

@ARTICLE{chen2024,
       author = {{Chen}, Yingtian and {Gnedin}, Oleg Y.},
        title = "{Galaxy assembly revealed by globular clusters}",
      journal = {The Open Journal of Astrophysics},
         year = 2024,
        month = mar,
       volume = {7},
          eid = {23},
        pages = {23},
          doi = {10.33232/001c.116169},
archivePrefix = {arXiv},
       eprint = {2401.17420},
 primaryClass = {astro-ph.GA},
       adsurl = {https://ui.adsabs.harvard.edu/abs/2024OJAp....7E..23C}
}

@ARTICLE{roberts2025,
       author = {{Roberts}, Daniel and {Gieles}, Mark and {Erkal}, Denis and {Sanders}, Jason L.},
        title = "{Stellar streams from black hole-rich star clusters}",
      journal = {\mnras},
         year = 2025,
        month = mar,
       volume = {538},
       number = {1},
        pages = {454-469},
          doi = {10.1093/mnras/staf321},
archivePrefix = {arXiv},
       eprint = {2402.06393},
 primaryClass = {astro-ph.GA},
       adsurl = {https://ui.adsabs.harvard.edu/abs/2025MNRAS.538..454R}
}

@article{fukushige2000,
    author = {Fukushige, T. and Heggie, D. C.},
    title = "{The time-scale of escape from star clusters}",
    journal = {Monthly Notices of the Royal Astronomical Society},
    volume = {318},
    number = {3},
    pages = {753-761},
    year = {2000},
    month = {11},
    issn = {0035-8711},
    doi = {10.1046/j.1365-8711.2000.03811.x},
    url = {https://doi.org/10.1046/j.1365-8711.2000.03811.x},
    eprint = {https://academic.oup.com/mnras/article-pdf/318/3/753/3471471/318-3-753.pdf},
}

@ARTICLE{bovy2017,
       author = {{Bovy}, Jo and {Erkal}, Denis and {Sanders}, Jason L.},
        title = "{Linear perturbation theory for tidal streams and the small-scale CDM power spectrum}",
      journal = {\mnras},
         year = 2017,
        month = apr,
       volume = {466},
       number = {1},
        pages = {628-668},
          doi = {10.1093/mnras/stw3067},
archivePrefix = {arXiv},
       eprint = {1606.03470},
 primaryClass = {astro-ph.GA},
       adsurl = {https://ui.adsabs.harvard.edu/abs/2017MNRAS.466..628B}
}

@article{miyamoto1975,
    author = "Miyamoto, M. and Nagai, R.",
    title = "{Three-dimensional models for the distribution of mass in galaxies}",
    journal = "Publ. Astron. Soc. Jap.",
    volume = "27",
    pages = "533--543",
    year = "1975"
}

@ARTICLE{navarro1996,
       author = {{Navarro}, Julio F. and {Frenk}, Carlos S. and {White}, Simon D.~M.},
        title = "{The Structure of Cold Dark Matter Halos}",
      journal = {\apj},
         year = 1996,
        month = may,
       volume = {462},
        pages = {563},
          doi = {10.1086/177173},
archivePrefix = {arXiv},
       eprint = {astro-ph/9508025},
 primaryClass = {astro-ph},
       adsurl = {https://ui.adsabs.harvard.edu/abs/1996ApJ...462..563N}
}

@ARTICLE{grillmair2006,
       author = {{Grillmair}, C.~J. and {Dionatos}, O.},
        title = "{Detection of a 63{\textdegree} Cold Stellar Stream in the Sloan Digital Sky Survey}",
      journal = {\apjl},
         year = 2006,
        month = may,
       volume = {643},
       number = {1},
        pages = {L17-L20},
          doi = {10.1086/505111},
archivePrefix = {arXiv},
       eprint = {astro-ph/0604332},
 primaryClass = {astro-ph},
       adsurl = {https://ui.adsabs.harvard.edu/abs/2006ApJ...643L..17G}
}

@article{flamary2021,
  author  = {RÃ©mi Flamary and Nicolas Courty and Alexandre Gramfort and Mokhtar Z. Alaya and AurÃ©lie Boisbunon and Stanislas Chambon and Laetitia Chapel and Adrien Corenflos and Kilian Fatras and Nemo Fournier and LÃ©o Gautheron and Nathalie T.H. Gayraud and Hicham Janati and Alain Rakotomamonjy and Ievgen Redko and Antoine Rolet and Antony Schutz and Vivien Seguy and Danica J. Sutherland and Romain Tavenard and Alexander Tong and Titouan Vayer},
  title   = {POT: Python Optimal Transport},
  journal = {Journal of Machine Learning Research},
  year    = {2021},
  volume  = {22},
  number  = {78},
  pages   = {1--8},
  url     = {http://jmlr.org/papers/v22/20-451.html}
}

@INPROCEEDINGS{lam2015,
       author = {{Lam}, Siu Kwan and {Pitrou}, Antoine and {Seibert}, Stanley},
        title = "{Numba: A LLVM-based Python JIT Compiler}",
    booktitle = {Proc. Second Workshop on the LLVM Compiler Infrastructure in HPC},
         year = 2015,
        month = nov,
        pages = {1-6},
          doi = {10.1145/2833157.2833162},
       adsurl = {https://ui.adsabs.harvard.edu/abs/2015llvm.confE...1L}
}

@ARTICLE{malhan2019a,
       author = {{Malhan}, Khyati and {Ibata}, Rodrigo A. and {Carlberg}, Raymond G. and {Bellazzini}, Michele and {Famaey}, Benoit and {Martin}, Nicolas F.},
        title = "{Phase-space Correlation in Stellar Streams of the Milky Way Halo: The Clash of Kshir and GD-1}",
      journal = {\apjl},
         year = 2019,
        month = nov,
       volume = {886},
       number = {1},
          eid = {L7},
        pages = {L7},
          doi = {10.3847/2041-8213/ab530e},
archivePrefix = {arXiv},
       eprint = {1911.00009},
 primaryClass = {astro-ph.GA},
       adsurl = {https://ui.adsabs.harvard.edu/abs/2019ApJ...886L...7M}
}

@ARTICLE{mukherjee2021,
       author = {{Mukherjee}, Diptajyoti and {Zhu}, Qirong and {Trac}, Hy and {Rodriguez}, Carl L.},
        title = "{Fast Multipole Methods for N-body Simulations of Collisional Star Systems}",
      journal = {\apj},
         year = 2021,
        month = jul,
       volume = {916},
       number = {1},
          eid = {9},
        pages = {9},
          doi = {10.3847/1538-4357/ac03b2},
archivePrefix = {arXiv},
       eprint = {2012.02207},
 primaryClass = {astro-ph.GA},
       adsurl = {https://ui.adsabs.harvard.edu/abs/2021ApJ...916....9M}
}

@ARTICLE{grudic2023,
       author = {{Grudi{\'c}}, Michael Y. and {Hafen}, Zachary and {Rodriguez}, Carl L. and {Guszejnov}, D{\'a}vid and {Lamberts}, Astrid and {Wetzel}, Andrew and {Boylan-Kolchin}, Michael and {Faucher-Gigu{\`e}re}, Claude-Andr{\'e}},
        title = "{Great balls of FIRE - I. The formation of star clusters across cosmic time in a Milky Way-mass galaxy}",
      journal = {\mnras},
         year = 2023,
        month = feb,
       volume = {519},
       number = {1},
        pages = {1366-1380},
          doi = {10.1093/mnras/stac3573},
archivePrefix = {arXiv},
       eprint = {2203.05732},
 primaryClass = {astro-ph.GA},
       adsurl = {https://ui.adsabs.harvard.edu/abs/2023MNRAS.519.1366G}
}

@PHDTHESIS{stadel2001,
       author = {{Stadel}, Joachim Gerhard},
        title = "{Cosmological N-body simulations and their analysis}",
       school = {University of Washington, Seattle},
         year = 2001,
        month = jan,
       adsurl = {https://ui.adsabs.harvard.edu/abs/2001PhDT........21S}
}

@ARTICLE{chen2010,
       author = {{Chen}, C.~W. and {Chen}, W.~P.},
        title = "{Morphological Distortion of Galactic Globular Clusters}",
      journal = {\apj},
         year = 2010,
        month = oct,
       volume = {721},
       number = {2},
        pages = {1790-1819},
          doi = {10.1088/0004-637X/721/2/1790},
       adsurl = {https://ui.adsabs.harvard.edu/abs/2010ApJ...721.1790C}
}

@ARTICLE{kuzma2025,
       author = {{Kuzma}, Pete B. and {Ishigaki}, Miho N. and {Kirihara}, Takanobu and {Ogami}, Itsuki},
        title = "{Constructing a Pristine View of Extended Globular Cluster Structure}",
      journal = {\aj},
         year = 2025,
        month = sep,
       volume = {170},
       number = {3},
          eid = {157},
        pages = {157},
          doi = {10.3847/1538-3881/aded8e},
archivePrefix = {arXiv},
       eprint = {2507.05590},
 primaryClass = {astro-ph.GA},
       adsurl = {https://ui.adsabs.harvard.edu/abs/2025AJ....170..157K}
}

@ARTICLE{malhan2019b,
       author = {{Malhan}, Khyati and {Ibata}, Rodrigo A.},
        title = "{Constraining the Milky Way halo potential with the GD-1 stellar stream}",
      journal = {\mnras},
         year = 2019,
        month = jul,
       volume = {486},
       number = {3},
        pages = {2995-3005},
          doi = {10.1093/mnras/stz1035},
archivePrefix = {arXiv},
       eprint = {1807.05994},
 primaryClass = {astro-ph.GA},
       adsurl = {https://ui.adsabs.harvard.edu/abs/2019MNRAS.486.2995M}
}

@ARTICLE{penarrubia2006,
       author = {{Pe{\~n}arrubia}, Jorge and {Benson}, Andrew J. and {Mart{\'\i}nez-Delgado}, David and {Rix}, Hans Walter},
        title = "{Modeling Tidal Streams in Evolving Dark Matter Halos}",
      journal = {\apj},
         year = 2006,
        month = jul,
       volume = {645},
       number = {1},
        pages = {240-255},
          doi = {10.1086/504316},
archivePrefix = {arXiv},
       eprint = {astro-ph/0512507},
 primaryClass = {astro-ph},
       adsurl = {https://ui.adsabs.harvard.edu/abs/2006ApJ...645..240P}
}

@ARTICLE{carlberg2012,
       author = {{Carlberg}, R.~G.},
        title = "{Dark Matter Sub-halo Counts via Star Stream Crossings}",
      journal = {\apj},
         year = 2012,
        month = mar,
       volume = {748},
       number = {1},
          eid = {20},
        pages = {20},
          doi = {10.1088/0004-637X/748/1/20},
archivePrefix = {arXiv},
       eprint = {1109.6022},
 primaryClass = {astro-ph.CO},
       adsurl = {https://ui.adsabs.harvard.edu/abs/2012ApJ...748...20C}
}

@ARTICLE{erkal2016,
       author = {{Erkal}, Denis and {Belokurov}, Vasily and {Bovy}, Jo and {Sanders}, Jason L.},
        title = "{The number and size of subhalo-induced gaps in stellar streams}",
      journal = {\mnras},
         year = 2016,
        month = nov,
       volume = {463},
       number = {1},
        pages = {102-119},
          doi = {10.1093/mnras/stw1957},
archivePrefix = {arXiv},
       eprint = {1606.04946},
 primaryClass = {astro-ph.GA},
       adsurl = {https://ui.adsabs.harvard.edu/abs/2016MNRAS.463..102E}
}

@ARTICLE{yoon2011,
       author = {{Yoon}, Joo Heon and {Johnston}, Kathryn V. and {Hogg}, David W.},
        title = "{Clumpy Streams from Clumpy Halos: Detecting Missing Satellites with Cold Stellar Structures}",
      journal = {\apj},
         year = 2011,
        month = apr,
       volume = {731},
       number = {1},
          eid = {58},
        pages = {58},
          doi = {10.1088/0004-637X/731/1/58},
archivePrefix = {arXiv},
       eprint = {1012.2884},
 primaryClass = {astro-ph.GA},
       adsurl = {https://ui.adsabs.harvard.edu/abs/2011ApJ...731...58Y}
}

@ARTICLE{amorisco2017,
       author = {{Amorisco}, N.~C.},
        title = "{Contributions to the accreted stellar halo: an atlas of stellar deposition}",
      journal = {\mnras},
         year = 2017,
        month = jan,
       volume = {464},
       number = {3},
        pages = {2882-2895},
          doi = {10.1093/mnras/stw2229},
archivePrefix = {arXiv},
       eprint = {1511.08806},
 primaryClass = {astro-ph.GA},
       adsurl = {https://ui.adsabs.harvard.edu/abs/2017MNRAS.464.2882A}
}

@article{katz2013,
    author = {Katz, Harley and Ricotti, Massimo},
    title = {Two epochs of globular cluster formation from deep field luminosity functions: implications for reionization and the Milky Way satellites},
    journal = {Monthly Notices of the Royal Astronomical Society},
    volume = {432},
    number = {4},
    pages = {3250-3261},
    year = {2013},
    month = {05},
    issn = {0035-8711},
    doi = {10.1093/mnras/stt676},
    url = {https://doi.org/10.1093/mnras/stt676},
    eprint = {https://academic.oup.com/mnras/article-pdf/432/4/3250/18598816/stt676.pdf},
}

@ARTICLE{kruijssen2020a,
       author = {{Kruijssen}, J.~M. Diederik and {Pfeffer}, Joel L. and {Chevance}, M{\'e}lanie and {Bonaca}, Ana and {Trujillo-Gomez}, Sebastian and {Bastian}, Nate and {Reina-Campos}, Marta and {Crain}, Robert A. and {Hughes}, Meghan E.},
        title = "{Kraken reveals itself - the merger history of the Milky Way reconstructed with the E-MOSAICS simulations}",
      journal = {\mnras},
         year = 2020,
        month = aug,
       volume = {498},
       number = {2},
        pages = {2472-2491},
          doi = {10.1093/mnras/staa2452},
archivePrefix = {arXiv},
       eprint = {2003.01119},
 primaryClass = {astro-ph.GA},
       adsurl = {https://ui.adsabs.harvard.edu/abs/2020MNRAS.498.2472K}
}

@ARTICLE{piatti2020,
       author = {{Piatti}, Andr{\'e}s E. and {Carballo-Bello}, Julio A.},
        title = "{The tidal tails of Milky Way globular clusters}",
      journal = {\aap},
         year = 2020,
        month = may,
       volume = {637},
          eid = {L2},
        pages = {L2},
          doi = {10.1051/0004-6361/202037994},
archivePrefix = {arXiv},
       eprint = {2004.11747},
 primaryClass = {astro-ph.GA},
       adsurl = {https://ui.adsabs.harvard.edu/abs/2020A&A...637L...2P}
}

@ARTICLE{harris2010,
       author = {{Harris}, William E.},
        title = "{A New Catalog of Globular Clusters in the Milky Way}",
      journal = {arXiv e-prints},
         year = 2010,
        month = dec,
          eid = {arXiv:1012.3224},
        pages = {arXiv:1012.3224},
archivePrefix = {arXiv},
       eprint = {1012.3224},
 primaryClass = {astro-ph.GA},
       adsurl = {https://ui.adsabs.harvard.edu/abs/2010arXiv1012.3224H}
}

@ARTICLE{hernquist1992,
       author = {{Hernquist}, Lars and {Ostriker}, Jeremiah P.},
        title = "{A Self-consistent Field Method for Galactic Dynamics}",
      journal = {\apj},
         year = 1992,
        month = feb,
       volume = {386},
        pages = {375},
          doi = {10.1086/171025},
       adsurl = {https://ui.adsabs.harvard.edu/abs/1992ApJ...386..375H}
}

@ARTICLE{barnes1986,
       author = {{Barnes}, Josh and {Hut}, Piet},
        title = "{A hierarchical O(N log N) force-calculation algorithm}",
      journal = {\nat},
         year = 1986,
        month = dec,
       volume = {324},
       number = {6096},
        pages = {446-449},
          doi = {10.1038/324446a0},
       adsurl = {https://ui.adsabs.harvard.edu/abs/1986Natur.324..446B}
}

@ARTICLE{renaud2015,
       author = {{Renaud}, Florent and {Gieles}, Mark},
        title = "{A flexible method to evolve collisional systems and their tidal debris in external potentials}",
      journal = {\mnras},
         year = 2015,
        month = apr,
       volume = {448},
       number = {4},
        pages = {3416-3422},
          doi = {10.1093/mnras/stv245},
archivePrefix = {arXiv},
       eprint = {1502.01268},
 primaryClass = {astro-ph.GA},
       adsurl = {https://ui.adsabs.harvard.edu/abs/2015MNRAS.448.3416R}
}

@article{grondin2022,
    author = {Grondin, Steffani M and Webb, Jeremy J and Leigh, Nathan W C and Speagle, Joshua S and Khalifeh, Reem J},
    title = "{Searching for the extra-tidal stars of globular clusters using high-dimensional analysis and a core particle spray code}",
    journal = {Monthly Notices of the Royal Astronomical Society},
    volume = {518},
    number = {3},
    pages = {4249-4264},
    year = {2022},
    month = {11},
    issn = {0035-8711},
    doi = {10.1093/mnras/stac3367},
    url = {https://doi.org/10.1093/mnras/stac3367},
    eprint = {https://academic.oup.com/mnras/article-pdf/518/3/4249/47614289/stac3367.pdf},
}

@ARTICLE{grondin2024,
       author = {{Grondin}, Steffani M. and {Webb}, Jeremy J. and {Lane}, James M.~M. and {Speagle}, Joshua S. and {Leigh}, Nathan W.~C.},
        title = "{A catalogue of Galactic GEMS: Globular cluster Extra-tidal Mock Stars}",
      journal = {\mnras},
         year = 2024,
        month = mar,
       volume = {528},
       number = {3},
        pages = {5189-5211},
          doi = {10.1093/mnras/stae203},
archivePrefix = {arXiv},
       eprint = {2310.09331},
 primaryClass = {astro-ph.GA},
       adsurl = {https://ui.adsabs.harvard.edu/abs/2024MNRAS.528.5189G}
}

@ARTICLE{renaud2011,
       author = {{Renaud}, Florent and {Gieles}, Mark and {Boily}, Christian M.},
        title = "{Evolution of star clusters in arbitrary tidal fields}",
      journal = {\mnras},
         year = 2011,
        month = dec,
       volume = {418},
       number = {2},
        pages = {759-769},
          doi = {10.1111/j.1365-2966.2011.19531.x},
archivePrefix = {arXiv},
       eprint = {1107.5820},
 primaryClass = {astro-ph.CO},
       adsurl = {https://ui.adsabs.harvard.edu/abs/2011MNRAS.418..759R}
}

@ARTICLE{weinberg1996,
       author = {{Weinberg}, Martin D.},
        title = "{High-Accuracy Minimum Relaxation N-Body Simulations Using Orthogonal Series Force Computation}",
      journal = {\apj},
         year = 1996,
        month = oct,
       volume = {470},
        pages = {715},
          doi = {10.1086/177902},
archivePrefix = {arXiv},
       eprint = {astro-ph/9511121},
 primaryClass = {astro-ph},
       adsurl = {https://ui.adsabs.harvard.edu/abs/1996ApJ...470..715W}
}

@ARTICLE{panithanpaisal2025,
       author = {{Panithanpaisal}, Nondh and {Sanderson}, Robyn E. and {Rodriguez}, Carl L. and {Starkenburg}, Tjitske and {Pearson}, Sarah and {Bonaca}, Ana and {Hopkins}, Philip F. and {Cook}, Brian T. and {Arora}, Arpit and {Weatherford}, Newlin C.},
        title = "{Breaking Down the $\textsf{CosmoGEMS}$: Toward Modeling and Understanding Globular Cluster Stellar Streams in a Fully Cosmological Context}",
      journal = {arXiv e-prints},
         year = 2025,
        month = sep,
          eid = {arXiv:2509.03599},
        pages = {arXiv:2509.03599},
          doi = {10.48550/arXiv.2509.03599},
archivePrefix = {arXiv},
       eprint = {2509.03599},
 primaryClass = {astro-ph.GA},
       adsurl = {https://ui.adsabs.harvard.edu/abs/2025arXiv250903599P}
}

@BOOK{fitzpatrick2012,
       author = {{Fitzpatrick}, Richard},
        title = "{An Introduction to Celestial Mechanics}",
         year = 2012,
       adsurl = {https://ui.adsabs.harvard.edu/abs/2012icm..book.....F}
}

@ARTICLE{bardeen1986,
       author = {{Bardeen}, J.~M. and {Bond}, J.~R. and {Kaiser}, N. and {Szalay}, A.~S.},
        title = "{The Statistics of Peaks of Gaussian Random Fields}",
      journal = {\apj},
         year = 1986,
        month = may,
       volume = {304},
        pages = {15},
          doi = {10.1086/164143},
       adsurl = {https://ui.adsabs.harvard.edu/abs/1986ApJ...304...15B}
}

@ARTICLE{vasiliev2019b,
       author = {{Vasiliev}, Eugene},
        title = "{AGAMA: action-based galaxy modelling architecture}",
      journal = {\mnras},
         year = 2019,
        month = jan,
       volume = {482},
       number = {2},
        pages = {1525-1544},
          doi = {10.1093/mnras/sty2672},
archivePrefix = {arXiv},
       eprint = {1802.08239},
 primaryClass = {astro-ph.GA},
       adsurl = {https://ui.adsabs.harvard.edu/abs/2019MNRAS.482.1525V}
}

@ARTICLE{arora2024,
       author = {{Arora}, Arpit and {Sanderson}, Robyn and {Regan}, Christopher and {Garavito-Camargo}, Nicol{\'a}s and {Bregou}, Emily and {Panithanpaisal}, Nondh and {Wetzel}, Andrew and {Cunningham}, Emily C. and {Loebman}, Sarah R. and {Dropulic}, Adriana and {Shipp}, Nora},
        title = "{Efficient and accurate force replay in cosmological-baryonic simulations}",
      journal = {arXiv e-prints},
         year = 2024,
        month = jul,
          eid = {arXiv:2407.12932},
        pages = {arXiv:2407.12932},
          doi = {10.48550/arXiv.2407.12932},
archivePrefix = {arXiv},
       eprint = {2407.12932},
 primaryClass = {astro-ph.GA},
       adsurl = {https://ui.adsabs.harvard.edu/abs/2024arXiv240712932A}
}

@ARTICLE{arora2022,
       author = {{Arora}, Arpit and {Sanderson}, Robyn E. and {Panithanpaisal}, Nondh and {Cunningham}, Emily C. and {Wetzel}, Andrew and {Garavito-Camargo}, Nicol{\'a}s},
        title = "{On the Stability of Tidal Streams in Action Space}",
      journal = {\apj},
         year = 2022,
        month = nov,
       volume = {939},
       number = {1},
          eid = {2},
        pages = {2},
          doi = {10.3847/1538-4357/ac93fb},
archivePrefix = {arXiv},
       eprint = {2207.13481},
 primaryClass = {astro-ph.GA},
       adsurl = {https://ui.adsabs.harvard.edu/abs/2022ApJ...939....2A}
}

@ARTICLE{rodriguez2023,
       author = {{Rodriguez}, Carl L. and {Hafen}, Zachary and {Grudi{\'c}}, Michael Y. and {Lamberts}, Astrid and {Sharma}, Kuldeep and {Faucher-Gigu{\`e}re}, Claude-Andr{\'e} and {Wetzel}, Andrew},
        title = "{Great balls of FIRE II: The evolution and destruction of star clusters across cosmic time in a Milky Way-mass galaxy}",
      journal = {\mnras},
         year = 2023,
        month = may,
       volume = {521},
       number = {1},
        pages = {124-147},
          doi = {10.1093/mnras/stad578},
archivePrefix = {arXiv},
       eprint = {2203.16547},
 primaryClass = {astro-ph.GA},
       adsurl = {https://ui.adsabs.harvard.edu/abs/2023MNRAS.521..124R}
}

@ARTICLE{takahashi2012,
       author = {{Takahashi}, K. and {Baumgardt}, H.},
        title = "{Tidal mass loss in star clusters and treatment of escapers in Fokker-Planck models}",
      journal = {\mnras},
         year = 2012,
        month = feb,
       volume = {420},
       number = {2},
        pages = {1799-1808},
          doi = {10.1111/j.1365-2966.2011.20183.x},
archivePrefix = {arXiv},
       eprint = {1111.3788},
 primaryClass = {astro-ph.GA},
       adsurl = {https://ui.adsabs.harvard.edu/abs/2012MNRAS.420.1799T}
}

@ARTICLE{giersz2008,
       author = {{Giersz}, Mirek and {Heggie}, Douglas C. and {Hurley}, Jarrod R.},
        title = "{Monte Carlo simulations of star clusters - IV. Calibration of the Monte Carlo code and comparison with observations for the open cluster M67}",
      journal = {\mnras},
         year = 2008,
        month = jul,
       volume = {388},
       number = {1},
        pages = {429-443},
          doi = {10.1111/j.1365-2966.2008.13407.x},
archivePrefix = {arXiv},
       eprint = {0801.3968},
 primaryClass = {astro-ph},
       adsurl = {https://ui.adsabs.harvard.edu/abs/2008MNRAS.388..429G}
}

@ARTICLE{chen2025c,
       author = {{Chen}, Yingtian and {Gnedin}, Oleg Y. and {Price-Whelan}, Adrian M.},
        title = "{StarStream on Gaia: Stream Discovery and Mass-loss Rate of Globular Clusters}",
      journal = {\apjs},
         year = 2026,
        month = apr,
       volume = {283},
       number = {2},
          eid = {60},
        pages = {60},
          doi = {10.3847/1538-4365/ae471f},
archivePrefix = {arXiv},
       eprint = {2510.14924},
 primaryClass = {astro-ph.GA},
       adsurl = {https://ui.adsabs.harvard.edu/abs/2026ApJS..283...60C}
}

@article{kullback1951,
author = {S. Kullback and R. A. Leibler},
title = {{On Information and Sufficiency}},
volume = {22},
journal = {The Annals of Mathematical Statistics},
number = {1},
publisher = {Institute of Mathematical Statistics},
pages = {79 -- 86},
year = {1951},
doi = {10.1214/aoms/1177729694},
URL = {https://doi.org/10.1214/aoms/1177729694}
}

@ARTICLE{amorisco2015,
       author = {{Amorisco}, N.~C.},
        title = "{On feathers, bifurcations and shells: the dynamics of tidal streams across the mass scale}",
      journal = {\mnras},
         year = 2015,
        month = jun,
       volume = {450},
       number = {1},
        pages = {575-591},
          doi = {10.1093/mnras/stv648},
archivePrefix = {arXiv},
       eprint = {1410.0360},
 primaryClass = {astro-ph.GA},
       adsurl = {https://ui.adsabs.harvard.edu/abs/2015MNRAS.450..575A}
}

@ARTICLE{white2025,
       author = {{White}, Ethan B. and {Vesperini}, Enrico and {Dalessandro}, Emanuele and {Varri}, Anna Lisa},
        title = "{Evolution of the kinematic properties of rotating multiple-population globular clusters}",
      journal = {arXiv e-prints},
         year = 2025,
        month = oct,
          eid = {arXiv:2510.15037},
        pages = {arXiv:2510.15037},
          doi = {10.48550/arXiv.2510.15037},
archivePrefix = {arXiv},
       eprint = {2510.15037},
 primaryClass = {astro-ph.GA},
       adsurl = {https://ui.adsabs.harvard.edu/abs/2025arXiv251015037W}
}

@article{weatherford2023,
doi = {10.3847/1538-4357/acbcc1},
url = {https://dx.doi.org/10.3847/1538-4357/acbcc1},
year = {2023},
month = {apr},
publisher = {The American Astronomical Society},
volume = {946},
number = {2},
pages = {104},
author = {Weatherford, Newlin C. and Kıroğlu, Fulya and Fragione, Giacomo and Chatterjee, Sourav and Kremer, Kyle and Rasio, Frederic A.},
title = {Stellar Escape from Globular Clusters. I. Escape Mechanisms and Properties at Ejection},
journal = {The Astrophysical Journal}
}

@ARTICLE{weatherford2024,
       author = {{Weatherford}, Newlin C. and {Rasio}, Frederic A. and {Chatterjee}, Sourav and {Fragione}, Giacomo and {K{\i}ro{\u{g}}lu}, Fulya and {Kremer}, Kyle},
        title = "{Stellar Escape from Globular Clusters. II. Clusters May Eat Their Own Tails}",
      journal = {\apj},
         year = 2024,
        month = may,
       volume = {967},
       number = {1},
          eid = {42},
        pages = {42},
          doi = {10.3847/1538-4357/ad39df},
archivePrefix = {arXiv},
       eprint = {2310.01485},
 primaryClass = {astro-ph.GA},
       adsurl = {https://ui.adsabs.harvard.edu/abs/2024ApJ...967...42W}
}

@ARTICLE{breivik2020,
       author = {{Breivik}, Katelyn and {Coughlin}, Scott and {Zevin}, Michael and {Rodriguez}, Carl L. and {Kremer}, Kyle and {Ye}, Claire S. and {Andrews}, Jeff J. and {Kurkowski}, Michael and {Digman}, Matthew C. and {Larson}, Shane L. and {Rasio}, Frederic A.},
        title = "{COSMIC Variance in Binary Population Synthesis}",
      journal = {\apj},
         year = 2020,
        month = jul,
       volume = {898},
       number = {1},
          eid = {71},
        pages = {71},
          doi = {10.3847/1538-4357/ab9d85},
archivePrefix = {arXiv},
       eprint = {1911.00903},
 primaryClass = {astro-ph.HE},
       adsurl = {https://ui.adsabs.harvard.edu/abs/2020ApJ...898...71B}
}

@ARTICLE{henon1971,
       author = {{H{\'e}non}, M.~H.},
        title = "{The Monte Carlo Method (Papers appear in the Proceedings of IAU Colloquium No. 10 Gravitational N-Body Problem (ed. by Myron Lecar), R. Reidel Publ. Co. , Dordrecht-Holland.)}",
      journal = {\apss},
         year = 1971,
        month = nov,
       volume = {14},
       number = {1},
        pages = {151-167},
          doi = {10.1007/BF00649201},
       adsurl = {https://ui.adsabs.harvard.edu/abs/1971Ap&SS..14..151H}
}

@ARTICLE{rodriguez2022,
       author = {{Rodriguez}, Carl L. and {Weatherford}, Newlin C. and {Coughlin}, Scott C. and {Amaro-Seoane}, Pau and {Breivik}, Katelyn and {Chatterjee}, Sourav and {Fragione}, Giacomo and {K{\i}ro{\u{g}}lu}, Fulya and {Kremer}, Kyle and {Rui}, Nicholas Z. and {Ye}, Claire S. and {Zevin}, Michael and {Rasio}, Frederic A.},
        title = "{Modeling Dense Star Clusters in the Milky Way and beyond with the Cluster Monte Carlo Code}",
      journal = {\apjs},
         year = 2022,
        month = feb,
       volume = {258},
       number = {2},
          eid = {22},
        pages = {22},
          doi = {10.3847/1538-4365/ac2edf},
archivePrefix = {arXiv},
       eprint = {2106.02643},
 primaryClass = {astro-ph.GA},
       adsurl = {https://ui.adsabs.harvard.edu/abs/2022ApJS..258...22R}
}

@ARTICLE{plummer1911,
       author = {{Plummer}, H.~C.},
        title = "{On the problem of distribution in globular star clusters}",
      journal = {\mnras},
         year = 1911,
        month = mar,
       volume = {71},
        pages = {460-470},
          doi = {10.1093/mnras/71.5.460},
       adsurl = {https://ui.adsabs.harvard.edu/abs/1911MNRAS..71..460P}
}

@ARTICLE{fouvry2021,
       author = {{Fouvry}, Jean-Baptiste and {Hamilton}, Chris and {Rozier}, Simon and {Pichon}, Christophe},
        title = "{Resonant and non-resonant relaxation of globular clusters}",
      journal = {\mnras},
         year = 2021,
        month = dec,
       volume = {508},
       number = {2},
        pages = {2210-2225},
          doi = {10.1093/mnras/stab2596},
archivePrefix = {arXiv},
       eprint = {2103.10165},
 primaryClass = {astro-ph.GA},
       adsurl = {https://ui.adsabs.harvard.edu/abs/2021MNRAS.508.2210F}
}

@ARTICLE{lowing2011,
       author = {{Lowing}, Ben and {Jenkins}, Adrian and {Eke}, Vincent and {Frenk}, Carlos},
        title = "{A halo expansion technique for approximating simulated dark matter haloes}",
      journal = {\mnras},
         year = 2011,
        month = oct,
       volume = {416},
       number = {4},
        pages = {2697-2711},
          doi = {10.1111/j.1365-2966.2011.19222.x},
archivePrefix = {arXiv},
       eprint = {1010.6197},
 primaryClass = {astro-ph.CO},
       adsurl = {https://ui.adsabs.harvard.edu/abs/2011MNRAS.416.2697L}
}

@BOOK{gough2009,
  title     = "{GNU} scientific library reference manual",
  editor    = "Gough, Brian",
  publisher = "Network Theory",
  edition   =  3,
  month     =  jan,
  year      =  2009,
  address   = "Bristol, England",
  language  = "en"
}

@BOOK{press2002,
       author = {{Press}, William H. and {Teukolsky}, Saul A. and {Vetterling}, William T. and {Flannery}, Brian P.},
        title = "{Numerical recipes in C++ : the art of scientific computing}",
         year = 2002,
       adsurl = {https://ui.adsabs.harvard.edu/abs/2002nrca.book.....P}
}

@ARTICLE{cluttonbrock1973,
       author = {{Clutton-Brock}, M.},
        title = "{The Gravitational Field of Three Dimensional Galaxies}",
      journal = {\apss},
         year = 1973,
        month = jul,
       volume = {23},
       number = {1},
        pages = {55-69},
          doi = {10.1007/BF00647652},
       adsurl = {https://ui.adsabs.harvard.edu/abs/1973Ap&SS..23...55C}
}

@article{Shea1988,
 ISSN = {00359254, 14679876},
 URL = {http://www.jstor.org/stable/2347328},
 author = {B. L. Shea},
 journal = {Journal of the Royal Statistical Society. Series C (Applied Statistics)},
 number = {3},
 pages = {466--473},
 publisher = {[Royal Statistical Society, Oxford University Press]},
 title = {Algorithm AS 239: Chi-Squared and Incomplete Gamma Integral},
 urldate = {2025-07-21},
 volume = {37},
 year = {1988}
}

@Article{Lau1980,
journal={Journal of the Royal Statistical Society Series C},
author={Chi Leung Lau},
title={A Simple Series for the Incomplete Gamma Integral},
year={1980},
month={March},
pages={113-114},
volume={29},
number={1},
doi={10.2307/2346431},
url={https://ideas.repec.org/a/bla/jorssc/v29y1980i1p113-114.html},
}

\appendix

% Counter for appendix
% https://tex.stackexchange.com/questions/85776/change-figure-numbering-for-appendix
\counterwithin{figure}{section}
\counterwithin{table}{section}
\renewcommand\thefigure{\thesection.\arabic{figure}}  

\section{MW Potential Models in \krios}
\label{app:static_host_potentials_krios}

\subsection{Spherical Components: Galactic Bulge, Hernquist Potential, and NFW Halo}

The density of the {\tt MWPotential2014} Galactic bulge is defined in the following way\footnote{\url{https://docs.galpy.org/en/latest/reference/potentialpowerspherwcut.html}}:
\begin{align}
\rho(r') &= \rho_{0} \left({r_{c}\over r'}\right)^{\alpha}\exp\left[-(r'/r_{c})^{2}\right],
\end{align}
where we set the reference radius equal to the cutoff radius for convenience. The enclosed mass, as a
function of $r'$, is
\begin{align}
M_{\rm enc}^{\rm Bulge}(r') &= 4\pi \int_{0}^{r'} \dd r'' \, r''^{\,2} \rho(r'')\notag \\
&= 4\pi \rho_{0} r_{c}^{3} \int_{0}^{r'/r_{c}} \hspace{-4mm}\dd u \, u^{-\alpha+2}\exp(-u^{2}).
\end{align}
The choice of reference radius, along with the bulge mass, dictates ${\rho_{0} \!=\! M_{b}/\left(4\pi r_{c}^{3}\int_{0}^{\infty}\! \dd u \, u^{-\alpha+2}\exp[-u^{2}]\right)}$. This means we can write the enclosed mass as a function of $M_{b}$:
%
%\begin{subequations}
\begin{align}
M_{\rm enc}^{\rm Bulge}(r') &= M_{b} {\int_{0}^{r'/r_{c}} \dd u \, u^{-\alpha+2}\exp(-u^{2}) \over \int_{0}^{\infty} \dd u \, u^{-\alpha+2}\exp(-u^{2})} \notag\\
&= M_{b}{\gamma(s, x)\over \Gamma(s)},
\end{align}
%\end{subequations}
%
where ${x\!=\!(r'/r_c)^2}$, ${s\!=\!{1\over 2}(3-\alpha)}$ and $\gamma(a,x)$ is the lower incomplete gamma function defined as %, and $\Gamma(a,x)$ is the upper incomplete gamma function. These gamma functions are defined in the following way:
%
%\begin{subequations}
\begin{align}
    \gamma(a,x) &= \int_0^x \dd t \, t^{a-1} e^{-t}.%, \\
 %   \Gamma(a,x) &= \int_x^{\infty} \dd t \, t^{a-1} e^{-t}.
\end{align}
%\end{subequations}
%
We use the associated potential implemented in \galpy\ and \nbody, which reads
\begin{align}
    \label{eq:psi_bulge} \psi_{\rm Bulge}(r') &= \frac{G M_b}{\Gamma(s)}\bigg(\frac{\gamma(1-\alpha/2,x)}{r_c}-\frac{\gamma(s,x)}{r'}\bigg).
\end{align}
The bulge potential tends to a non-zero value at spatial infinity:
\begin{align}
   \psi_{{\rm bulge},\infty} = {GM_{b} \over r_{c}} \left({\Gamma(1-\alpha/2) \over \Gamma(s)}\right)>0.
\end{align}
Therefore, if a component of this kind is included in a Galactic model (as it is in {\tt MWPotential2014}), there will be bound orbits with positive energies up to $ \psi_{{\rm bulge},\infty}$.

The bulge is spherically symmetric, so only $f_{v_{r'}}^{\rm Bulge}$ is non-zero and can be computed directly from the enclosed mass:
\begin{subequations}
\begin{align}
f_{v_{r'}}^{\rm Bulge} &= -{GM_{\rm enc}^{\rm Bulge}(r') \over r'^{2}} = -{GM_{b}\over r'^{2}}{\gamma(s,x) \over \Gamma(s)}.
\end{align}
\end{subequations}
The associated tidal tensor only has one non-zero term in the spherical coordinates of the host frame:
\begin{align}
\partial_{r'}f_{v_{r'}}^{\rm Bulge} &= {2GM_{b}\over r'^{3}}{\gamma(s,x) - x^{s}\exp(-x) \over \Gamma(s)}.
\end{align}
The gravitational potential and enclosed mass for a Hernquist model \citep{hernquist1990, bovy2026} of total mass $M_{\rm H}$ and scale radius $a$ are
\begin{subequations}
\begin{align}
\psi_{\rm H}(r') &=   - {GM_{\rm H} \over a} \, {1 \over (1+r'/a)}, \\
M_{\rm enc, \, H}(r') &= M_{\rm H} \, {(r'/a)^{2} \over \left(1+ r'/a\right)^{2}},
\end{align}
\end{subequations}
The spherically symmetric potential only has a non-zero gradient in the radial direction, i.e.
\begin{align}
f_{v_{r'}}^{\rm H} &= -{GM_{\rm enc, \, H}(r') \over r'^{2}}.
\end{align}
For this spherically symmetric potential, we only need to compute $\partial_{r'}f_{v_{r'}}^{\rm NFW}$ to get the tidal tensor:
\begin{align}
\partial_{r'}f_{v_{r'}}^{\rm H} %&= {2GM_{\rm enc, \, H}(r') \over r'^{3}} \left(1 - {r' \over 2 M_{\rm enc, \, H}(r')}{\partial M_{\rm enc, \, H}(r') \over \partial r'} \right), \\
%&= {2GM_{\rm enc, \, H}(r') \over r'^{3}} \left(1 - {r' \over 2 M_{\rm enc, \, H}(r')}\left[{2M_{\rm H}\over a}{r'/a\over (1+r'/a)^{3}}\right]\right), \\
&= {2GM_{\rm H} \over r'^{3}} {(r'/a)^{3}\over (1+r'/a)^{3}}.
\end{align}
The NFW potential \citep{navarro1996, price-whelan2024, bovy2026} can be written in terms of a scale mass $M_{\rm NFW}$ and scale radius $r_{s}$ in the following way:
\begin{align}
\psi_{\rm NFW}(r') &= -{GM_{\rm NFW} \over r'} \, \ln\left(1+{r'\over r_{s}}\right).
\end{align}
The enclosed mass, which does not converge for NFW halos, is
\begin{align}
M_{\rm enc, \, NFW}(r') &= M_{\rm NFW} \, g(r'/r_{s}),
\end{align}
where $g(x) \!\equiv\! \ln(1+x) \!-\! x/(1+x)$. The associated non-zero force and tidal tensor terms (in the spherical coordinates of the host frame) are
\begin{subequations}
\begin{align}
f_{v_{r'}}^{\rm NFW} &= -{GM_{\rm enc, \, NFW}(r') \over r'^{2}}, \\
%\partial_{r'}f_{v_{r'}}^{\rm NFW} &= {2GM_{\rm enc, \, NFW}(r') \over r'^{3}}\left(1-{r'\over 2M_{\rm enc, \, NFW}(r')} {\partial M_{\rm enc, \, NFW}(r') \over \partial r'}\right), \\
%&= {2GM_{\rm enc, \, NFW}(r') \over r'^{3}}\left(1-{r'\over 2M_{\rm NFW} g(r'/r_{s})} \left[M_{\rm NFW} \, {\partial g \over \partial x} {\partial x \over \partial r'}\right]\right), \\
\partial_{r'}f_{v_{r'}}^{\rm NFW} &= {2GM_{\rm enc}^{\rm NFW}(r') \over r'^{3}}\left(1-{1\over 2 g(r'/r_{s})}\left({r'/r_{s} \over 1 + r'/r_{s}}\right)^{2}\right).
\end{align}
\end{subequations}

\subsection{Cylindrical Components: Miyamoto-Nagai Disk}

The Miyamoto-Nagai disk potential \citep{miyamoto1975, bovy2026} can be written in terms of the mass $M_{\rm MN}$, scale length $a$, and scale height $b$: 
\begin{align}
\psi_{\rm MN}(\varrho', z') %&= -{GM_{\rm MN} \over \sqrt{\varrho'^{2}+(\sqrt{z'^{2}+b^{2}}+a)^{2}}},
&= -{GM_{\rm MN} \over \sqrt{\varrho'^{2}+z^{\prime 2}_{ab}}},
\end{align}
where $(\varrho', \varphi', z')$ are the cylindrical coordinates with respect to the host and ${z'_{ab}\!\equiv\!\sqrt{z^{\prime 2}\!+\!b^2}\!+\!a}$. The potential is axially symmetric, so we get 
\begin{subequations}
\begin{align}
f_{v_{\varrho'}}^{\rm MN} %&= -GM_{\rm MN}\, {\varrho' \over \left(\varrho'^{2}+(\sqrt{z'^{2}+b^{2}}+a)^{2}\right)^{3/2}}, \\
&= -GM_{\rm MN}\, {\varrho' \over \left(\varrho'^{2}+z^{\prime 2}_{ab}\right)^{3/2}}, \\
f_{v_{\varphi'}}^{\rm MN} &=0, \\ 
f_{v_{z'}}^{\rm MN} %&= -GM_{\rm MN}{z'(\sqrt{z'^{2}+b^{2}}+a) \over \sqrt{z'^{2}+b^{2}}\left(\varrho'^{2}+(\sqrt{z'^{2}+b^{2}}+a)^{2}\right)^{3/2}}.
&= -GM_{\rm MN}{z'z^{\prime}_{ab} \over \sqrt{z'^{2}+b^{2}}\left(\varrho'^{2}+z^{\prime 2}_{ab}\right)^{3/2}}.
\end{align}
\end{subequations}
There are four non-zero elements of the tidal tensor (in the cylindrical coordinates of the host frame) in this case: 
\begin{subequations}
\begin{align}
\partial_{\varrho'}f_{v_{\varrho'}}^{\rm MN} %&= GM_{\rm MN}\left({3\varrho'^{2} \over \left(\varrho'^{2}+(\sqrt{z'^{2}+b^{2}}+a)^{2}\right)^{5/2}} - {1\over \left(\varrho'^{2}+(\sqrt{z'^{2}+b^{2}}+a)^{2}\right)^{3/2}}\right), \\
&= {3GM_{\rm MN}\varrho'^{2} \over \left(\varrho'^{2}+z^{\prime 2}_{ab}\right)^{5/2}} - {GM_{\rm MN}\over \left(\varrho'^{2}+z^{\prime 2}_{ab}\right)^{3/2}}, \\
\partial_{\varrho'}f_{v_{z'}}^{\rm MN} %&= \partial_{z'}f_{v_{\varrho'}}^{\rm MN} = GM_{\rm MN}{3z'\varrho' (\sqrt{z'^{2}+b^{2}}+a) \over \sqrt{z'^{2}+b^{2}}\left(\varrho'^{2}+(\sqrt{z'^{2}+b^{2}}+a)^{2}\right)^{5/2}}, \\
&= \partial_{z'}f_{v_{\varrho'}}^{\rm MN} = {3GM_{\rm MN}z'\varrho' z^{\prime }_{ab} \over \sqrt{z'^{2}+b^{2}}\left(\varrho'^{2}+z^{\prime 2}_{ab}\right)^{5/2}}, \\
\partial_{z'}f_{v_{z'}}^{\rm MN} %&= GM_{\rm MN}\Bigg({3z'^{2}(\sqrt{z'^{2}+b^{2}}+a)^{2} \over (z'^{2}+b^{2})\left(\varrho'^{2}+(\sqrt{z'^{2}+b^{2}}+a)^{2}\right)^{5/2}} - {z'^{2} \over (z'^{2}+b^{2})\left(\varrho'^{2}+(\sqrt{z'^{2}+b^{2}}+a)^{2}\right)^{3/2}} \\ \nonumber &+ {z'^{2}(\sqrt{z'^{2}+b^{2}}+a) \over (z'^{2}+b^{2})^{3/2}\left(\varrho'^{2}+(\sqrt{z'^{2}+b^{2}}+a)^{2}\right)^{3/2}} - {(\sqrt{z'^{2}+b^{2}}+a)\over \sqrt{z'^{2}+b^{2}}\left(\varrho'^{2}+(\sqrt{z'^{2}+b^{2}}+a)^{2}\right)^{3/2}}\Bigg).
&=\! {3GM_{\rm MN}z'^{2}z^{\prime 2}_{ab} \over (z'^{2}\!+\!b^{2})\!\left(\varrho'^{2}\!+\!z^{\prime 2}_{ab}\right)^{5/2}} \!-\! {GM_{\rm MN}z'^{2} \over (z'^{2}\!+\!b^{2})\!\left(\varrho'^{2}\!+\!z^{\prime 2}_{ab}\right)^{3/2}} \nonumber \\ &+\! {GM_{\rm MN}z'^{2}z^{\prime }_{ab} \over (z'^{2}\!+\!b^{2})^{3/2}\!\left(\varrho'^{2}\!+\!z^{\prime 2}_{ab}\right)^{3/2}} \!-\! {GM_{\rm MN}z^{\prime }_{ab}\over \sqrt{z'^{2}\!+\!b^{2}}\!\left(\varrho'^{2}\!+\!z^{\prime 2}_{ab}\right)^{3/2}}.
\end{align}
\end{subequations}
A sum of three Miyamoto-Nagai disk potentials is approximately equivalent to a double-exponential disk \citep{smith2015, price-whelan2024}, with mass density ${\rho(\varrho', z') \!\propto\! \exp(-\varrho'/h_{\varrho'})\exp(-|z'|/h_{z'})}$. This form is used in the {\tt MilkyWayPotential2022} model; see Table~\ref{tab:mw2022_params}.

\subsection{{\tt MWPotential2014}}
\label{app:mwpot2014}

A standard static model for the potential of the MW is provided in \cite{bovy2015}, which is a sum of three potentials: a power-law density Galactic bulge that is exponentially cut off, a Miyamoto-Nagai disk, and an NFW halo. When reporting energies associated with this potential, we subtract off $\psi_{{\rm bulge}, \infty}$ to ensure that ${\rm sgn}(E)$ remains a reliable criterion for whether a orbit is bound to the host potential.

The bulge, disk, and halo masses need to be scaled such that their contributions to the total potential are consistent with the centripetal acceleration used by \nbody\ for a circular disk orbit at ${r_{0}'\!=\!8\,{\rm kpc}}$: ${f_{\rm cent}(r_{0}') \!=\! 6.32793804994 \, {\rm pc/Myr}^{2}}$. The centripetal acceleration can be broken down into its the contributions from the potential's constituent parts:
\begin{align}
f_{\rm cent}(r_{0}') &= \left|f_{v_{r'}}^{\rm Bulge}(r') \!+\! f_{v_{\varrho'}}^{\rm MN}(\varrho', z'\!=\!0) \!+\!f_{v_{r'}}^{\rm NFW}(r')\right|_{r'=\varrho'=r_{0}'} \notag\\
&\equiv {GM_{s} \over r_{0}'^{2}},
\end{align}
where we use a scale mass ${M_{s}\!=\!9.00273072 \times 10^{10}\,M_{\odot}}$ to ensure that each Galactic component can be scaled appropriately:
\begin{subequations}
\begin{align}
M_{b} &= f_{b} \, M_{s} \, {\Gamma(s) \over \gamma(s,x=(r_{0}'/r_{c})^{2})}, \\
M_{d} &= f_{d} \, M_{s} \, {(r_{0}'^{2}+(b+a)^{2})^{3/2} \over r_{0}'^{3}}, \\
M_{\rm h} &= f_{h} \, M_{s} \, {1 \over g(r_{0}'/R_{s})}.
\end{align}
\end{subequations}
%

\begin{comment}
\begin{table*}
\caption{The relevant physical quantities needed to re-cast the {\tt MWPotential2014} model into the SCF reference frame in H\'{e}non units. Each of the non-mass terms are copied directly from Table 1 in \cite{bovy2015} (except the power-law index; the sign is flipped so as to be consistent with our listed definition of the mass density). The masses are computed such that the \nbody\ centripetal acceleration at ${r_{0}'\!=\!8\,{\rm kpc}}$ is enforced.}
\centering
\begin{tabular}{ccccc}
\toprule
Physical Quantity & Symbol & Galactic Component & Value & Units \\
\midrule
Scale Mass & $M_{s}$ & Solar Orbit Scaling & 9.00273072  & $10^{10} \, M_{\odot}$ \\
\midrule
Bulge Contribution at ${r_{0}'\!=\!8}\, {\rm kpc}$ & $f_{b}$ & Solar Orbit Scaling & 0.05 & -- \\
Disk Contribution at ${r_{0}'\!=\!8}\, {\rm kpc}$ & $f_{d}$ & Solar Orbit Scaling & 0.6 & -- \\
Halo Contribution at ${r_{0}'\!=\!8}\, {\rm kpc}$ & $f_{h}$ & Solar Orbit Scaling & 0.35 & -- \\
\midrule
Power Law Index & $\alpha$ & Galactic Bulge & 1.8 & -- \\
Exponential Cutoff Radius & $r_{c}$ & Galactic Bulge & 1.9 & kpc \\
\midrule
Scale Length & $a$ & Miyamoto-Nagai Disk & 3.0 & kpc\\
Scale Height & $b$ & Miyamoto-Nagai Disk & 280 & pc \\
\midrule
Halo Scale Radius & $R_{s}$ & NFW Halo & 16.0 & kpc \\
Halo Concentration & $c$ & NFW Halo & 15.3 & -- \\
\bottomrule
\end{tabular}
\label{tab:mw2014_params}
\end{table*}
\end{comment}

\subsection{{\tt MilkyWayPotential2022}}

We use a standard static model for the MW \citep[{\tt MilkyWayPotential2022},][]{price-whelan2024} with spherical and cylindrical components. The Galactic nucleus and bulge are modeled using a Hernquist potential \citep{hernquist1990}, the dark matter halo with an NFW potential \citep{navarro1996}, and the disk with a sum of three Miyamoto-Nagai potentials  \citep{miyamoto1975}. The disk model is inspired by \cite{smith2015}, who showed that the sum of comparatively tractable terms can be used to approximate a double-exponential disk where the potential, and its associated gradients, are considerably more difficult to compute directly \citep[Section 2.6,][]{binney2008}.

\begin{table*}
\caption{The relevant physical quantities needed to recast the {\tt MilkyWayPotential2022} model into the SCF reference frame in H\'{e}non units. Each of the mass and scale radius values for the spherical host potential components are copied directly from the {\tt gala} source code. The exponential disk (with scale lengths ${h_{\varrho'}\!=\!2.6\, \rm{kpc}}$, ${h_{z'}\!=\!0.3\,{\rm kpc}}$) is approximated using three Miyamoto-Nagai disks, whose properties are found using the {\tt gala.potential.MN3ExponentialDiskPotential().get\_three\_potentials()} method.}
\centering
\begin{tabular}{ccccc}
\toprule
Physical Quantity & Symbol & Galactic Component & Value & Units \\
\midrule
Nucleus Mass & $M_{n}$ & Galactic Nucleus (Hernquist) & $1.8142\times 10^{9}$ & $M_{\odot}$ \\ 
Scale Radius & $a$ & Galactic Nucleus (Hernquist) & 68.8867 & pc \\
\midrule
Bulge Mass & $M_{b}$ & Galactic Bulge (Hernquist & $5\times10^{9}$ & $M_{\odot}$ \\
Scale Radius & $a$ & Galactic Bulge (Hernquist) & 1.0 & kpc \\
\midrule
Disk Mass & $M_{d}$ & First Miyamoto-Nagai Disk & $7.872306998700792\times10^{9}$ & $M_{\odot}$ \\
Scale Length & $a$ & First Miyamoto-Nagai Disk & 1.5259431976529216 & kpc\\
Scale Height & $b$ & First Miyamoto-Nagai Disk & 0.20663742603550295 & kpc \\
\midrule
Disk Mass & $M_{d}$ & Second Miyamoto-Nagai Disk & -$2.7562522194433154\times10^{11}$ & $M_{\odot}$ \\
Scale Length & $a$ & Second Miyamoto-Nagai Disk & 6.782764436261113 & kpc\\
Scale Height & $b$ & Second Miyamoto-Nagai Disk & 0.20663742603550295 & kpc \\
\midrule
Disk Mass & $M_{d}$ & Third Miyamoto-Nagai Disk & $3.206184188979487 \times 10^{11}$ & $M_{\odot}$ \\
Scale Length & $a$ & Third Miyamoto-Nagai Disk & 5.894799616164217 & kpc\\
Scale Height & $b$ & Third Miyamoto-Nagai Disk & 0.20663742603550295 & kpc \\
\midrule
Halo Scale Mass & $M_{h}$ & NFW Halo & $5.5427\times10^{11}$ & $M_{\odot}$ \\
Halo Scale Radius & $r_{s}$ & NFW Halo & 15.626 & kpc \\
\bottomrule
\end{tabular}
\label{tab:mw2022_params}
\end{table*}

For reproducibility, we provide each orbit's initial conditions in phase space accurate to 8 decimal places in Table~\ref{tab:exact_phase_space}. With these initial conditions, as well as the properties of the progenitor provided in Section \ref{sec:results} and host potential provided in Appendix~\ref{app:static_host_potentials_krios}, \krios\ can be compared against existing particle-spray \citep[e.g.,][]{fardal2015, grondin2022, roberts2025, chen2025} and direct {$N$-body} models.

\begin{table*}
    \centering
    \caption{The phase-space initial conditions needed to replicate the various orbits discussed in Section~\ref{sec:methods}, \ref{sec:results}. The validation orbits use the {\tt MWPotential2014} potential and the remaining orbits use the {\tt MilkyWayPotential2022} potential.} 
    \begin{tabular}{ccccccc}
    \toprule
    Orbit ID & $x_{\rm init}$ [kpc] & $y_{\rm init}$ [kpc] & $z_{\rm init}$ [kpc] & $v_{x, {\rm init}}$ [km/s] & $v_{y, {\rm init}}$ [km/s] & $v_{z, {\rm init}}$ [km/s] \\
    \midrule
    Circular 0 Test &5.0&0&0&0&225.55900747&0 \\
    Circular 1 Test &20.0&0&0&0&197.61111164&0 \\
    Eccentric Test &4.04231677&0&19.58670971&0&81.49205330&0 \\
    \midrule
    0 & -8.26114834 & 4.31927900 & -6.25723160 & -74.46836937 & 14.72551502 & -137.10220626 \\
1 & 6.65417846 & -6.77625060 & -6.89249860 & 131.88487388 & -54.38017150 & 4.34224828 \\
2 & 7.90677816 & 19.00502796 & -7.29228303 & -38.33588368 & -66.85081211 & 111.04181836 \\
3 & 9.73591969 & 8.80768907 & 0.73284850 & -59.71003689 & 91.24008328 & -39.80505238 \\
4 & -7.75511077 & 6.32704860 & 23.10803486 & 1.54688377 & -69.83998228 & 18.65659155 \\
5 & -9.14496546 & -2.86774379 & 4.55846674 & 25.93998870 & -146.50950043 & -246.21359733 \\
6 & -20.83691936 & -15.77790335 & -20.69545845 & -13.04785291 & -35.40337893 & -181.82805977 \\
7 & -18.82672221 & 27.56963834 & 27.40177203 & 25.20438668 & -112.02637209 & 29.55096514 \\
8 & 22.45554180 & 35.25629955 & -2.75494642 & -10.73021364 & -7.94047313 & -128.19547342 \\
9 & -23.95107022 & 9.24097888 & -14.06736155 & 27.31176344 & -167.54952847 & -128.11428538 \\
    \bottomrule
    \end{tabular}
    \label{tab:exact_phase_space}
\end{table*}

\section{Comparisons to {$N$-body} Simulations}
\label{app:nbody6++}

\subsection{Technical Details}

The {$N$-body} simulations described in Section~\ref{subsec:validation_with_nbody} were performed using  the direct {$N$-body} code \nbody\ \citep{wang2015}, which we parametrized to include the {\tt MWPotential2014} external field. 
\begin{figure}
\centering
\fbox{\parbox{8.0cm}
{\tt
1 1.0E8 5000.0 40 40 0 \\
50000 1 1 1 400 1 1 \\
0.02 0.01 0.10 1.0 1.0 1.0E5 2.0E-04 12.3084 2.0 \\
0 0 1 0 1 0 2 0 0 0 \\
0 0 0 5 0 1 0 0 0 0 \\
2 2 0 0 0 3 0 0 0 0 \\
0 0 0 0 0 0 0 0 0 0 \\
0 0 0 0 0 0 0 0 0 0 \\
1.0E-5 2E-4 0.1 1.0 1.0E-06 0.01 0.125 \\
0.0 1.0 1.0 0 0 0.001 0 1.0 \\
0.5 0.0 0.0 0.0 \\
20.0 0.0 0.0 0.0 202.0992878224309 0.0
}
}
\caption{Generic input file used for \nbody\ runs. This input file sets up a simple cluster of ${N\!=\!5\cdot 10^4}$ stars, initialized by a \texttt{dat.10} IC file (in H\'{e}non units), running until ${t_{\max}\!=\!5000\,\Myr}$ with outputs every $1 \,\rHU$. We set the virial radius to ${R_{\rm v}\!=\!12.3084\,\pc}$ and the individual masses to ${m\!=\!2 \,\Msun}$ (last two entries of the third line). In the final line, the cluster is initially placed at the position $(20.0,0,0)\,\kpc$, with initial velocity $(0,202.0992878224309,0)\,\mathrm{\pc/\Myr}$ within a Galactic potential centered at $(0,0,0)$ and modeled by the \texttt{MWPotential2014} external potential \citep{bovy2015}, as given by the parameter ${\tt{KZ}(14)\!=\!5}$. Contrary to the \nbody\ documentation, the initial cluster's velocity is to be given in $\mathrm{\pc/\Myr}$ and not in $\mathrm{\km/\s}$.
}
\label{fig:input_nbody}
\end{figure}
The initial conditions for the King spheres (with ${W_0\!=\!5}$) were generated from \cosmic\ \citep{breivik2020}. We show in Figure~\ref{fig:input_nbody} the typical input file of the runs made with \nbody.
Each {$N$-body} realization was composed of ${N\! =\!50000}$ particles, for a total mass of ${M\!=\!10^5 \Msun}$, and integrated up to ${t_{\max} \!=\!5\, {\rm Gyr}}$. We saved snapshots every ${\Delta t\!=\! 1 \,\rHU}$. On a 40-core node with a single V100 GPU, one simulation typically required about 29 hours for the 20-kpc circular orbit, 38 hours for the eccentric orbit, and 56 hours for the 5-kpc circular orbit.

\subsection{Corrections to the \nbody\ Source Files}
 
We note that a few corrections  had to be made to the  \nbody\ source files in order to obtain an accurate evolution of clusters subject to the {\tt MWPotential2014} external field, which we will list now.

There is a mistake in the documentation, which instructs to  provide the cluster's velocity in $\km/\s$. However, a careful study of the source files show that cluster's velocity should be given in $\pc/\Myr$ instead.

The implementation of the potential and the force induced by the bulge component requires the calculation of the increasing incomplete gamma function, $\gamma(a,x)$, which is handled in the original code by the file {\sf{asa147.f}} \citep{Lau1980}. However, this implementation breaks down for high enough values of $x$ (sometimes even as low as ${x\!=\!4}$), which almost always occurs in our runs. To remedy this issue, we replaced this file by the more recent  implementation {\sf{asa239.f90}} \citep{Shea1988} and updated the file {\sf{Makefile.in}} to include this change.

We also noticed that \nbody\ was using an older conversion rule between $\km/\s$ and $\pc/\Myr$ for the velocities in the file {\sf{mwpotinit.f}}. We changed that value using the one provided by \galpy: 
\begin{align}
    1\, \km/\s=1.022712165045695\, \pc/\Myr.
\end{align}
Finally, we modified the files {\sf{chdata.f}} and {\sf{output.F}} in order to save the snapshot data using {\tt{double}} precision instead of {\tt{float}} precision in order to be able to compute the integrals of motion with enough precision.

\clearpage
\onecolumngrid 

\begin{center}
{\large\bfseries Erratum: ``Modeling Globular Cluster Stellar Streams with a Basis-Expansion $N$-body Code'' (2026, {\it ApJ}, 1010, 155)}\\[1.5ex]
{\scshape Brian T.~Cook,$^{1}$ Kerwann~Tep,$^{1,2}$ Carl L.~Rodriguez,$^{1}$ Leah English,$^{3}$ Tjitske~Starkenburg,$^{4,5,6}$ Robyn~Sanderson,$^{7}$ Newlin~C.~Weatherford,$^{8}$ Sarah~Pearson,$^{9,10}$ and Nondh~Panithanpaisal$^{8,11}$}\\[1.5ex]
{\footnotesize\itshape
$^{1}$Department of Physics and Astronomy, University of North Carolina at Chapel Hill, 120 E. Cameron Ave, Chapel Hill, NC, 27514, USA\\
$^{2}$Observatoire Astronomique de Strasbourg, UMR 7550, 11 rue de l'Universit\'e, Strasbourg 67000, France\\
$^{3}$Department of Physics, Illinois Institute of Technology, 3105 S Dearborn St, Chicago, IL, 60616, USA\\
$^{4}$Department of Physics and Astronomy, Northwestern University, 2145 Sheridan Rd, Evanston IL 60208, USA\\
$^{5}$Center for Interdisciplinary Exploration and Research in Astrophysics (CIERA), Northwestern University, 1800 Sherman Ave, Evanston IL 60201, USA\\
$^{6}$NSF-Simons AI Institute for the Sky (SkAI), 172 E. Chestnut St., Chicago, IL 60611, USA\\
$^{7}$Department of Physics \& Astronomy, University of Pennsylvania, 209 S 33rd St, Philadelphia, PA 19104, USA\\
$^{8}$Observatories of the Carnegie Institution for Science, 813 Santa Barbara St, Pasadena, CA 91101, USA\\
$^{9}$DARK, Niels Bohr Institute, University of Copenhagen, Jagtvej 155A, 2200 Copenhagen, Denmark\\
$^{10}$DTU Space, Technical University of Denmark, Elektrovej 327, DK- 2800 Kgs. Lyngby, Denmark\\
$^{11}$TAPIR, Mailcode 350-17, California Institute of Technology, Pasadena, CA 91125, USA\\[1ex]
}
\end{center}

We have discovered an error in how we determined the ${(L_{z'}, E)}$ values for known Milky Way globular clusters \citep[][see also \url{https://people.smp.uq.edu.au/HolgerBaumgardt/globular/ }]{vasiliev2021} and stellar streams \citep{malhan2022} assuming the {\tt MilkyWayPotential2022} static host potential \citep{price-whelan2024}. The heliocentric-to-galactocentric reference frame transformation routine we wrote only considered the Sun's peculiar motion with respect to the Local Standard of Rest (LSR) and not the LSR's rotation about the Milky Way. Our routine also omitted the higher-order effects due to ${z_{\odot}'\!\neq\!0}$. Additionally, a $\cos\delta$ term was included when collating $\mu_{\alpha}^{\star}$ proper motions, which already contained the declination correction. We can now confirm agreement with \cite{bonaca2021, vasiliev2021, malhan2022, grondin2024} when our pipeline uses the built-in \cite{astropy2018} coordinate transformation methods (modulo choices in slightly different host potential models). There were also small bugs in our analysis scripts related to computing the circularity for certain orbits (Equation~\ref{eq:circularity}) and reporting the Earth Mover's Distance ratios in Figure~\ref{fig:normalized_emd_comps} that are now fixed. The primary results of the paper, namely the properties of stellar stream models produced by \krios\ and comparing those to other stream-modeling techniques, are unaffected \citep[See Section~5.3 of][]{facey2026}.

\section*{Section~3.3 Corrections}

\begin{itemize}
\itemsep0em
\item {\it Streams like the Fj\"{o}rm \citep{ibata2019b} system are less susceptible to integrals-of-motion errors, but, at apogalacticon, sky-plane variations will be pronounced.} The Fj\"{o}rm stream is in actuality less susceptible to sky-plane variations due to its large apogalacticon distance compared to M92, Gunnthr\'{a}, Pal 5, and M5. The commentary about integrals-of-motion errors remains true, but its location in integrals-of-motion space is now closer to the effective-potential barrier.
\item {\it The Gunnthr\'{a} stream \citep{ibata2021}, which may be debris from ${\omega\,\,{\rm Cen}}$ \citep{cloud2024}, appears to lie beyond the effective-potential barrier and is tightly bound to the Milky Way \citep[as is its sister Fimbulthul stream,][]{ibata2019}...} The Gunnthr\'{a} stream is indeed near the effective-potential barrier and tightly bound to the Milky Way, but the corrected integrals-of-motion data reveals it is more tightly bound to the Milky Way than previously reported.
\item {\it The Kshir stream, which has properties similar to GD-1 \citep{malhan2019a}, is shown in the bottom row of figures at high $|L_{z'}|$, similar to Orbit 9.} The Kshir stream is less similar to Orbit 9 after our corrections. There is no good direct comparison for Kshir in terms of apo/perigalacticon values. It is closer to Orbit 5 in integrals-of-motion space than before. Kshir is closer to a circular, in-plane orbit than both Orbit 3 (slightly radial) and Orbit 9 (slightly polar).
\end{itemize}

\section*{Section~5.1 Corrections}

\begin{figure}[b]
    \centering
    % Number shown in the caption/refs, local to this float only
    \renewcommand{\thefigure}{3, Corrected}
    \renewcommand{\theHfigure}{erratum.fig3}
    \includegraphics[width=0.5\linewidth]{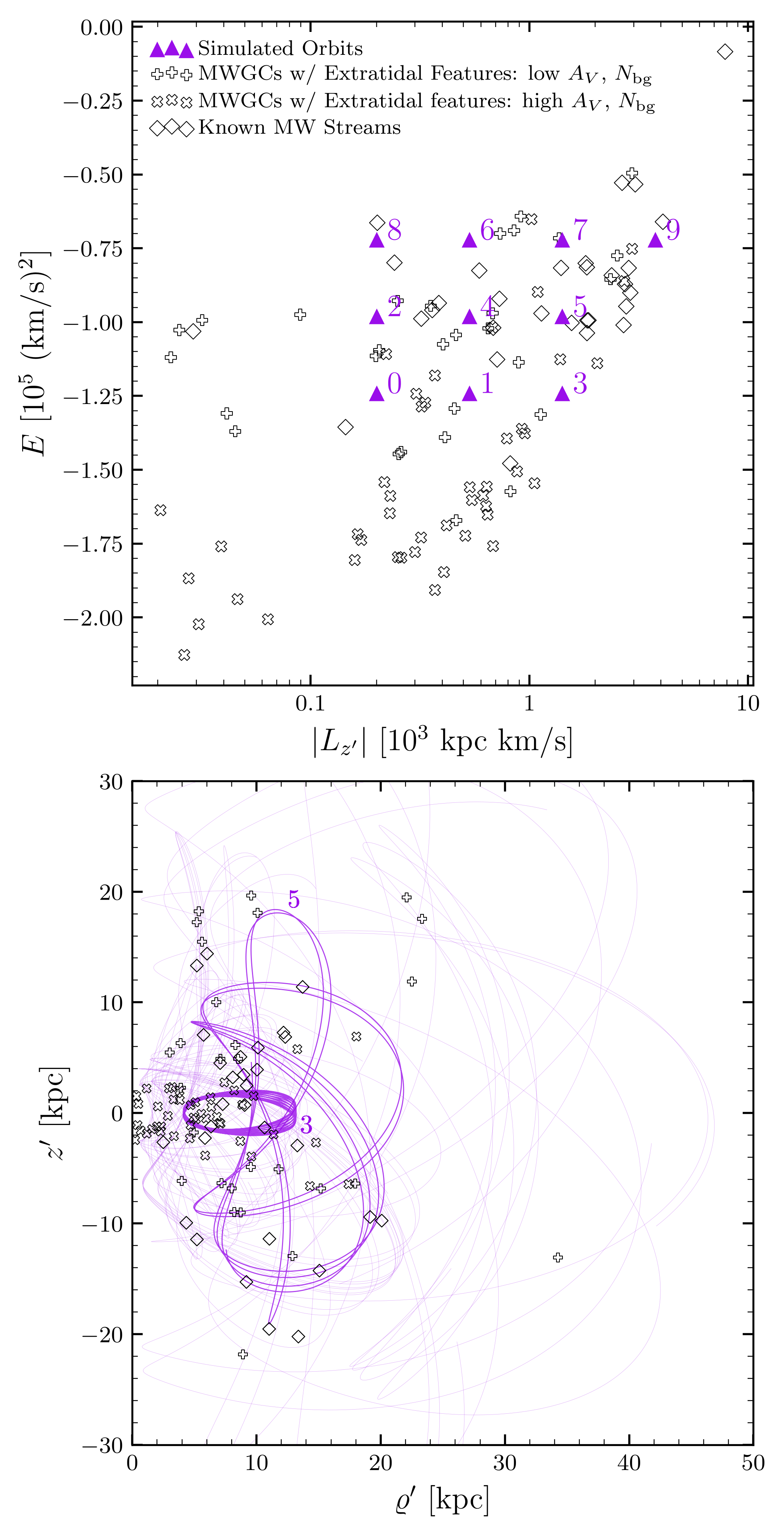}
    \caption{{\it Top panel}: The corrected integrals-of-motion distribution of Milky Way GCs \citep{vasiliev2021, chen2025c} and stellar streams
    \citep[Table~1 of][]{malhan2022}. The simulated orbits are unchanged; reaching the $E\!\lesssim\!-1.5\times10^{5}\,{\rm (km/s)^{2}}$ region was not a priority given peri/apogalacticon constraints. {\it Bottom panel}: The sampled orbits compared to the MWGCs and streams in the meridional plane, which were unaffected by the coordinate transformation errors.}
    \label{err:fig3}
\end{figure}

\begin{itemize}
\itemsep0em 
\item {\it We use a collection of {\tt MilkyWayPotential2022} sample orbits, spanning the region of orbit space occupied by known MW streams, to test how \krios\ and particle-spray models might disagree in scenarios similar to those found in the observational data.} The sample orbits were produced based on the faulty integrals-of-motion data for real Milky Way systems. In Figure~\ref{err:fig3} of this erratum, however, we show that the agreement with the corrected distribution remains satisfactory.
\item {\it The Fj\"{o}rm and M92 streams are examples of [GCs subject to strong tidal forces during their orbits, characterized by small perigalacticon and apogalacticon distances]. Differences in the stream's distribution on the sky plane are most apparent when they are near apogalacticon.} This remains true for M92, but not for Fj\"{o}rm. Streams for which this comment about sky-plane variations apply now includes Gunnthr\'{a} (small peri/apogalacticon) and M5 (small perigalacticon).
\item {\it [Systems] tightly bound to the host galaxy [lead to disagreements in how well the integrals of motion of the stream stars are conserved]. This applies to circular (M92, Gunnthr\'{a}) and radial (Fj\"{o}rm) orbits alike.} GD-1, Kshir, and Gunnthr\'{a} are closer to circular, in-plane orbits than M92, which has a comparatively small $|L_{z'}|$ value. Fj\"{o}rm is not on a radial orbit \citep[$|L_{z'}|\!\simeq \!0.6\,J_{\rm tot}$, see Figure~2 of][for comparison]{grondin2024} and the only streams with radial orbits, Hrid and {\it Gaia}-12, are beyond the scope of this study. The comment about being tightly bound to the host galaxy is robust, but the additional comment about orbit types was not properly tested.
\end{itemize}

\begin{figure*}
    \centering
    % Number shown in the caption/refs, local to this float only
    \renewcommand{\thefigure}{8, Corrected}
    \renewcommand{\theHfigure}{erratum.fig8}
    \includegraphics[width=\linewidth]{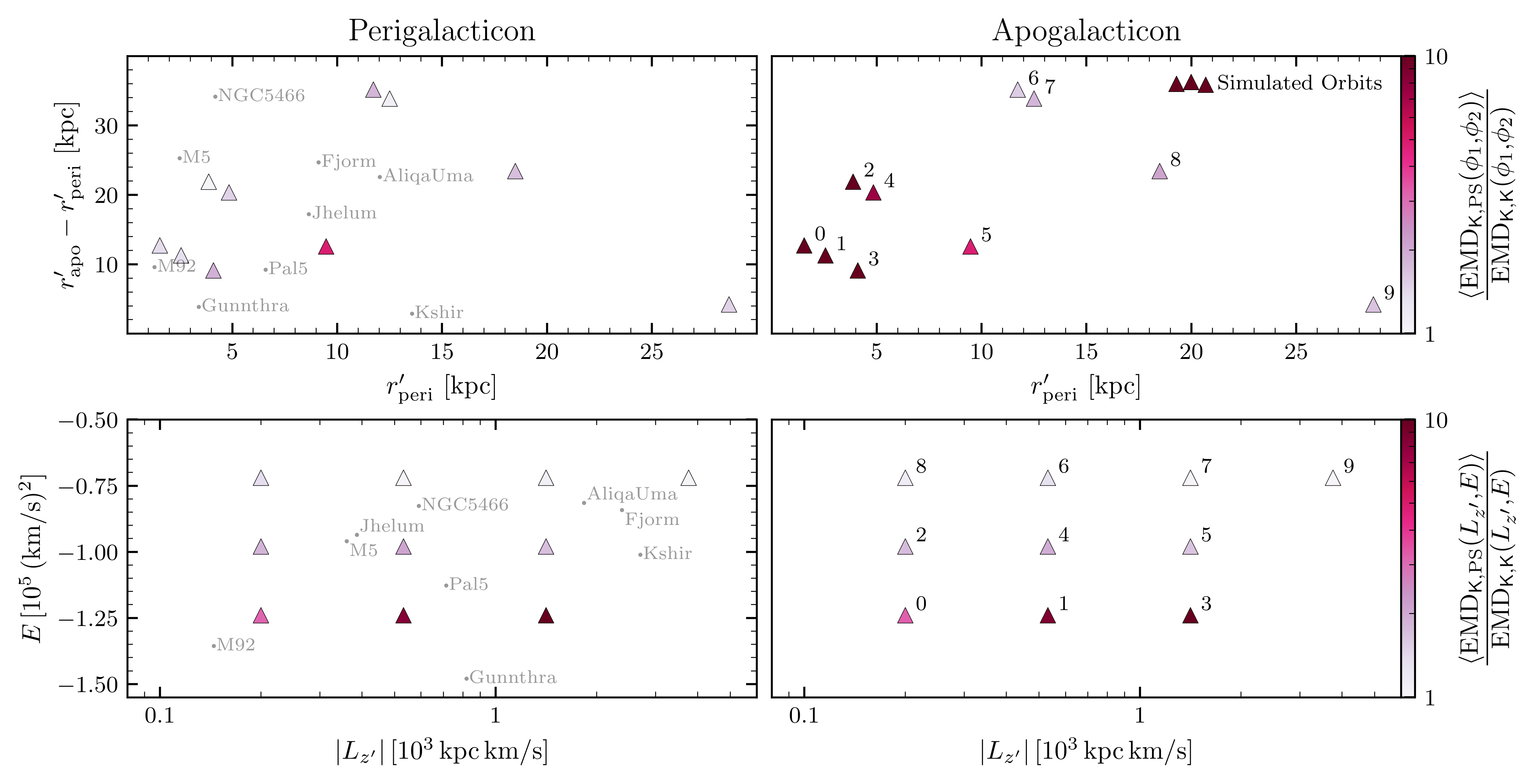}
    \caption{A corrected comparison between the MW stellar streams \citep[labeled in the left column,][]{malhan2022} and simulated orbits (labeled in the right column) in apogalacticon/perigalacticon space (top row) and integrals-of-motion space (bottom row). Each orbit is color coded by the ratio ${\langle{\rm EMD}_{{\sf K},{\rm PS}}\rangle/{\rm EMD}_{\sf{K},\sf{K}}}$, where we fix a small indexing error in the plotting script and now report the correct ratio in all cases. As before, particle spray models show poorer agreement to \krios\ when the progenitor is subject to strong tidal fields or is tightly bound to the host.}
    \label{err:fig8}
\end{figure*}

\begin{table*}[]
    \centering
    % Number shown in the caption/refs, local to this float only
    \renewcommand{\thetable}{1 Circularities, Corrected}
    \renewcommand{\theHtable}{erratum.tab1}
    \caption{The corrected table of circularity ${\varepsilon\!=\!L_{z'}/\Bigg[\varrho'\sqrt{2(E-\psi_{\rm host}(\varrho',z\!=\!0))}\Bigg]}$ values for each orbit. We use the cylindrical coordinate where a circular orbit in the disk has the same energy ${E\!=\!\half (L_{z',{\rm circ}}/\varrho')^{2} + \psi_{\rm host}(\varrho',z\!=\!0)}$, rather than the initial $\varrho'$ value from each orbit (as we did originally). Each orbit with circularity error greater than $1\%$ is in bold.}
    \begin{tabular}{lcccc}
    \toprule
    Orbit & $\varrho'_{\rm init}$ [kpc] & $\varrho'_{\rm circ}(E)$ [kpc] & $\varepsilon$, published & $\varepsilon$, Eq.~\eqref{eq:circularity} \\
    \midrule
    \multicolumn{5}{l}{Validation (\texttt{MWPotential2014})} \\
    Circular 0 Test & 5.0000000 & 5.0000000 & 1.000 & 1.000 \\
    Circular 1 Test & 20.000000 & 20.000000 & 1.000 & 1.000 \\
    \textbf{Eccentric Test} & \textbf{4.0423168} & \textbf{13.382119} & \textbf{0.2047} & \textbf{0.1188} \\
    \midrule
    \multicolumn{5}{l}{\texttt{MilkyWayPotential2022}} \\
    0 & 9.3221641 & 8.9936651 & 0.09761 & 0.09749 \\
    1 & 9.4971397 & 8.9936651 & 0.2600 & 0.2592 \\
    \textbf{2} & \textbf{20.584174} & \textbf{16.067428} & \textbf{0.06262} & \textbf{0.05762} \\
    \textbf{3} & \textbf{13.128729} & \textbf{8.9936651} & \textbf{0.9134} & \textbf{0.6894} \\
    \textbf{4} & \textbf{10.008661} & \textbf{16.067428} & \textbf{0.1743} & \textbf{0.1532} \\
    \textbf{5} & \textbf{9.5840674} & \textbf{16.067428} & \textbf{0.4727} & \textbf{0.4074} \\
    \textbf{6} & \textbf{26.136554} & \textbf{30.752659} & \textbf{0.08659} & \textbf{0.08488} \\
    7 & 33.384584 & 30.752659 & 0.2273 & 0.2257 \\
    \textbf{8} & \textbf{41.800215} & \textbf{30.752659} & \textbf{0.03692} & \textbf{0.03192} \\
    \textbf{9} & \textbf{25.671959} & \textbf{30.752659} & \textbf{0.6149} & \textbf{0.6002} \\
    \bottomrule
    \end{tabular}
    \label{err:tab1}
\end{table*}

\clearpage

\end{document}